\documentclass{aa} 

\usepackage{blindtext}
\usepackage{txfonts} 
\usepackage{color}    

\usepackage{graphicx}
\usepackage{amsmath,amssymb}
\usepackage{nccmath}

\usepackage{epstopdf}

\usepackage{epsfig}
\usepackage[usenames,dvipsnames]{pstricks}

\usepackage{grffile} 
\usepackage{hyperref} 

\usepackage{comment} 

\usepackage{natbib}
\bibpunct{(}{)}{;}{a}{}{,} 

\newcommand{\GaSPH}{\texttt{GaSPH}}
\newcommand{\HiGPUs}{\texttt{HiGPUs}}
\newcommand{\gadget}{\texttt{Gadget-2}}
\newcommand{\treesph}{\texttt{TREESPH}}
\newcommand{\fargo}{\texttt{FARGO}}
\newcommand{\omp}{\texttt{OpenMP\textregistered}}
\newcommand{\ompsemplice}{\texttt{OpenMP}}
\newcommand{\cpumia}{\texttt{Intel\textregistered Core$^{TM}$~i7-4710HQ}}
\newcommand{\rammia}{\texttt{DDR3L}}
\newcommand{\nbodysix}{\texttt{Nbody-6}}

\begin{document}
 
\title{Selfgravitating disks in binary systems: an SPH approach }

\subtitle{I. Implementation of the code and reliability tests.}

\titlerunning {}

\author{L.D. Pinto\inst{1}\inst{2} \and R. Capuzzo-Dolcetta\inst{1} \and G. Magni\inst{2}}

\institute{\inst{1} Dipartimento di Fisica, Sapienza, Universitá di Roma, P.le Aldo Moro, 5, 00185 – Rome, Italy \\ 
\inst{2} INAF-IAPS, Istituto di Astrofisica e Planetologia Spaziali, Area di Ricerca di Tor Vergata, via del Fosso del Cavaliere, 100, 00133 – Rome, Italy \\ }


\abstract { The study of the stability of massive gaseous disks around a star in a non isolated context is not a trivial issue and becomes a more complicated task for disks hosted by binary systems.
The role of self-gravity is thought to be significant, whenever the ratio of the disk to the star mass is non negligible. 
To tackle these issues we implemented, tested and applied our own Smoothed Particle Hydrodynamics (SPH) algorithm.
The code (named \GaSPH) passed various quality tests and shows good performances, so to be reliably applied to the study of disks around stars accounting for self-gravity. 
This work aims to introduce and describe the algorithm, making some performance and stability tests. 
It constitutes the first part of a series of studies in which self-gravitating disks in binary systems will be let evolve in larger environments such as Open Clusters.
\ \\ }


\maketitle 
\section{Introduction}

The study of protoplanetary disks in non isolated system has become a relevant topic in numerical astrophysics due to the recent observations of several planetary systems and disks around stars inside Open Clusters.

The recent discovery of a Neptune-sized planet hosted by a binary star in the nearby Hyades Cluster (\cite{ciardi&al18}) has opened new perspectives on the study of the evolution of primordial disks interacting with binary stars in a non isolated environment. 
Regarding the context of the study of isolated binary systems, it is worth noted that most of the classical models deal with low mass disks, with the result that, even considering their self-gravity, no appreciable change is observed in the time evolution of the star orbital parameters. 
Additionally, low-mass disks give a poor feedback on the hosting stars, which means a time scale for their orbital parameters variation which is large in comparison with the other dynamical time-scales involved.

To investigate such systems, we built our own Smoothed Particle Hydrodynamics (SPH) code to integrate the evolution of the composite star+gas system.
Following the scheme described in \cite{capuzzo&miocchi98}, our code treats gravity by means of a classical tree-based scheme \citep[see also][]{barnes86,barnes&hut86,miocchi&capuzzo02} with the addition of a proper treatment of the close gravitational interactions of the gas particles.
The evolution of a limited number of point-mass objects (which may represent both stars or planets) is treated with a high order explicit method.

Such preliminary work aims at providing an instrument suited not only to the modelization of heavy protoplanetary disks interacting with single and binary stars, but also to the study of such systems not in isolation but, rather, in stellar systems such as open star clusters. 

In the next section we describe our code, after a preliminary introduction of the numerical framework, while in section 3 we present and discuss some physical and performance tests.
In section 4 we describe the disk model adopted and the application of our code to the study of heavy keplerian disks.
Section 5 is dedicated to the conclusions.
 
 \section{The numerical algorithm}    

\subsection{Basic theory}

In order to study a self-gravitating gas, we developed an SPH code, coupled with a tree-based scheme for the Newtonian force integration.
Introduced for the first time by \citet{lucy77} and \citet{gingold&monaghan77}, the SPH scheme has been widely adopted to investigate of a huge set  of astronomical problems involving fluid systems.
An SPH scheme allows to integrate the fluid dynamical equations in a Lagrangian approach, by representing the system through a set of points, or `pseudo-particles'.
For each particle, a set of fundamental quantities (such as density $\rho$, pressure $P$, internal energy $u$, velocity $\vec{v}$) are calculated by means of an interpolation with a proper kernel function over a suitable neighbour.
For an exhaustive explanation of the method we refer to various papers in the literature as, e.g.,  those by \citet{monaghan&lattanzio85}, \citet{monaghan88}, \citet{monaghan05}.
Here we just recall some basic aspects.
Interpolations are performed with a  continuous  kernel function $W(\vec{r},h)$, whose spread scale is determined by a characteristic length $h$, called \textit{smoothing length}.
It can be easily shown \citep[see for example][]{hernquist&katz89} that under some additional constraints, interpolation errors are limited to the order $O(h^2)$.
In our work we used, as kernel function, the cubic spline adopted for the first time by \citet{monaghan&lattanzio85}, developing a formalism introduced by \cite{hockney&eastwood81}. 
This kernel function has the following form: 
\begin{equation}\label{eq_kernel}
W(r,h) = \frac{1}{\pi h^3} \cdot
\begin{cases}
~~ 1 - \dfrac{3}{2} \left( \dfrac{r}{h} \right)^2 +  \dfrac{3}{4} \left( \dfrac{r}{h}  \right)^3 , ~~~~~~ 0\leq r < h \\ \\
~~   \dfrac{1}{4} \left( 2 - \dfrac{r}{h} \right)^3 , ~~~~~~~~~~~~~~~~~~~~~ h\leq r < 2h  \\ \\
~~ 0 ~, ~~~~~~~~~~~~~~~~~~~~~~~~~~~~~~~~~~~~~~  r \geq 2h
\end{cases}
\end{equation} 
The SPH interpolation involves only a limited set of $N'$ neighbours particles enclosed within the range $2h$, thus the computational effort is expected to scale linearly with the total particle number $N$.
On the other hand, when long-range interactions, such as gravity, are considered the computational effort grows up because each particle interacts with the whole system.
A classical direct N-Body code would, hence, require a computational weight scaling as $N^2$.
However, a suitable gravitational tree-based  scheme allows to evaluate efficiently the Newtonian force by approximating the potential with a harmonic expansion \citep[see][for a full explanation]{barnes&hut86,barnes86}.
For each particle, only the contribution given by a local neighbourhood is calculated through a direct particle-particle coupling, while the contribution from farther particles is suitably approximated.
The following expressions (Eqs. \eqref{eq_sph_1a} and \eqref{eq_sph_1b}) represent the approximated potential $\Phi(\vec{r})$ and the force (per unit mass) $\vec{a}(\vec{r})  = - \vec{\nabla} \Phi(\vec {r}) $ given by a far `cluster' of particles:
\begin{equation}\label{eq_sph_1a}   
\Phi(\vec {r})  =  - \dfrac{ G M }{r}  - \dfrac{1}{2}   \dfrac{G \  \vec r \bar{\bar{\vec{Q}}} \vec{r} }{r^5}  
\end{equation}  
\begin{equation} \label{eq_sph_1b}  
\vec{a}(\vec {r})  =  - \dfrac{ G M }{r^3}   \vec{r}  \ + \ \dfrac{G \bar{\bar{\vec{Q}}}\cdot \vec{r}}{r^5}   \vec{r} \ - \ \dfrac{5}{2}   \dfrac{G \ \vec{r} \bar{\bar{\vec{Q}}} \vec{r} }{r^7}    \vec{r}  
\end{equation}  
$M$ is the total mass of such ensemble, $r = |\vec{r}|$ is the distance of the particle under study to the centre of mass of the cluster.  
The symbol $ \ \bar{\bar{\vec{Q}}} \ $ represents the so called `quadrupole tensor' associated to the specific cluster.
In indexed form, it is given by:
 \begin{equation}\label{eq_sph_1c}
Q_{ij} = \sum\limits_{k=1}^{N_C} \left( 3 x_{i}^{(k)}  x_{j}^{(k)}  - r^2_k \delta_{ij} \right) m_k 
\end{equation}
where $x_i^{(k)}$ and $x_i^{(k)}$ ($i,j=1,2,3$) refer to the Cartesian coordinates of the $k-th$ particle of mass $m_k$. 
The summation is performed over all the $N_C$ particles included in the cluster.

In the following section we are going to describe the main structure and formalism used by \GaSPH, further computational details related to the implementation of the algorithm can be found in the appendix \ref{sec_appendiceA} and \ref{sec_appendiceB}.
  
\subsection{The main structure of the algorithm}  

A single step for computing the acceleration contains two preliminary phases. 
One cycle is dedicated to  map the particles into an octal grid domain.
A further cycle, linear in $ N $, is needed to evaluate some key parameters like the density, $\rho$, the smoothing length, $h$, and pressure, $P$. 
Then a third set of operations, the heaviest one, is that of evaluating the gravitational and hydrodynamical forces, in addition to the gas internal energy rate $\dot{u}$.

\GaSPH~ can easily treat also a system made of a set of point masses, simply by turning off the part of SPH computations and using just the tree scheme for gravity interactions.
On the other hand, a gas can be treated with pure hydrodynamics, by turning off the gravitational field and using only the SPH formalism.

After the main computations of the acceleration $\vec{a}$ and the energy rate $\dot{u}$, the algorithm updates in time the velocity, the position, and gas internal energy with a 2nd order \textit{Verlet} method.
Due to the structure of the 2nd order technique, the three main computational cycles should be performed twice into a single time iteration, in order to have two estimations of $\vec{a}$ and $\dot{u}$.

In addition, the smooth particles may interact with a small number $ N _{\textup{ob}} $ of additional objects, an ensemble of point masses which mimic stars and/or planets.
Differently from the other particles, the motion of such few objects is integrated with a 14th order \textit{Runge-Kutta} method, by direct particle-to-particle N-body interactions without any approximation for the gravitational field.
Provided that $ N_{\textup{ob}} $ is sufficiently small, such operations request a little additional computational effort which scales roughly linearly with respect to the total number of points (including both the SPH particles number $ N$ and the objects number $ N_{\textup{ob}}$).
For the specific purpose of our investigation where we have $ N_{\textup{ob}} \leq 2$, there is no relevant impact on the global efficiency of the code.

 \subsubsection{Particle mapping and density computation. }  
Given a set of $N$ equal-mass points, in order to apply the multipole approximation for the Newtonian field contribution given by a `cluster', we need preliminarly to subdivide the system into a hierarchical series of sub-groups of points.
To do that, we use a classical Barnes-Hut tree-code to map the particles into an octal grid space, according to their positions.
We follow, in particular, the technique adopted by \citet{miocchi&capuzzo02}, by mapping the points through a 3-bit-based codification (see section \ref{par_2_3_2} for further details).  

Before computing the accelerations, SPH particles need a preliminary stage in which densities and  smoothing lengths are computed.
To perform a good interpolation, we need to keep, for each point, a fixed number of neighbours.
Thus, for inhomogeneous fluids, we must use a smoothing length $h \equiv h(\vec{r},t)$ which varies in space and in time. 

Individual smoothing lengths should be chosen in such a way that the higher the local number density $n= \rho/m$, the smaller the interpolation kernel radius: $h \varpropto n^{- 1/3 } $, in order to have a roughly constant number of neighbours of the given particle.
At such purpose, we adopt a commonly used prescription  \citep{hernquist&katz89,monaghan05}.
For each particle, we start from an initial guess for $h$, then we vary it until the number of particles lying within the kernel dominion reaches a fixed value $N_{0}$. 
We iterate a process in which each time the number of neighbour points, $N'$, are counted using a certain smoothing length $h_{\textup{prev}}$, then  we update the latter to a new value $h_{\textup{new}}$ according to the following formula:
\begin{equation} \label{eq_sph_2_03}  
h_{\textup{new}} = h_{\textup{prev}}   \dfrac{1}{2} \left[ 1+\left( \dfrac{N_0}{N'}\right)^{1/3} \right] 
\end{equation}

If the fluid was homogeneous, $h_{\textup{prev}}   \left(  N_0 / N' \right)^{1/3}$ would provide immediately the correct value of the smoothing length, without any further iteration.
The addend 1 lets the program perform an average of the old smoothing length, damping any excessive oscillation error due to non-homogeneities in the spatial distribution of particles. 
The iteration is stopped when a convergence is reached according to the criterion: $ |N'-N_0| \leq \Delta N$, where $\Delta N$ is a tolerance number.
At this regard, \cite{attwood&al07} investigated the acoustic oscillations of some models of polytrope around the equilibrium, by imposing a constant neighbour number $N'$ and letting $\Delta N$ vary.
They found that the fluctuation of $N'\pm \Delta N$ introduced an additional numerical noise able to break the stability of such system, giving rise to errors.
To prevent errors, the authors found that $\Delta N$ should be set to zero, which is the choice we adopt in this paper.
Moreover, they showed that the calculation of $h$ according to the iterative process illustrated above, and with $\Delta N = 0$, is equivalent to solve, for all the particles, the $2N$-equations system described by the two following equations:
\begin{equation}\label{eq_hrho}
\begin{cases}
~~ \rho_i = \sum\limits_{j=1}^{N'} m_j W(r_{ij},h_i)  \\ 
~~ h_i = \delta \left(\dfrac{m_i}{\rho_i}\right)^{1/3}
\end{cases}
\end{equation}
and to find the exact solutions of density and smoothing length $\{~\rho_i,h_i ~ | ~ i \in [1,N] ~\} $, with $\delta \approx 0.31 ~N_0^{1/3} $ a suitable constant, and $r_{ij}$ the mutual distance between the the i-th and the j-th particles.
 
We typically use a number of neighbors $N'=60$, such as $\delta \approx 1.2 $.

Once the density $\rho_i$ is evaluated, the corresponding pressure $P_i$ can be computed by means of a suitable equation of state. 
The appendix \ref{sec_appendiceB1} illustrates further technical details about the neighbour searching procedure.

\subsubsection{Force calculation and softened Interactions}\label{paragrafo_softened}  

For a generic $i-th$ particle, the acceleration $\vec{a}_i$ is computed by adding both the SPH terms and the Newtonian terms in the same iteration. 
Together with the acceleration, the particle internal energy rate is also computed.

In treating a self-gravitating gas with an SPH scheme, a proper treatment of the gravitational potential is necessary, in order to avoid an overestimation of the gravity field.
Indeed, particles can be considered as point sources of the Newtonian field as far as their mutual distance is larger than $2h$.
Otherwise, their Newtonian interaction is, in consistency with the assumed kernel function \citep{gingold&monaghan77}, such that it vanishes at inter-particle distance approaching to zero. 
 
Using the cubic spline kernel, a different form of the Newtonian interaction between two particles can be obtained, such that the classical term is softened if the particles approach within a distance of the order of a \textit{softening length} $\epsilon= 2h$.
See the appendix in \citet{hernquist&katz89} for more details, and the appendix in \citet{price&monaghan07}, for an explicit expression of the force and the potential.
When SPH interaction are turned off, a constant value $\epsilon$, in place of $2h$ is generally used for the softening length.
In such case, the total energy is conserved within numerical error. 
On the other hand, with SPH systems, due to the time variation of the softening length, the Hamiltonian becomes time dependent and so the energy is no more conserved. 
To fix this problem, equations of motion must be rewritten in a conservative form, taking into account the variation of $h$.
We follow the Hamiltonian formalism adopted for the first time by \cite{springel&hernquist02} for the hydrodynamical interactions, and further developed by \citet{price&monaghan07} for the gravitational field. 
The SPH equation assume thus the following form:
\begin{equation}\label{eq_sph_18a}
\begin{aligned}
 \dfrac{d \vec v_i}{dt} =- \sum\limits_{j}   \frac{1}{2} \left(  g_{\textup{soft}} (r_{ij},h_i) + g_{\textup{soft}} (r_{ij},h_j) \right)    \frac{\vec{r_{ij}}}{r_{ij}}   \\
 - \sum\limits_{j} m_j  \frac{G}{2} \left(  \frac{\zeta_i}{\Omega_i} \vec{\nabla_i} W(r_{ij},h_i)  +  \frac{\zeta_j}{\Omega_j} \vec{\nabla_i} W(r_{ij},h_j) \right)\\ 
-\sum\limits_{j}  m_j\left( \dfrac{P_i}{ \rho_i^2 \Omega_i} \vec{\nabla_i} W(r_{ij},h_i) + \dfrac{P_j}{ \rho_j^2 \Omega_j} \vec{\nabla_i} W(r_{ij},h_j) \right)      \\
-\sum\limits_{j}  m_j \Pi_{ij} \left[\frac{  \vec{\nabla_i} W(r_{ij},h_i) + \vec{\nabla_i} W(r_{ij},h_j) }{2}\right]  \\
+\dfrac{d \vec{v}_{i}^{[stars]}}{dt}
\end{aligned}
\end{equation}
 
\begin{equation}\label{eq_sph_19a}
\begin{aligned}
\dfrac{d u_i}{dt} =  \sum\limits_{j} m_j \left(\dfrac{P_i}{ \rho_i^2 \Omega_i} + \dfrac{1}{2}\Pi_{ij}\right)\vec{v_{ij}} \cdot \vec{\nabla_i} W(h_i)\\
\end{aligned}
\end{equation}
where the index \textit{i} refers to a generic particle, while the index $j$ in the sums refers to the generic $j-th$ particle enclosed within the range $2h_M = 2\cdot \max(h_i,h_j)$.
The term $g_{\textup{soft}}$ represents the softened gravitational force per unit mass mentioned above: it is function only of the mutual particle distance $r_{ij}$ and of the smoothing length $h$.
It tends to zero as $r_{ij} \rightarrow 0$ and assumes the classical Newtonian form $m_j G r_{ij}^{-2}$ for $r_{ij} \geq 2h$.
The operator $\vec{\nabla}_i$ represents the gradient with respect to the coordinates of the \textit{i-th} particle.
The gradient is performed over two different expressions of the Kernel $W$, with two different lengths $h_i$ and $h_j$.
The terms $\zeta_i$, and $\Omega_i $ are suitable functions which account for the variation of the smoothed Newtonian potential with respect to the softening length.
They assume, for a generic particle of index i, the following form:
\begin{equation}\label{eq_omega}
  \Omega_i = 1 + \dfrac{h_i}{3\rho_i}  \sum\limits_{j}  m_j \vec{\nabla_i} W(r_{ij},h_i)   
\end{equation}
\begin{equation}\label{eq_zeta}
  \zeta_i = - \dfrac{h_i}{3\rho_i}  \sum\limits_{j }  m_j \dfrac{\partial ~\phi_{\textup{soft}} (r_{ij},h_i) }{\partial h_i}  , 
\end{equation}
where, as for the system \eqref{eq_hrho}, the sum extends over the particles enclosed within the range $2h_i$.
In the Eq. \eqref{eq_zeta}, the function $\phi_{\textup{soft}}$ represents the softenend gravitational potential, such that $ \vec \nabla \phi_{\textup{soft}} = -g_{\textup{soft}} \vec{r_{ij}} / r_{ij} $.
The potential reaches a constant value as $r_{ij} \rightarrow 0 $ and becomes equal to the Newtonian potential for $r_{ij} \geq 2h$ (for an explicit expression see for instance \cite{price&monaghan07}).
Terms $\Omega$ and $\zeta$ are computed in the same neighbour searching iterative loop where $\rho$ and $h$ are worked out.

Only if the gas interacts with stars, in the equation \eqref{eq_sph_18a} the last term $d \vec{v}_{i}^{[stars]}/dt$ (discussed in the section \ref{sec_star_objects}) represents a non-null acceleration, accounting for the Newtonian interaction between particle $i$ and the point masses.

The function $\Pi_{ij}$, which we will discuss in the following section, characterize the well-known `artificial viscosity'.
The expression of the equation of motion \eqref{eq_sph_18a} guarantees a symmetric exchange of linear momentum between the particles.
\subsubsection{Artificial viscosity} \label{sec_SPH_artificial_viscosity}
In high compression regions, such as shock wavefronts, the velocity gradient may be so strong that two layers may interpenetrate and the hydrodynamical equations may not be integrated correctly, generating unphysical effects.
Additional artificial pressure terms are a possible cure for this problem.
In our code, we added an artificial term adopting the same classical schematization of \cite{monaghan89}, which corresponds to introducing a suitable \textit{artificial viscosity} aimed at damping the velocity gradient when two particles approach.
Practically, a viscous-pressure term $\Pi_{ij}$ is included in the equations \eqref{eq_sph_18a} and \eqref{eq_sph_19a}.
It assumes the following expression:
\begin{equation}\label{eq_3_20_tesi} 
\Pi_{ij}=
\begin{cases}
\dfrac{-\alpha \bar{c} \mu_{ij} + \beta \mu_{ij}^2}{\bar{\rho}_{ij}} 
& \ \ , \ \ \text{if} \ \ \ \ \vec{v_{ij}} \cdot \vec{r_{ij}} < 0,  \\

0  
& \ \ , \ \ \text{if} \ \ \ \  \vec{v_{ij}} \cdot \vec{r_{ij}} > 0,
\end{cases}
\end{equation}
where $\mu_{ij} = \dfrac{h \vec{v_{ij}} \cdot \vec{r_{ij}} }{r_{ij}^2 + \eta^2 \bar{h}^2}$.
The dot product $\vec{v_{ij}} \cdot \vec{r_{ij}}$ involves the relative velocity and the distance of a pair of particles $i-j$.
Only the particles which move in, for which $\vec{v_{ij}} \cdot \vec{r_{ij}}<0$, give a contribution to the artificial viscosity.
The parameter $\eta$ is a suitable term to prevent singularities when two particles get very close (we use the typical value of $\eta$ = 0.1).
The terms $\bar{h}, \bar{\rho}$ and $\bar{c}$ represent respectively the average values of the smoothing length $\frac{1}{2} \left(h_i+h_j\right)$, the density $\frac{1}{2} \left(\rho_i+\rho_j\right)$ and the speed of sound $\frac{1}{2} \left(c_{si}+c_{sj}\right)$.
We set $\beta=2\alpha$.
In this simple formulation, the artificial viscosity is activated all over the fluid; nevertheless, there are two circumstances in which it should be damped to prevent unphysical effects.
Artificial viscosity must be damped in regions where shear dominates, and where the velocity gradient is low.

Actually, when we have two shearing layers of fluid, the relative velocity between the particles leads to an approach which is `interpreted' by the artificial viscosity \eqref{eq_3_20_tesi} as a compression.
Such wrong interpretation leads the code to overestimate the strength of the viscous interaction.
To prevent false compressions, \cite{balsara95} multiplied the term $\mu_{ij}$ by a proper switching coefficient:

\begin{equation} \label{eq_3_21_tesi} 
f = \frac{ |\vec{\nabla}\cdot\vec{v}| }{ |\vec{\nabla}\cdot\vec{v}| + |\vec{\nabla} \times \vec{v}| + 10^{-4}c_{\textup{s}} h^{-1}},
\end{equation}
with the divergence of velocity and the velocity curl evaluated, for a particle of index $i$, as:
\begin{equation}\label{eq_3_22_tesi} 
\begin{cases}
\left(\vec{\nabla}\cdot\vec{v}\right)_i = \rho^{-1}_i \sum\limits_j m_j ~ \vec{v}_{ij}\cdot \vec{\nabla}_i W(r,h_i)   \\
\left(\vec{\nabla} \times \vec{v}\right)_i = \rho^{-1}_i \sum\limits_j m_j ~ \vec{v}_{ij} \times \vec{\nabla}_i W(r,h_i) 
\end{cases}
\end{equation}
We implemented the term $f$ by multiplying $\mu_{ij}$ for an average value $\bar{f} = \frac{1}{2} \left(f_i+f_j\right)$.
Further problems may arise far away from high compression regions.
In the classical formulation of $\Pi_{ij}$, $\alpha = 1 = cost.$ (like e.g. in \cite{monaghan92}). 
In such a scheme, the viscosity acts in every region with the same effectiveness, while we would expect the artificial term to be efficient just where it is needed, i.e. close to the shock fronts.
To solve such issue, we use the same formalism introduced by \cite{morris&monaghan97} and further developed by \cite{rosswog&al00} by considering, for each particle, an individual $\alpha_i$ which follows the time variation equation:

\begin{equation}\label{eq_3_23_tesi} 
\dfrac {d\alpha_i}{dt} = - \dfrac{\left(\alpha_i - \alpha_{\textup{min}}\right)}{\tau_{\textup{$\alpha$}}} + S_i,
\end{equation}
where $S_i = \max(- (\vec{\nabla} \cdot \vec{v})_i,0) \ (\alpha_{\textup{max}} - \alpha_{\textup{min}})$ represents a 'source' term, which increases in the proximity of the shock front; $\alpha_{\textup{min}}$ represents a minimum threshold value for $\alpha$, while $\alpha_{\textup{max}}$ represents its maximum.
The (increasing) rate of the viscosity coefficient is driven by a characteristic time-scale $\tau_\alpha = h_i /b c_s$ which depends on how the fluid lets the perturbations propagate through the resolution length.
The individual viscosity coefficients, $\alpha_i$ and $\alpha_j$, when referred to a generic \textit{i}-\textit{j} particle pairing, are averaged in the same way as done with the other quantities.

For a gas with $\gamma=5/3$, a good value for the $b$ coefficient can be set such that ~ $5 \leq b^{-1} \leq 10$ \citep{morris&monaghan97}.
For our tests, we set $\alpha_{\textup{max}} = 2$, $\alpha_{\textup{min}} = 0.1$ and $b^{-1} = 5$.
These are the most common values adopted in literature to face a wide class of problems involving collapse, stars merging or protoplanetary disks (see, for instance, \citet{rosswog&price07,stamatellos11,hosono&al16}).
The implementation of the artificial viscosity term (equations \eqref{eq_3_21_tesi},\eqref{eq_3_22_tesi} and \eqref{eq_3_23_tesi}), together with its form implemented in the equations \eqref{eq_sph_18a} and \eqref{eq_sph_19a},  may affect the accuracy of the code in preserving the total angular momentum.
In sect. \ref{sec_Lconservation} we will discuss how this form of viscosity, with different choices of the coefficients $\alpha_{min}$ and $b$, guarantees the conservation of the angular momentum.
\subsubsection{ Additional star objects}\label{sec_star_objects}
We calculate a direct point-to-point interaction both both for the mutual interaction between stars and to couple stars with SPH particles.
The equation of motion of a generic $p-th$ star takes the following form:
 \begin{equation}\label{eq_sph_20}
  \begin{aligned}
  \dfrac{d \vec v_{p}}{dt} = - \sum\limits_{j} \frac{1}{2} \left( g_{\textup{\textup{soft}}} (r_{pj} ,\epsilon_{p}) + g_{\textup{soft}} (r_{pj} ,h_j) \right)    \frac{\vec{r_{pj} }}{r_{pj} } \\
  - \sum\limits_{s}  \frac{1}{2} \left( g_{\textup{soft}} ( r_{ps},\epsilon_{p}) + g_{\textup{soft}} (r_{ps} ,\epsilon_{s}) \right)   \frac{\vec{r_{ps} }}{r_{ps} },
   \end{aligned}
\end{equation}
where $g_{\textup{soft}}(r,\epsilon)$ represents the Newtonian acceleration which takes the form discussed above in section \ref{paragrafo_softened}.
The force softening is accounted for the stars, too, according to a constant softening length $\epsilon_{s} = cost$. 
The gravity is thus softened when the mutual distance approaches $\epsilon_{s}$. 
The first summation is extended over all SPH particles, while the index $s$ in the second sum refers to the generic stars. 

Similarly, the equation of motion \eqref{eq_sph_18a}, referred to a gas particle $i$, contains the following sum:  
 \begin{equation}\label{eq_sph_21}
  \dfrac{d \vec{v}_i^{[stars]}}{dt} = - \sum\limits_{s}   \frac{1}{2} \left(  g_{\textup{soft}} ( r_{is} ,\epsilon_{s}) + g_{\textup{soft}}  (r_{is} ,h_i) \right)   \frac{ \vec{r_{is}}}{r_{is} }, 
\end{equation}

where, again, the index $s$ refers to the stars, and $\vec{r_{is}}$ is the distance vector between a gas particle and a star.

\subsubsection{Time integration and time-stepping}

To evolve in time the gas system, we adopt a 2nd order integration method, similar to a classical 2nd order \textit{Runge-Kutta} scheme but, at the same time, very similar to a \textit{Leap-Frog} integrator: the well-known \textit{Velocity-Verlet} method (see \cite{andersen83}, and \cite{allen&tildesley89}, chap. 3, for detailed references).
The Verlet method is based on a trapezoidal scheme coupled with a predictor-corrector technique for the estimation of $\vec{v}$ and $u$.
The  structure of such a scheme is very similar to that of classical symplectic leap-frog algorithms, although it requires two computation of the force every time iteration (see appendix \ref{appendiceA1}).
Nevertheless, the general Velocity-Verlet method applied to a gas evolution shows some advantages compared to the symplectic algorithm of same order.
Actually, like a standard Runge-Kutta method, velocity and positions are updated in synchronized steps, without the  $\Delta t/2$ shift.
Such a feature provides a good flexibility in problems approached with non uniform time-step and which involve the interaction of the gas component with other components integrated with different methods, as in our case.
Various applications of Velocity-Verlet methods in SPH schemes are found in the literature as, for example, in \cite{hubber&al13} or in \cite{hosono&al16}.

The additional point masses are `ballistic' elements, whose equation of motion needs to be integrated with a very high precision, in order to avoid secular trends which are typical of few-body gravitational problems. 
Although the SPH precision is just at 2nd order, we decided to integrate the Newtonian motion of the (few) stars and planets in the system with a 14th order Runge-Kutta method, recently developed by \cite{feagin12} through the so-called \textit{m-symmetry} formalism.
The method consists in 35 force computations per time-step and, in analogy with the well-known 2nd and 4th order RK methods, it updates the velocities and the positions by suitable linear combinations of 35 different $\vec{Kr}$ and $\vec{Kv}$ coefficients (see Appendix \ref{appendiceA2} for further details). 

For the gas, we chose the time-step $\Delta t$ following a criterion similar to the standard Courant-Friedrichs-Lewy (CFL), commonly adopted for SPH systems \citep[see for example][]{monaghan92}, added by some additional criteria.
A global time-step $\Delta t_{\textup{min}}$ can be determined by taking the minimum between the following two quantities:


\begin{equation}\label{eq_sph_22_A1}
\Delta t_{\textup{term}} = \min\limits_i \left(  \dfrac{C ~h }{c_{s i} + h_i  |\vec{\nabla}\cdot\vec{v}|_i  + \varphi \alpha_i \left[ c_{s i} + 2 \max\limits_{j}(\mu_{ij}) \right]  } ~,~ C_{\textup{u}} \dfrac{u_i }{\dot{u}_i } \right) 
\end{equation}

\begin{equation}\label{eq_sph_22_A2}
\Delta t_{\textup{dyn}} = \min\limits_i  \left( C_{\textup{a}} \sqrt{\dfrac{h_i}{a_i} }  \ ,  \ C_{\textup{d}}\dfrac{v_i}{a_i} \right),  
\end{equation}

where $c_{s}$ is the sound speed, $C$ is a coefficient whose typical value lies between 0.1 and 0.4, we usually choose 0.15. 
Moreover, $C_{\textup{u}}$, $C_{\textup{a}}$, and $C_{\textup{d}}$ are coefficients to be set $<1$.
We choose $C_{\textup{u}}=0.04$, $C_{\textup{a}}=0.15$ and $C_{\textup{d}} = 0.02$. Finally, $\varphi$ is a coefficient typically ranging from 0.6 to 1.2 (throughout this paper we will adopt  $\varphi = 1.2$).
Similarly to the control of kinetic energy variation, we control the time variation of the thermal energy $u/\dot{u}$ in a single time-step, by its limitation to a certain fraction $C_{\textup{u}}=0.04$.
The index i refers to an individual time-step $\Delta t_i$ related to a specific particle.

For the point particles phase in the system (i.e. stars or planets), we choose a characteristic time-step, $\Delta t$, defined as
\begin{equation} \label{eq_sph_22_A3}
\Delta t_{\textup{stars}} = \min\limits_{s}  \left( C_{ob} \sqrt{\dfrac{ \epsilon_{s} }{a} }   , C_d \dfrac{ v_{s} }{a_{s}} \right)
\end{equation}
where we use $ C_{ob} = 0.15 $.
The various quantities with the index $s$ of course characterize a specific star particle. 

For a homogeneous medium, the integration can be performed with a global time-step, i.e. the smallest value among gas and stars. 
Generally, the particles have different resolutions $h_i$ and different accelerations, which lead to a wide class of typical evolution time-scales. 
Thus, for some particles, the integration could be done with different $\Delta t_i$, avoiding the explicit force calculation at every time iteration, saving some computing time.
We adopt a technique implemented in several $N$-body algorithms, like, for instance, in the classical \treesph~ \citep{hernquist&katz89} or in the multi-GPU-parallelized $N$-body code \HiGPUs ~\citep{capuzzo&spera&punzo13}.
We assign to each point a time-step as a negative $2$-power fraction of a reference time $ \Delta t_{\textup{max}} =\max\limits_{i}(\Delta t_i) $ (it can be a fixed quantity or it may change periodically during the simulation). 
The particles motion is updated periodically according to their $\Delta t_i$, in such a way that, after an integration time $\Delta t_{\textup{max}}$, all of them are synchronized(further details are explained in Appendix \ref{appendiceA1}). 
Particles mapping and sorting are performed every time for every particle as well, independently of their individual time-step.
Thus, the configuration of the tree grid, together with total mass and quadrupole momentum of the boxes, are computed every single step $\Delta t_{min}$.
Similarly, at each minimum time step iteration, the gravitational interactions between gas and star are computed even during `non active' stages of the gas.
Thus, the non active SPH particles contain an acceleration splitted in two terms: one is given by a fixed non-updated hydrodynamical and selfgravitaty term, while the other is given by a constantly updated gas-star gravitational force.

In our scheme, the stars and planets do not follow an individual time-step scheme and their mutual interactions are computed every single step $\Delta t_{min}$, even in case $\Delta t_{\textup{stars}} \neq \Delta t_{min}$.
Furthermore, we force the particles close to the stars within to a tolerance distance, to be integrated every time iteration.
Practically, for a generic $i-th$ particle, we compute its distance from the stars and, furthermore, predict such distance at the following time iteration.
If such values are smaller than a tolerance of $\kappa ~ \epsilon_{ob}$ (with a constant $\kappa \geq 2 $), the particle time-step drops to $\Delta t_{\textup{min}}$. 
For our practical purposes, a small number of objects is used (in the current investigation, $N_{\textup{ob}} \leq 2$), thus, the 35-stage RK scheme turns out to require a relatively small CPU-time (less than 2\% of the total). 

In gas problems involving strong shocks, the use of individual time-steps may lead to strong errors.
Indeed, even though CFL conditions are satisfied, the strong velocity gradients may determine a great discrepancy of time-step between close particles.
Consequently, close particles may evolve with excessively different time-scales.
This may create too many asymmetries in the mutual hydrodynamical interactions, causing unphysical discontinuities of velocity and pressure.
Following the idea of  \cite{saitoh&makino09}, for each couple of neighbour particles $i$ and $j$, we limit the ratio of time-steps $\dfrac{\Delta t_i} {\Delta t_j} \leq A $.
The above investigators have shown that a good compromise is given by the choice of $A=4$, which gives good results without affecting abruptly the efficiency of the code.


\subsubsection{Approximation of gravitational field: opening criterion.}\label{par_2_3_2}
\begin{figure}   
\resizebox{\hsize}{!}{ \includegraphics{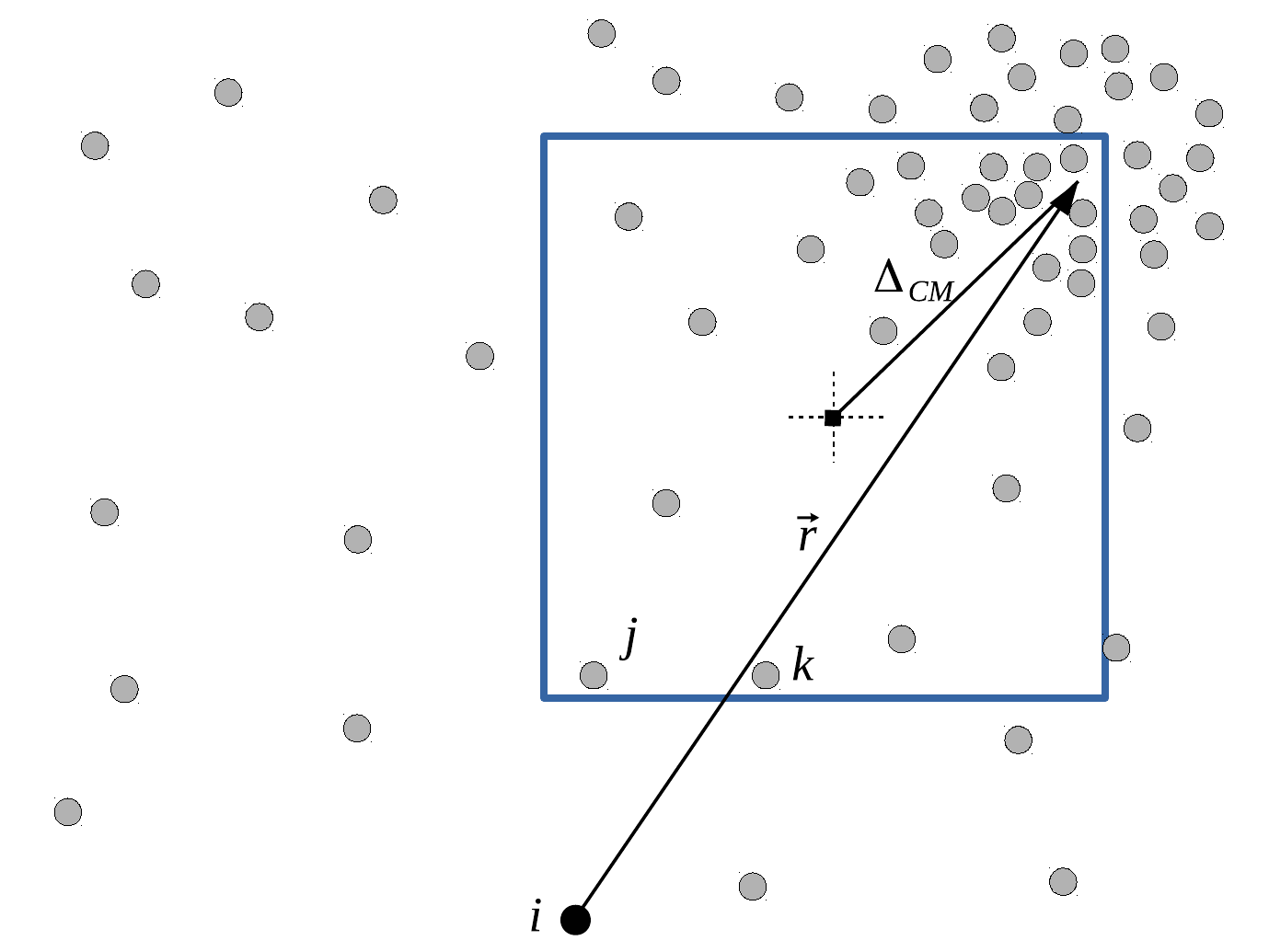} }
\caption{Schematic 2D example of lack of accuracy in the field computation, due to a large offset $\Delta_{\textup{CM}}$.
The center of gravity is faraway from the $i-th$ particle and the cube is not opened.
Nevertheless, some particles lying near the edge of the box, like $j$ and $k$, are very close to $i$, but their direct contribution is missed, resulting in a loss of accuracy.}  
\label{fig_tree01}
\end{figure}

The decomposition of the system into a series of clusters is performed, by the Tree algorithm, through a recursive octal cube subdivision of the entire ensemble.
Starting with the so-called `root box' with side length $D_0$, particle are indexed into a first L=1 subdivision order of 8 sub-boxes, each one subdivided in a further 8 sub-cubes of order L=2, and so on.
A tree structure is thus constituted, made of several nested boxes, each one containing a group of particles.
To calculate the acceleration of a particle $i$, the algorithm walks along the tree, starting from the low order cubes towards the highest-order cubes(containing just 1 particle), evaluating the distance between the particle and the center of mass of the boxes.
Each time a box is probed, the code decides to open it and probes its internal cubes only if the well-known \textit{Opening Criterion} is satisfied:
\begin{equation}\label{eq_sph_2}
 \dfrac{D_{\textup{L}}}{r} > \theta,
\end{equation}
where $\theta$ is the so-called \textit{Opening Angle} parameter for which reasonable values range among 0.3 and 1 (see section \ref{subsec_3_2}, dedicated to performance tests) and $D_{\textup{L}} = D_0 \times 2^{-L}$ is the box side length.
In the opposite case, the algorithm decides to approximate the gravitational field by adding, for to acceleration of the particle $i$, just the contribution of the box (given by the equation \ref{eq_sph_1b}).
With such a scheme,  the net amount of computation scales down to $  N \log  N  $, far less than $ N ^2 $ for large $ N $.
 
Given $\vec{r}_a$ the geometrical center of a certain cube and $\vec{r}_{\textup{CM}}$ its center of mass position, under particular circumstances we may find a very large offset $\Delta_{\textup{CM}} \equiv |\vec{r}_a - \vec{r}_{\textup{CM}} |$.
With a center of mass faraway from the box center, some errors may arise in the force approximation, since a cube can be considered `far enough' from a particle according to the Opening Criterion, despite some of the point enclosed in the box may be still very close to the particle (see figure \ref{fig_tree01}).
Thus, those close particles are ignored and the whole box gives the multi-polar approximated contribution to the particle acceleration.
The acceleration will thus be calculated with less accuracy than it would be expected.
A first key to avoid this errors should be adopted by checking whether the particle lies very close to a box \citep[like it is done for example by][]{springel05}. 
If the test particle is inside a cube, or close to its borders according to a certain tolerance, the box is always opened, independently of the truthfulness of the \eqref{eq_sph_2}.

Optionally, in our code we can furthermore modify the opening criterion by taking into account the offset term, we thus may use the following rule to open a box:
\begin{equation}\label{eq_sph_23}
  r < ~\dfrac{D_{\textup{L}}}{\theta } + \Delta_{\textup{CM}}
\end{equation}
Such prescription is equivalent to the classic opening criterion, but with an effective opening angle $\theta' < \theta$, to guarantee that every close box is opened.
In some peculiar cases in which $\Delta_{\textup{CM}}$ is large (i.e. comparable with the length of the semi-diagonal of the box), like in the example of figure \ref{fig_tree01}, the effective opening angle is considerably shorter than $\theta$.
In section \ref{subsec_3_2} we will show that, for a typical value of $\theta=0.6$, the adoption of the new criterion doesn't require too many additional computational efforts, especially for large number of particles involved.

\section{Code testing.}\label{sessione_test}

We illustrate here some basic physics tests (Sec. \ref{subsec_3_1}) and a series of performance tests (Sec. \ref{subsec_3_2}).

In sect. \ref{subsec_3_1} we apply \GaSPH~to two basic problems: (i) a non-hydrodynamical system, characterized by a cluster of point mass particles distributed according to a Plummer profile, and (ii) a classical shock-wave problem.
Such quality tests are followed by some applications to hydrodynamical systems at equilibrium.
First, we treat some polytropes with finite radius. 
Then, we compare our algorithm with a well-known hydrodynamical tree-based code (\gadget), in the case of a gaseous Plummer sphere.

In section \ref{subsec_3_2}, we analyze the computational efficiency and the accuracy of our code in different contexts.

\subsection{ Tests with gas and pressureless systems}\label{subsec_3_1}
 
\subsubsection{ Turning off the SPH: the evolution of a pressureless system}\label{par_3_plummer_test}
\begin{figure}   %
\resizebox{\hsize}{!}{ \includegraphics{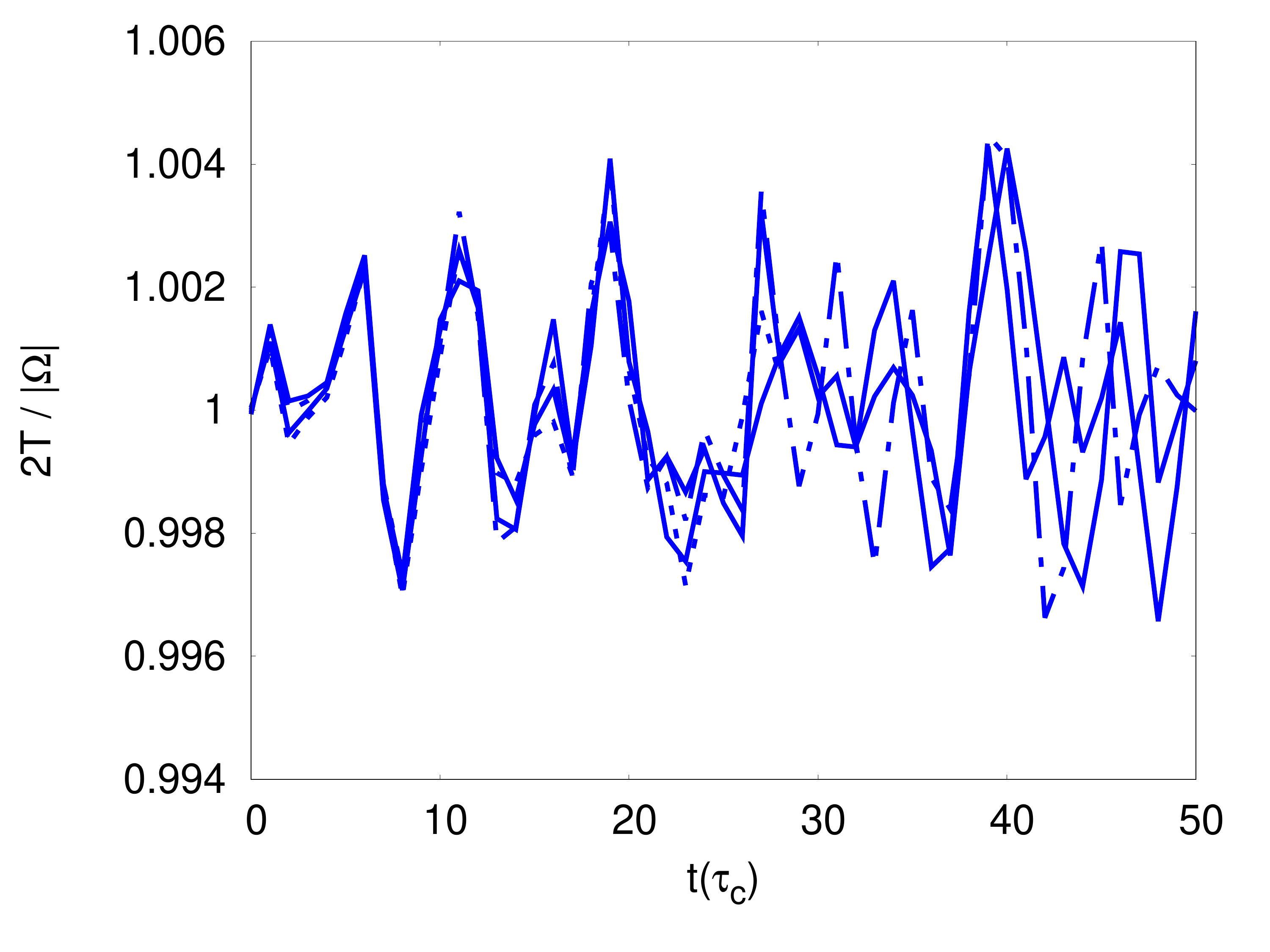} }
\caption{Oscillation of virial ratio in function of time, for different choice of code configuration parameters for the simulation of a Plummer distribution of $10^5$ equal-mass particles.
Continuous line: $\theta=0.6$, $\epsilon=0.2$; dashed line: $\theta=1.0$, $\epsilon=0.2$ ;  dotted line: $\theta=0.6$, $\epsilon=0.5$ ; dashed-dotted line: $\theta=1.0$, $\epsilon=0.5$.}  
\label{fig:fig_test01}
\end{figure}

In order to test the stability of our numerical method, we performed a series of simulations placing a set of points according to the standard Plummer configuration \citep{plummer11}, often adopted to study the star distribution in globular clusters. 
The Plummer sphere is pressureless, so the particles interact only though gravity, and the SPH interaction is missing.
Choosing, as units of measurement, the total mass $M$, and the gravitational constant $G$, we placed an ensemble of $N=10^5$ particles in a Plummer distribution with core radius $R=1$ and cutoff radius $R_{out}=10 R$.
Particles have equal masses $m= N^{-1}$ and equal softening length $\epsilon$, chosen as a fraction of the central mean interparticle distance: $\epsilon =  \alpha_{\textup{s}}  \left(\frac{m}{\rho_0}\right)^{1/3}  = \alpha_{\textup{s}}   \left(\frac{4 \pi}{3N}\right)^{1/3}$ (with $\alpha_{\textup{s}}\in [0.2 , 1.0]$). 
Starting the Plummer distribution at the virial equilibrium, we integrated its time evolution for 50 mean crossing-times $\tau_c$.
Such parameter is defined as the initial ratio between the half mass radius and the mean dispersion velocity $ \frac{R_{1/2}}{\sqrt{<v^2>}} $.
Figure  \ref{fig:fig_test01} shows the virial ratio $\frac{2T}{|\Omega|}$ in function of time, comparing four runs made by using different combinations of opening angle $\theta ~(0.6 ; 1.0)$ and $\epsilon ~(0.2 ; 0.5)$.
The four results illustrated in figure \ref{fig:fig_test01} do not show any relevant difference: virial ratios oscillates within a small fraction < 0.5\%, especially for the configuration with $\theta=1$ and $\alpha_{s}=0.5$, which was expected to be the worst case. 
 
 It is worth to note that here we are not dealing with a classical high precision N-body code as \nbodysix, for example, \citep[see][]{aarseth99} 
The Newtonian force is approximated by means of both the multipolar expansion, occurring when particles are sufficiently far, and the softening length damping, occurring when the particles approach within a distance of the order of $\epsilon$.
Despite such approximations, in a non-collisional system like our Plummer distribution, acceptable results can be obtained, lying within reasonable errors. 
 
\subsubsection{Sedov-Taylor blast wave}  

To test the code with strong shock waves, we simulated the effects of a point explosion on a homogeneous infinite hydrodynamical medium having constant density $\rho_0$ and null pressure.
If an amount of energy $E_0$ is injected at a certain point $r_0$, an explosion occurs and then a radial symmetric shock wave propagates outwards.
\cite{sedov59}  investigated such problem  and found a simple analytical law for the time evolution of the shock front:
\begin{equation}\label{eq_test_01}
r_{\textup{s}}(t) = \left( \dfrac{E_0}{\rho_0   a }   \right) ^ {1/5}    t ^{2/5}
\end{equation}
where $r_s$ is the radial position of the front, relative to the point of the explosion $r_0$, while $a$ is a function of the adiabatic constant $\gamma $ ( it is close to 0.5 for $\gamma= 5/3$, and it approaches $1$ for $\gamma=7/5$).
Furthermore, the fluid density right behind the shock front ($r<=r_s$) has the following  radial profile:
\begin{equation}\label{eq_test_02}
\rho(r,t) = \dfrac{\gamma+1}{\gamma-1} \rho_0   G_{\gamma}\left( \dfrac{r}{r_{\textup{s}}} \right)  
\end{equation}
being $G_{\gamma}$ an analytical function of the relative radial coordinate $r / r_{\textup{s}}$.
Similarly as many previous works (see for example \cite{rosswog&price07} or \cite{tasker&al08}), we set the initial conditions for a homogeneous and static medium ($\rho_0 = 1$, $v=0$) by placing $10^6$ equal-mass particles in a cubic lattice structure, confined in a box with x,y, and z coordinates ranging, each one, from -1 to 1.
$\gamma$ was set to $5/3$, and the explosion was simulated by giving an amount of energy $E_0=1$ to the origin of the system.
Actually, we could not reproduce a point explosion with an SPH system, since its spatial resolution is determined by the kernel support.
Hence, in our case, we needed to inject the energy in a small region with the same scale as $2h$.
We thus gave, at a time $t_{*} $, the energy $E_0$ to those particles enclosed in a sphere having radius $R=2h$.
\begin{figure}   %
\resizebox{\hsize}{!}{ \includegraphics{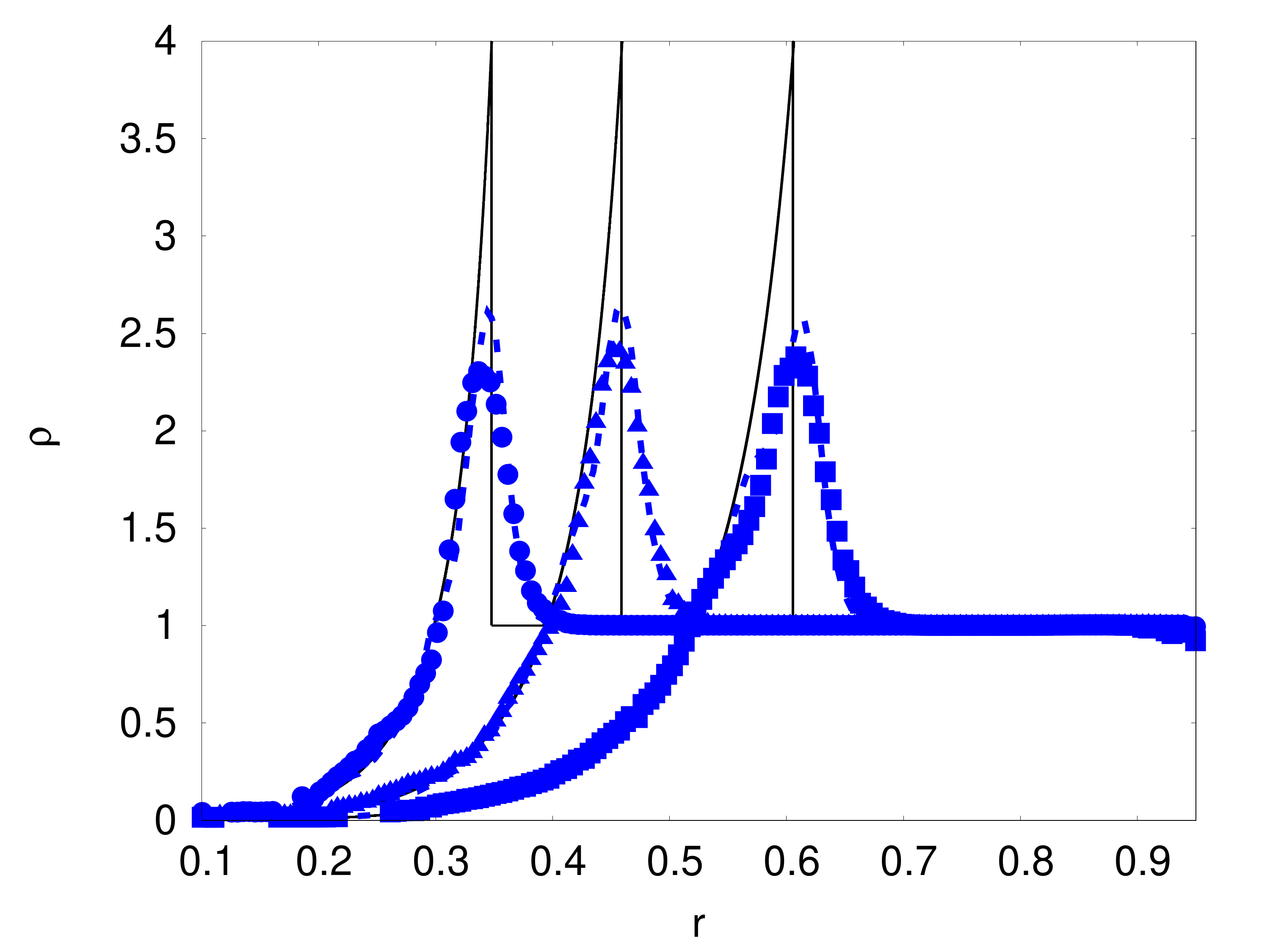} }
\caption{
 Radial density profiles of the Sedov-Taylor blast wave, at several times (increasing rightward): t= 0.05 (circles), t=0.1 (triangles), t=0.2 (squares), in a simulation with $N=10^6$.
Results obtained using a higher resolution ($N=3,375,000$) are plotted with dotted lines.
The full lines represent the classical Sedov-Taylor auto-similar solutions. 
} 
\label{fig:fig_test03}
\end{figure}
Figure \ref{fig:fig_test03} shows three different radial average density profiles $\rho(r,t')$, corresponding respectively to the time t=0.05, t=0.1 and t=0.2.
The results are compared with the analytical solution.
Despite the position of the front follows the expected law \eqref{eq_test_01}, the peak does not reach the expected value $\frac{\gamma +1}{\gamma-1}\rho_0 = 4 \rho_0$. 

Intrinsic errors in approximating the physical quantities, given by the smoothing kernel, let the density spread out and follow a wider distribution than the true profile.
 This corresponds to a smoothing of the vertical discontinuity and so a lower peak of the density.
The same figure shows a comparison with results obtained from a further test, made with the same system but using a better resolution ($N=3,375,000$).
As can be seen, the peak of the curve reaches an higher value indeed. 

\subsubsection{{Polytropes at equilibrium}}  

We tested our code in the case of hydrodynamic self-gravitational systems by building static polytropes with different indexes (n=1, n=3/2, and n=2). 
A generic polytrope of index $n$ constitutes a radially symmetric system whose equation of state follows the expression: 
\begin{equation}\label{eq_stato_politropica}
P(r) = K_n ~\rho(r)^{1+\frac{1}{n}}
\end{equation}
{where the density is parametrized as $\rho(r)/\rho_0 = \theta^{~n}(r)$, being $\rho_0$ the central value.
The static radial solution $\theta(r)$  can be found by writing an equilibrium condition between the hydrostatic pressure gradient and the gravitational forces, from which it can be obtained the well-known Lane-Emden equation (an exhaustive treatment can be found, for example, in \cite{chandrasekhar58}):}
\begin{equation}\label{eq_lane_emden}
\dfrac{\alpha^2}{r^2} \dfrac{d}{dr} \left( r^2 \dfrac{d\theta}{dr} \right) = -\theta^{~n}
\end{equation}
{with $\alpha^2 = (n+1) ~K_n ~\rho_0^{\frac{1}{n}-1} / 4\pi G$, and $K_n$ a suitable normalization coefficient. 
For an index $ n \in (0,5) $, the system has a finite radius and the coefficient $K_n$ depends, through $\alpha^2$, both on the radius R and on the total mass M.
We set both them to 1 in our tests, implying $K_1 \approx 0.637$ , $K_{3/2} \approx 0.424 $ and $K_{2} \approx 0.365 $.}

{We tested the ability of our code to let a system spontaneously relax in a polytrope configuration, following the prescription adopted in \cite{price&monaghan07}.
Starting from a homogeneous sphere of particles placed in a lattice structure, the system was let evolve by forcing the pressure to follow the equation \eqref{eq_stato_politropica}.
We forced the SPH system to evolve by damping the velocities with an additional acceleration $\vec{a}_{damp} = -0.05~ \vec{v}$, until the kinetic energy decreases down to a small fraction (1\%) of the total energy.
A standard non constant $\alpha$ was chosen for the artificial SPH viscosity, with $\alpha_0 = 0.1$, and a number of neighbours of 110 was set for the particles.}

\begin{figure}   %
\resizebox{\hsize}{!}{ \includegraphics{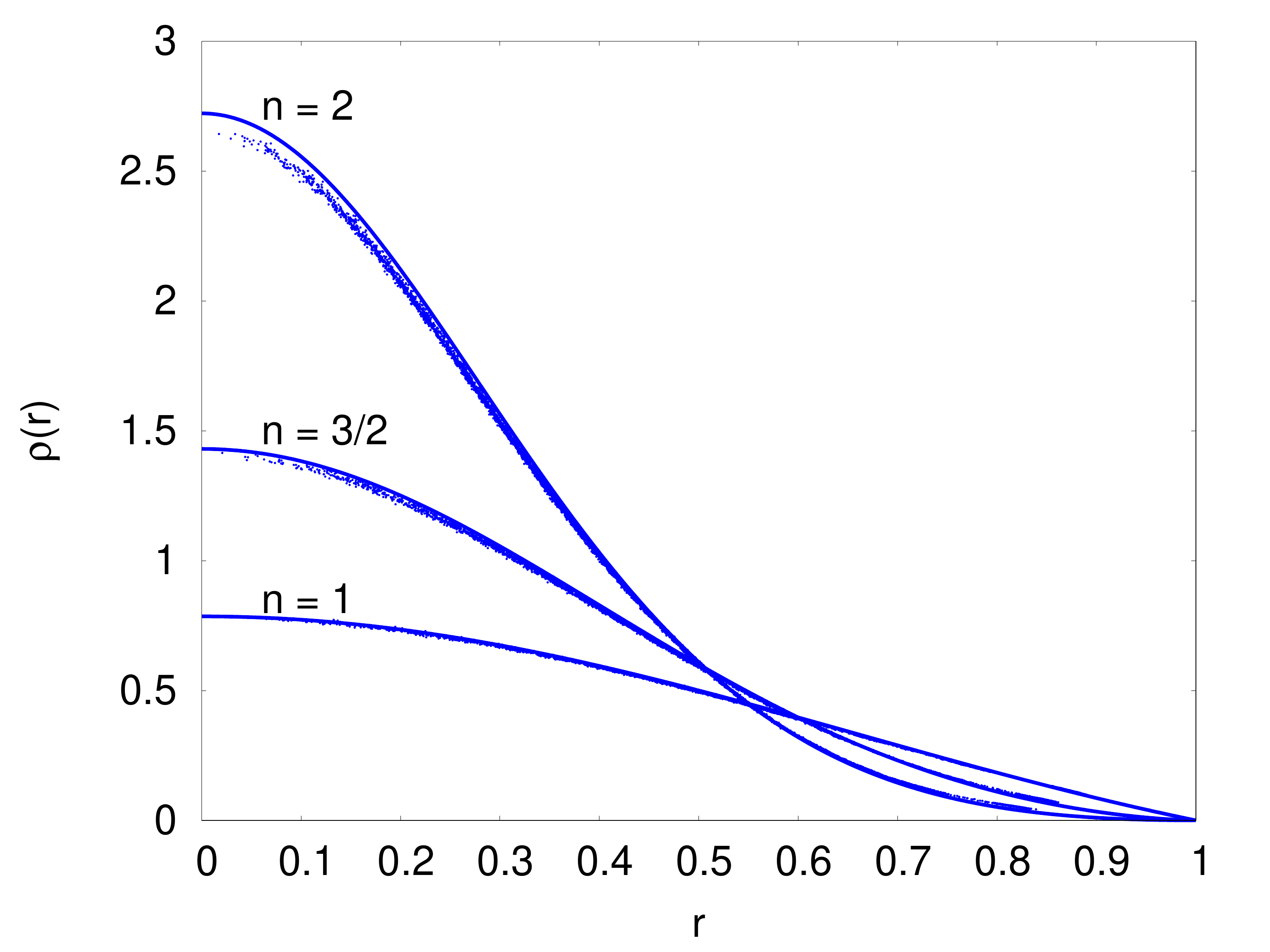} }
\caption{  
Equilibrium analytical solutions of the density profiles $\rho(r)$ related to three different models of polytropes, respectively with index $n=1$, $n=3/2$, $n=2$, drawn as solid curves.
Comparison with the computed profiles (dots). 
}
\label{fig_test_politropiche}
\end{figure}

{A correct treatment of self-gravity and hydrodynamic interactions among SPH particles, and the choice of equation of state \eqref{eq_stato_politropica}, allows the system to acquire the density profile $\rho_0\theta^{~n}$, solution of the equation \eqref{eq_lane_emden}.
Figure \ref{fig_test_politropiche} shows the three radial density profiles obtained for the different polytropic indexes.
The resolution, related to the particles number, affects the accuracy of the code in sampling correctly the profile $\rho(r)$, especially in the central denser regions.
Mainly for the higher index n=2, a higher particle number is needed to let the numerical density approach the theoretical expected value at a specific accuracy level.
10,000 particles have been used for the models with n=1 and n=3/2, while the polytrope with index n=2 has been built with 20,000 particles.}
 
\subsubsection{  {Gaseous Plummer distribution}}\label{sec_politropica_plummer}  
\begin{figure}   %
\resizebox{\hsize}{!}{ \includegraphics{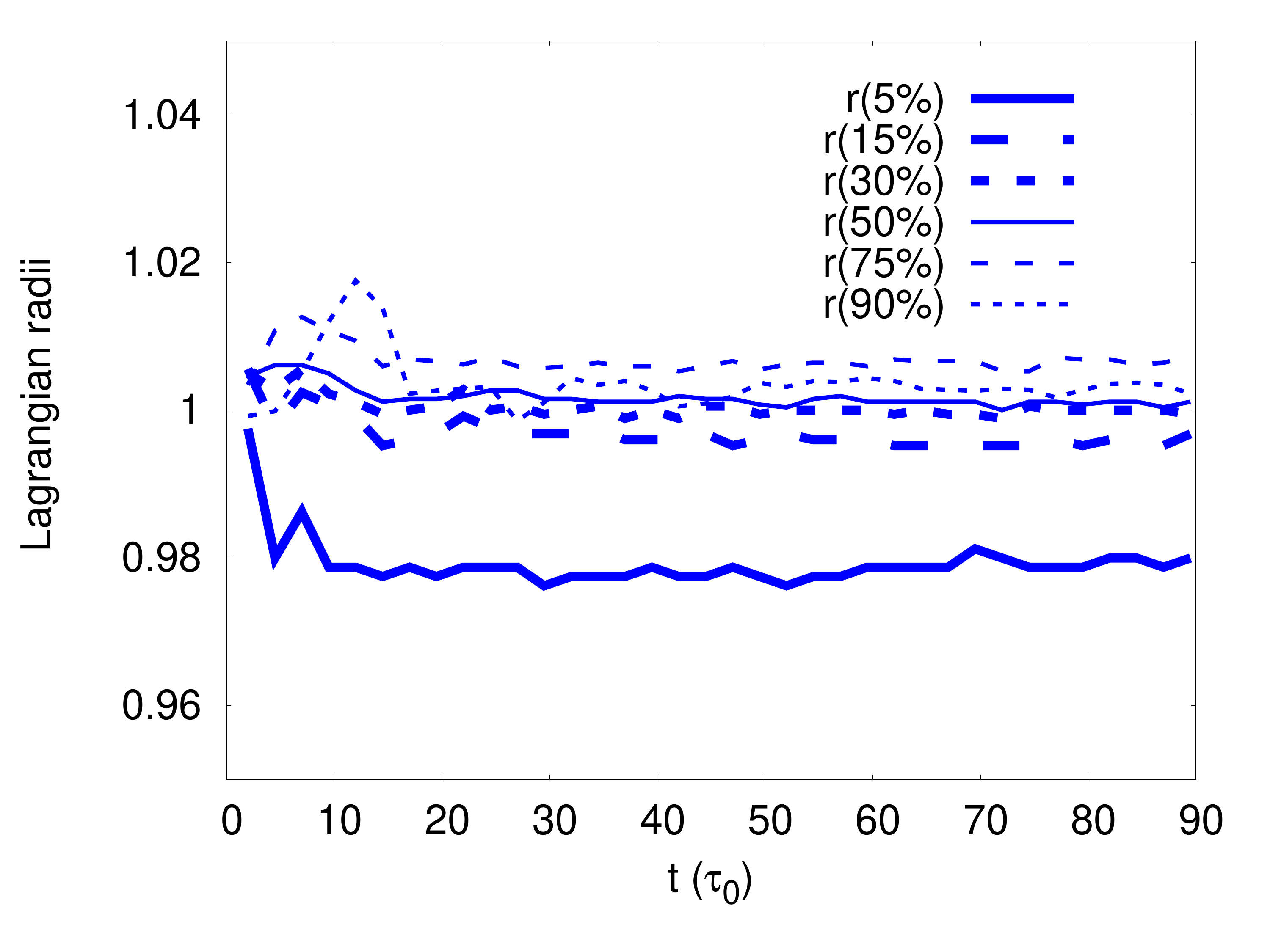} }
\caption{  Lagrangian radii in function of time for a hydro-Plummer distribution at equilibrium (see sect. \ref{sec_politropica_plummer}). The radii are normalized to their respective initial values.}
\label{fig_test_rlag}
\end{figure}

{We tested the equilibrium of a static gas density distribution according to the Plummer function:}
\begin{equation}\label{eq_rho_plummer}
\rho(r) = \rho_0 \left[ 1 + \left( \dfrac{r}{a} \right)^2 \right]^{-5/2}
\end{equation}
{with the central density defined by $\rho_0 =  3M / 4\pi a^3$, being $M$ and $a$ are respectively the total mass of the system and a characteristic length.}
{We set $M=1$ and $a=1$ ($G=1$ in code internal units) such as the half mass radius of the system turns out to be $r(50\%) \approx 1.3$.
In a static configuration with a null velocity field, the gas SPH particles compensate the mutual self-gravity with a pressure gradient resulting from a temperature distribution $T(r) = \kappa ~ \rho(r) ^ {1/5} $.
$\kappa$ represents a constant calibrated taking into account both the equation state of a perfect gas $P=(\gamma - 1) \rho u $ and imposing the Virial equilibrium between gravitational energy W and total thermal energy $U = \sum\limits_i u_i m_i$, i.e. $|W| = 2U$.  
We placed 50,000 particles according to a Montecarlo sampling of the distribution \eqref{eq_rho_plummer}.
A realistic distribution has an infinite radius, thus, a cut-off was used at a proper radial distance $r\approx22$, such that the distribution contained the 99.8 \% of the mass of a realistic infinite-extended Plummer sphere.
Figure \ref{fig_test_rlag}  shows the time variation of some Lagrangian radii, containing respectively the 5\%, 15\%, 30\%, 50\%, 75\%, 90\% of the total system mass, for an integration time of 90 central free-fall time-scales ( $  \tau_0 = \left( 3\pi ~/~32 G \rho_0\right)^{1/2} \approx 1 $ in our code units).}
 
\begin{figure}  %
\resizebox{\hsize}{!}{ \includegraphics{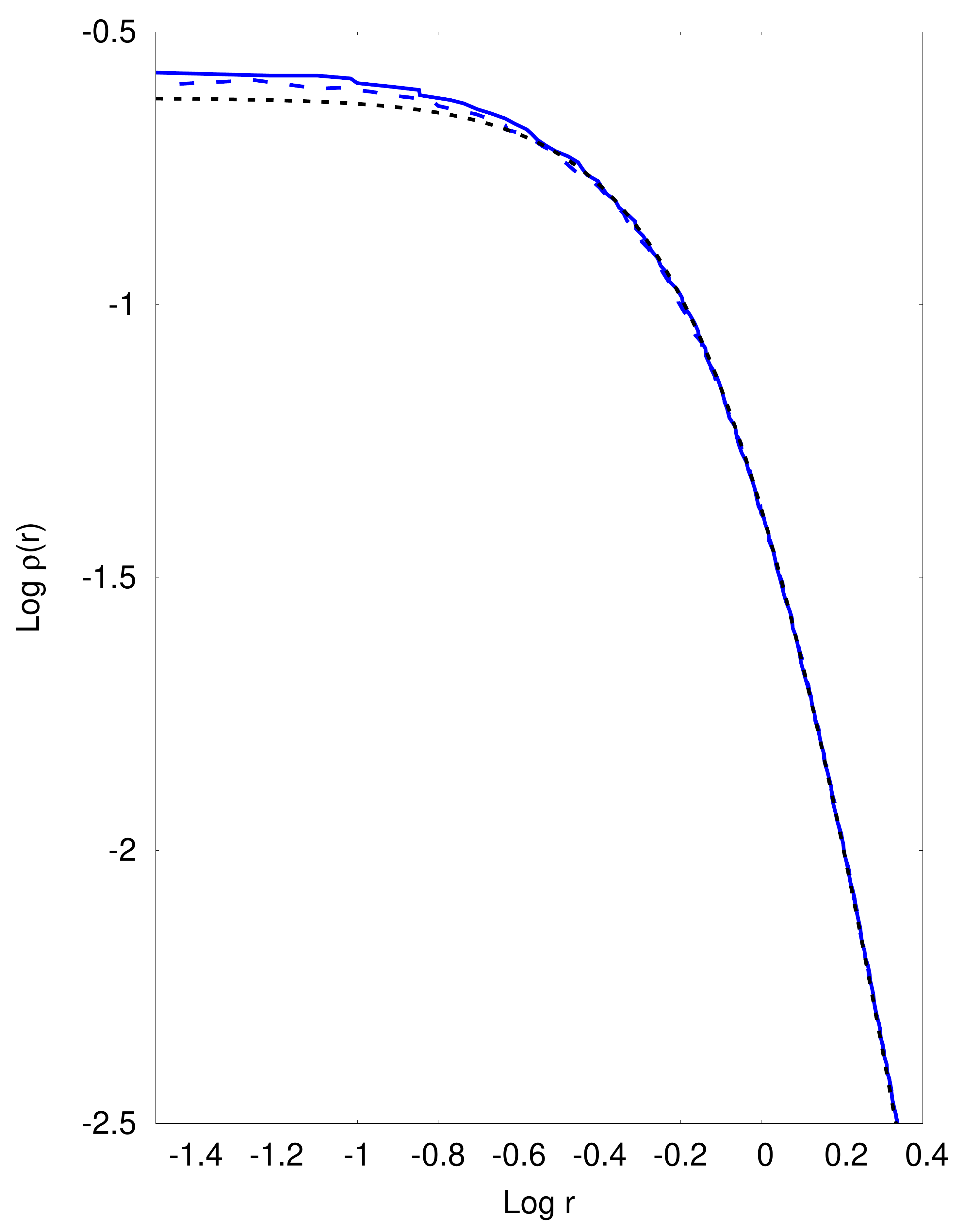} }
 \caption{ 
 Radial density profile for a Plummer distribution with 50,000 SPH particles (solid line). 
 The results obtained with \gadget~are also shown (dashed line). 
 The analytical Plummer profile is plotted with a dotted line.
 Density is in units such that $\rho_0=3/(4\pi)$.
 }
\label{fig_test_plummerrho_A}
\end{figure}

\begin{figure}  %
\resizebox{\hsize}{!}{ \includegraphics{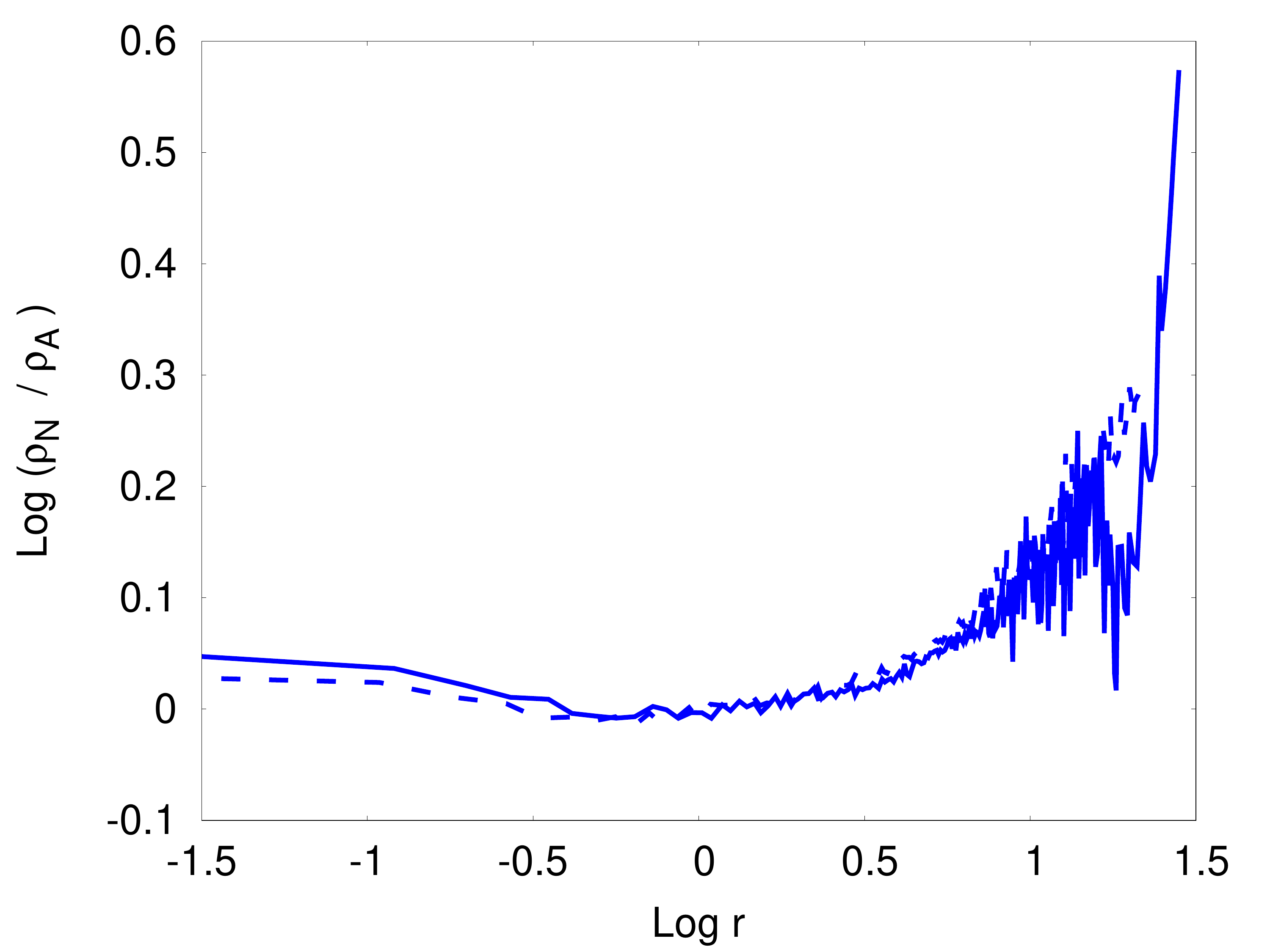} }
 \caption{Logarithmic ratio $\rho_N / \rho_A$ of the numerical to analytical radial density profile, for a Plummer distribution with 50,000 SPH particles (solid line).
 The results obtained with \gadget~are also shown (dashed line).}
\label{fig_test_plummerrho_B}
\end{figure}

We compare the results with the well-known gravitational SPH code \gadget ~(\cite{springel05}).
Figures \ref{fig_test_plummerrho_A} and \ref{fig_test_plummerrho_B} show a comparison between the radial density profiles obtained by the two algorithms with the same choice of the main parameters.
The $\alpha$ viscosity coefficient was set constant and equal to 1. 
The density reported is computed at $t=90~\tau_0$, despite the system reaches an acceptable equilibrium state already within few units of $\tau_0$, after several slight oscillations.

Observing the figure \ref{fig_test_plummerrho_B}, we can distinguish three main radial zones, respectively for $r \leq 0.3$, $0.3 < r < 2$ and $r\geq2$. 
In the middle zone, the codes are in good agreement, providing a density profile with an accuracy less than 2\% with respect to the analytical model. 
For $r<0.3$, \gadget~ describes a density which deviates up to the 6\% from the expected value, while our program has a maximum deviation of 11\%. 
For both models, such higher errors can be ascribed to the fact that the system contains only about the 2\% of the total mass (and thus the 2\% of the total particles) within the radial distance $r=0.3$, and consequently to a poor sampling of the potential inside the sphere.  
Consequently, the system tends to shrink slightly.
In the outward zone, the deviations can reach significantly higher values with both codes, due to the extremely low values of the density with respect to the central zone.
So we can conclude that, in the context of a standard physical environment, the two codes show, on whole, a satisfactory agreement. 
\subsubsection{ Artificial viscosity and angular momentum conservation}\label{sec_Lconservation}      
A non-constant artificial viscosity may lead to a non-conservation of the angular momentum, $\vec{L}$.
The actual conservation of this quantity has been checked by letting evolve a system similar the one described in the previous section, with different settings of the artificial viscosity parameters in the equation \eqref{eq_3_23_tesi}.
We use the same plummer distribution (see Eq. \eqref{eq_rho_plummer}), with $M=1$ and $a=1$, made of 50,000 SPH particles.
The same thermal energy profile was adopted but scaled down by a factor 1/2, so that $T(r) = \frac{\kappa}{2} ~ \rho(r) ^ {1/5} $.
We converted the (subtracted) thermal energy into kinetic energy, by assigning to each $i-th$ particle a clockwise azimuthal velocity, with absolute value $v_i = \sqrt{u_i}$ (where $u_i$ was the original specific thermal energy characterizing the Plummer system used in sect. \ref{sec_politropica_plummer}), and direction parallel to the X,Y plane.
The system thus acquires a non zero vertical component of the angular momentum, $L_z = \sum\limits_i m_i \left( x_i v_{yi} - y_i v_{xi}\right)$.
The virial equilibrium is still formally preserved because gravitational potential energy and thermokinetic energy keep such to give $|W| = 2(K + U)$, but the (new) angular rotation triggers changes in the density distribution.
We integrated in time for about 100 initial central free-fall timescales $\tau_0$ (which is of the same order of the azimuthal dynamical timescale, taken as the ratio $r_c / v(r_c) \approx 1.4$, being $r_c$ the initial radius at which the density drops by a factor $1/2$).
\begin{figure}   %
\resizebox{\hsize}{!}{ \includegraphics{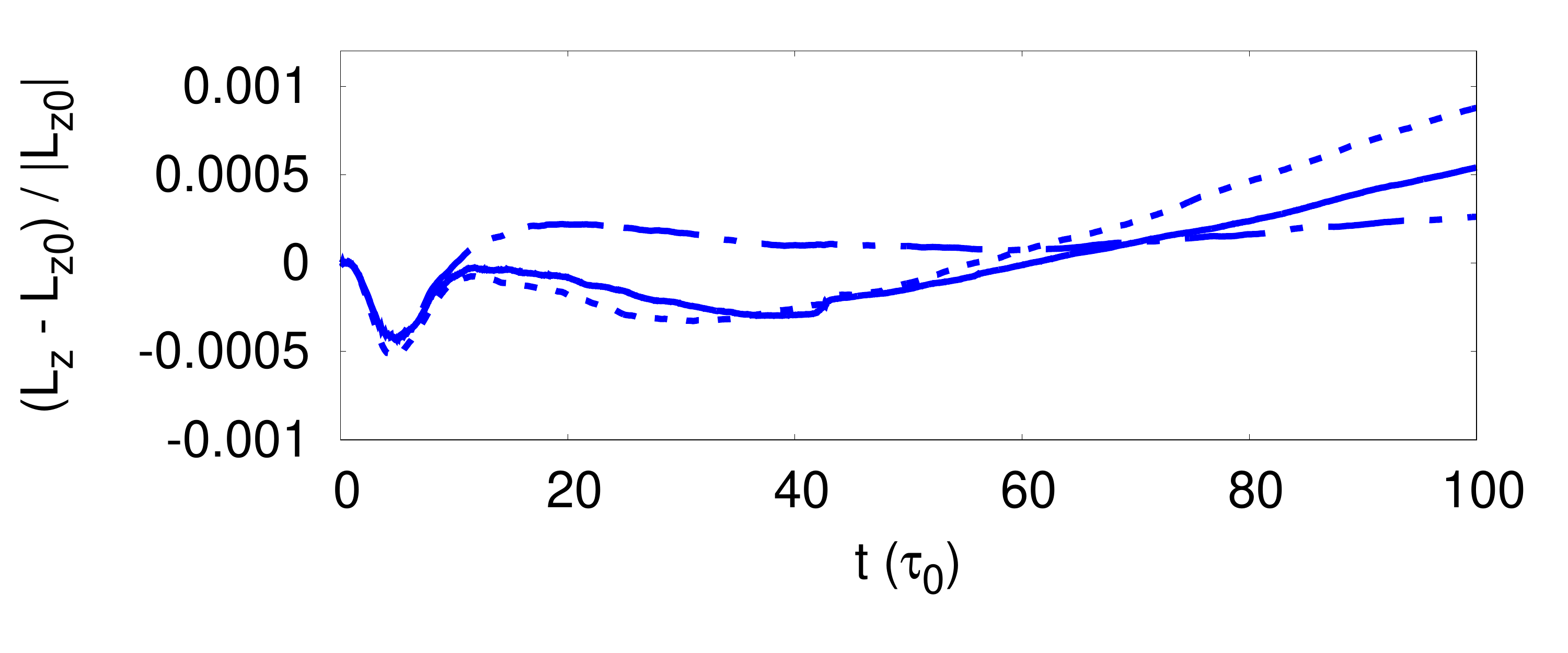} }
\caption{ 
Fractional variation of the $z$-component of the angular momentum for our simulated, rotating Plummer model (see sect. 3.1.5).
Results at varying some configuration parameters are plotted: $\alpha_{min}=0.1, b^{-1} = 5$ (full line), $\alpha_{min}=0.02, b^{-1} = 5$ (dotted line), $\alpha_{min}=0.1, b^{-1} = 7$ (dashed-dotted line).
}\label{fig_alpha_L}
\end{figure}
We performed three different simulations by varying, in the $\alpha$ rate equation \eqref{eq_3_23_tesi}, the parameters $\alpha_{min}$ and $b$.
Respectively, we set $\alpha_{min}=0.1, b^{-1} = 5$, $\alpha_{min}=0.02, b^{-1} = 5$, $\alpha_{min}=0.1, b^{-1} = 7$.
The angular rotation changes the configuration of the system leading the initial Plummer density distribution to get flatter perpendicularly to the z-axis, while the whole system expands.
During an initial phase of the order of $20 \tau_0$ the distribution undergoes some rapid variation followed by a slow, secular, evolution. 
Figure \ref{fig_alpha_L} shows the quantity $\left(L_z - L_{z0}\right) / |L_{z0}|$ which represents the variation, in function of time, of the component $L_z$ compared to its initial value $L_{z0}$.
The three lines refer to the different choices of the parameters $\alpha_{min}$ and $b$.
The curves show a conservation of the angular momentum within $10^{-3}$ up to 100 evolution timescales.
In particular, the choice of $\alpha_{min}=0.1, b^{-1} = 7$, compared with the other configurations, gives a larger variation of $L_z$ during the first phases, while it shows a smaller rate of change during the secular evolution of the system.
On the other hand, a small value $\alpha_{min}=0.02$ gives rise to a better conservation in the initial phases and a higher deviation during later stages.
We have observed, in all the three simulations, that the two components $L_x$ and $L_y$ keep negligible values compared to $L_z$, within a relative error of $10^
{-3}$.
\ \\

\subsection{Code performance} \label{subsec_3_2}
In order to analyse the computational efficiency of our algorithm in function of the particle number, we performed several tests by measuring the average CPU-time spent, for a single run, by the main routines.
So, we have studied the performances of the \GaSPH~ code in three different contexts:

1) System with pure self-gravity and zero pressure, adding a comparison with the results of \gadget.

2) System with self-gravity and SPH pressure.

3) System similar to that of case 2) but with the addition of 20 point star-like external objects.

We performed the tests by placing a set of $N$ particles with the same Plummer density profile distribution as adopted in section \ref{par_3_plummer_test}.  
The program has been tested on an \cpumia ~architecture with 6MB of Cache memory, and with 16GB of RAM-memory \rammia~with a data transferring speed of 1600 MHz.

\begin{figure}     
\resizebox{\hsize}{!}{ \includegraphics{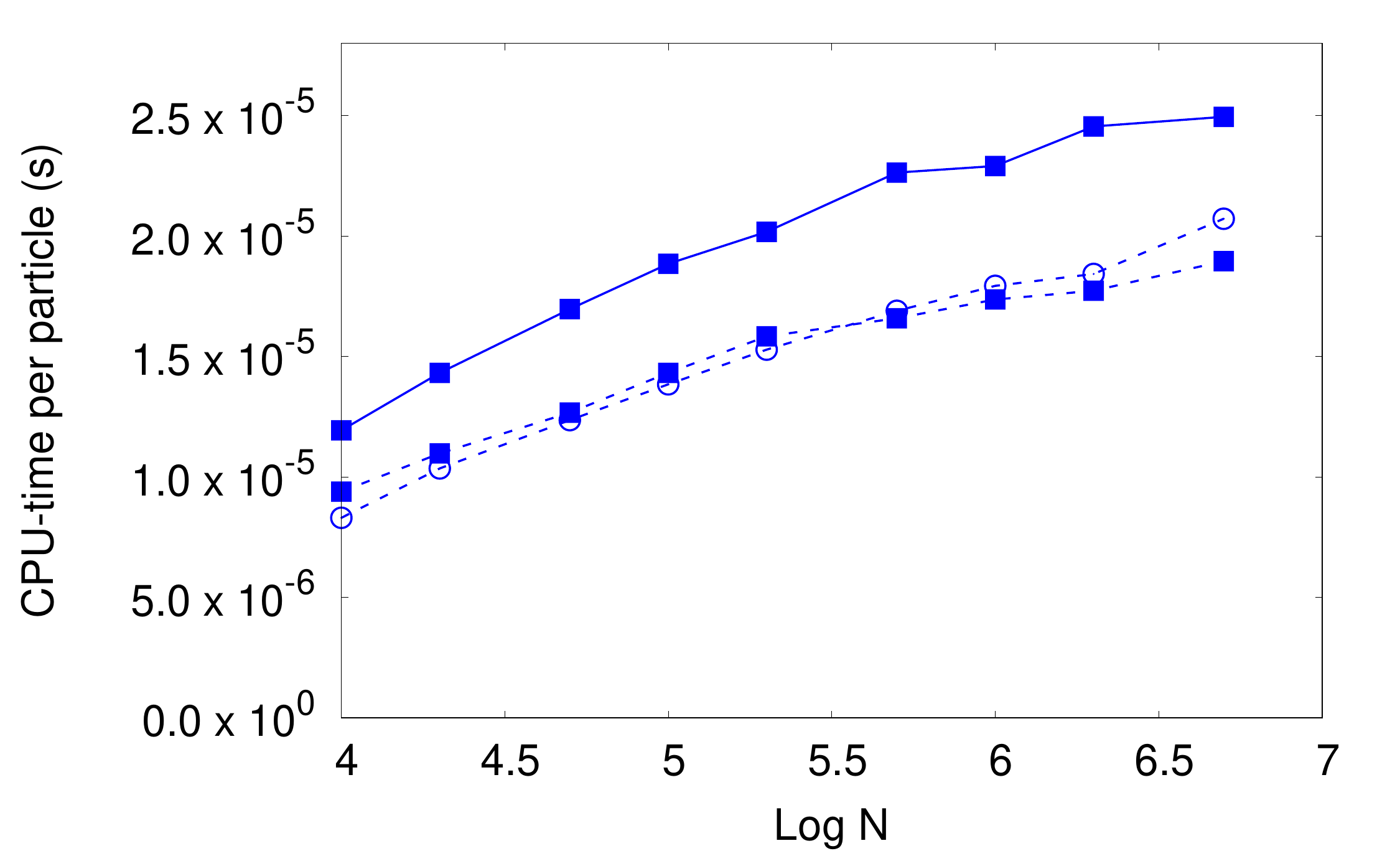} }
 \caption {CPU-time per particle for the pure gravitational force calculation in monopole and quardupole approximations, at different $N$.
 \gadget ~ results (empty circles) are compared to our code results (squares connected with  dashed lines) in the same monopole approximation.
Continuous line refers to the performance of \GaSPH~with the quadrupole term included in the field.}
\label{fig3_1A}
\end{figure}

For a standard tree-code without SPH, the computational time per particle is expected to be linear in $\log N$, since the overall time scales as $ N \log N$.
To increase the efficiency and save considerable memory resources, we can also use a simple formalism made by considering only the first `monopole' term $-MG r^{-3} \vec{r}$ appearing in the right member of the equation \eqref{eq_sph_1b}.
This is a technique, adopted also in \gadget, that simplifies the complexity of the algorithm by neglecting the efforts for the quadrupole tensor computation.
The suppression of the quadrupole term decreases the computational time, with a minor cost in terms of accuracy.
Considering a pure self-gravitating system, figure \ref{fig3_1A} shows the CPU-time needed for a single particle force calculation, in function of the ten-based logarithm of the particle number $N$ (ranging from $10^4$ to $5 \times 10^6$).
Choosing an opening angle $\theta=0.6$, we performed a series of force evaluation by considering a simple pressure-less system, with particles interacting only with the Newtonian field.
The computational times measured by using our code (averaged over a reasonable number $\geq 30$ of equal tests) show to be comparable with the average CPU-times measured by using \gadget. 
The figure shows also the results based on a second series of runs with \GaSPH~performed with the quadrupole term included in the gravitational field.
Including such term, an additional CPU-time of the order of 30\% is requested.  
\begin{figure}  
\resizebox{\hsize}{!}{ \includegraphics{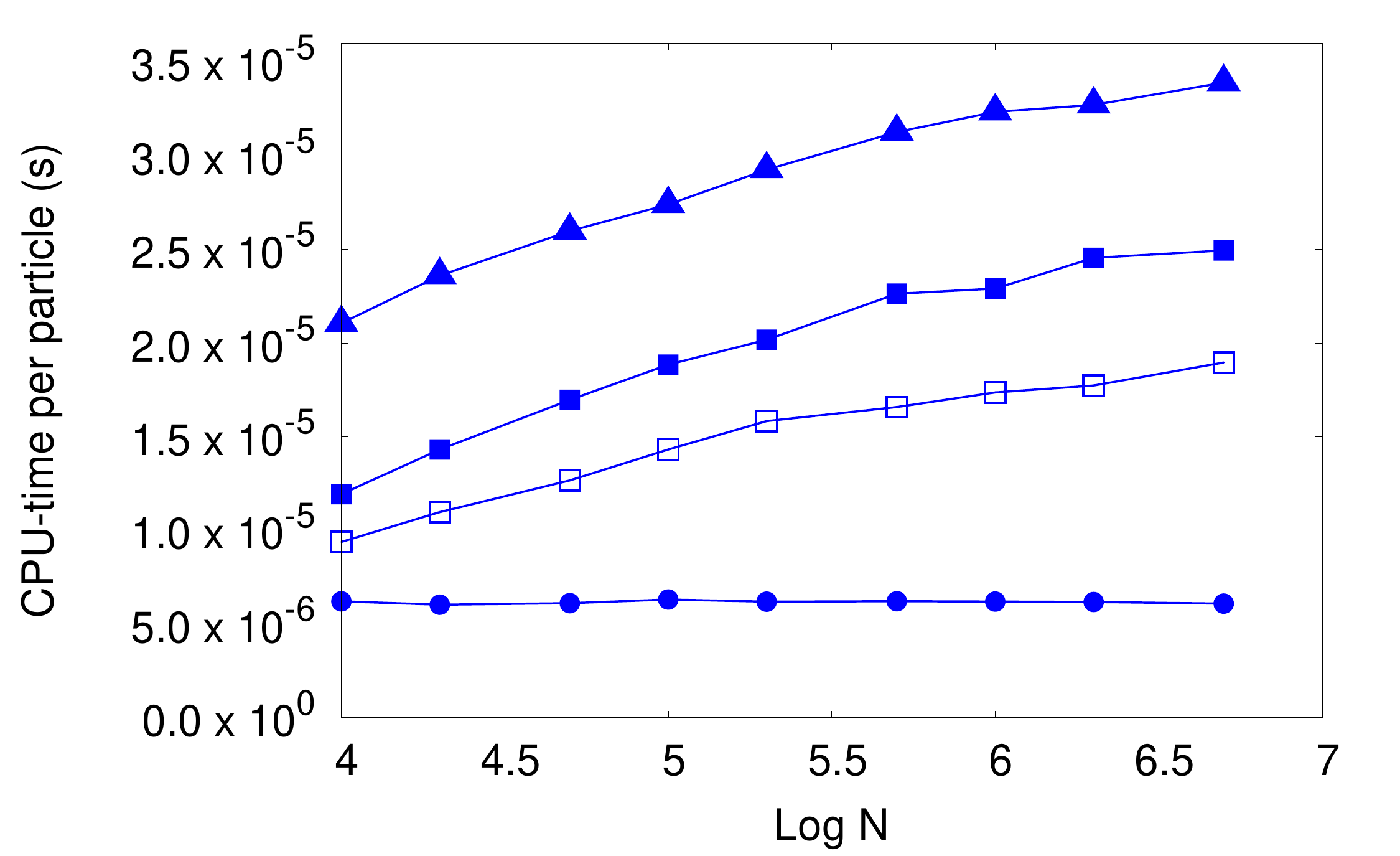} }
 \caption{ \GaSPH ~ average CPU-times per particle in function of $N$. 
Comparison among different routines: SPH neighbour searching routine averaged for a single iteration (circles), pure gravity computation with monopole term (empty squares) and with quadrupole term (filled squares), self-gravity computation up to quadrupole and including the SPH terms (triangles). 
Units: same as figure \ref{fig3_1A}. }
\label{fig3_2A}
\end{figure} 
Figure \ref{fig3_2A} compares the previous CPU-times with the times needed by \GaSPH~ for a full self-gravitating SPH system, with the quadrupole term included in the computation, keeping the same value of $\theta=0.6$.
Times per particle for the density computation routine are also shown.
In computing the acceleration, the additional time per particle is fairly independent of $N$, as can be seen in the figure, since the close SPH interactions are always made over a fixed number of neighbours points, which we set, in this example, to 60.
For the same reasons, also the average time per particle needed to calculate the density is expected to be constant, like the figure \ref{fig3_2A} shows indeed.
Actually, the calculus of $\rho$ and $h$ requests an iterative process in which, for each particle, the routine is called several times.
The CPU-times illustrated by the figure are the average values per single iteration.
Typically, in finding the optimal value of $h$ the code requires, on the average, no more than 2 iterations.

\begin{figure}    %
\resizebox{\hsize}{!}{ \includegraphics{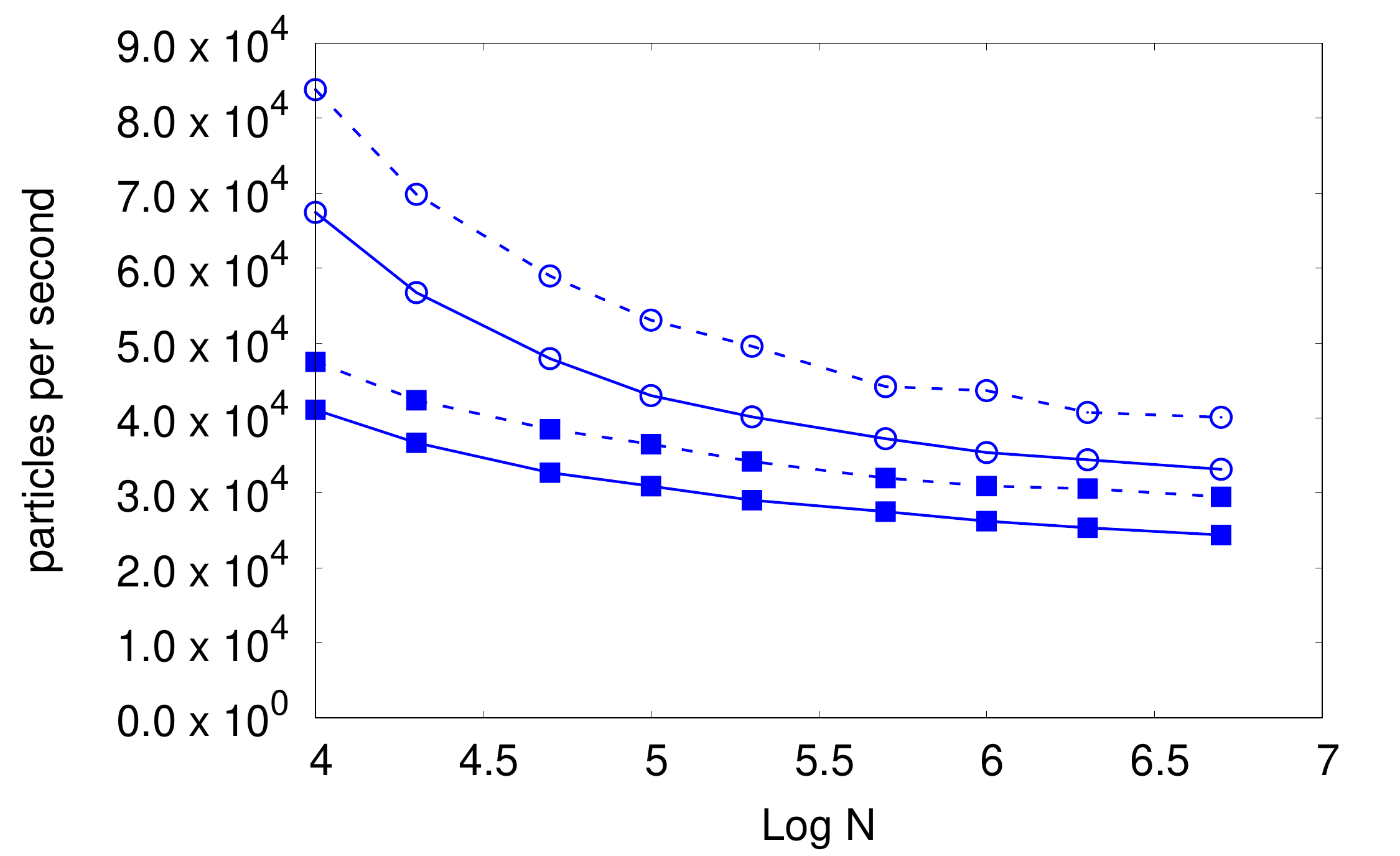} }
 \caption{  Number of processed particles per second.
 The full lines and dashed lines refer, respectively, to results with the correction term $\Delta_{CM}$ included and excluded.
 Simple gravity field calculation (empty circles), full self-gravity routine with hydrodynamics (squares).
 Quadrupole term is considered for the gravity field. 
 $\theta$ = 0.6. }
\label{fig3_3A}
\end{figure}

The optional introduction of the offset $\Delta_{CM}$ term in the opening criterion (as discussed in section \ref{par_2_3_2}) {causes, in some cases, a considerable reduction the effective angle $\theta$.}
Consequently, the number of direct particle-to-particle interactions increases, lowering the code performance.
{Figure \ref{fig3_3A}} illustrates the code efficiency in terms of number of particles processed in a second. 
The results, related to the two accelerations routines (pure self-gravity and self-gravity with SPH) shown in the previous graph, are compared with other result obtained by including the offset term $\Delta_{CM}$ in the opening criterion \eqref{eq_sph_23}.
A substantial, but not drastic, worsening in performance can be observed.
For instance, using $5 \times 10^6$ SPH particles and including $\Delta_{CM}$, the code computes the accelerations at a rate of $\approx$ 24,000 particles per second (about 17\% slower than the case without $\Delta_{CM}$).
Computations have been made with $\theta=0.6$ and the quadrupole terms included.

\begin{figure}   %
\resizebox{\hsize}{!}{ \includegraphics{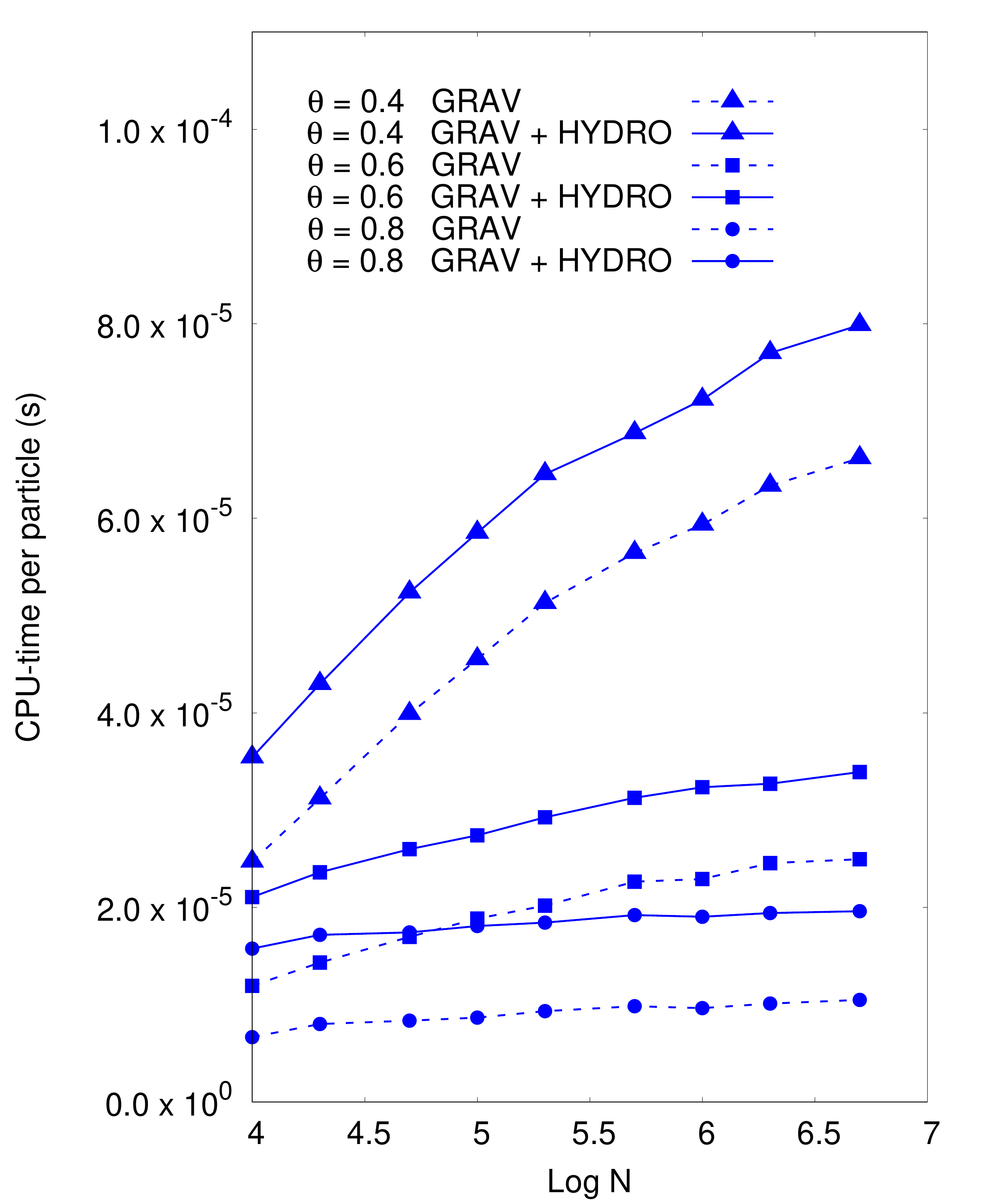} } 
\caption{Computational time per particle vs. $\log N$, for various values of the opening angle, $\theta$.
Pure tree gravitational algorithm (dashed line) and full SPH+gravity algorithm are shown as dashed and solid lines, respectively.
Quadrupole term is included in the force evaluation.
The $\Delta\vec{r}_{CM}$ offset term is not considered by the opening criterion.
  }  
\label{fig3_4A}
\end{figure}

The performance of the same two force subroutines (pure gravity and gravity plus hydrodynamics) are also studied at different values of $\theta$ ({CPU-times per particle in function of $N$ }are shown in figure \ref{fig3_4A}).

\begin{figure}   %
\resizebox{\hsize}{!}{ \includegraphics{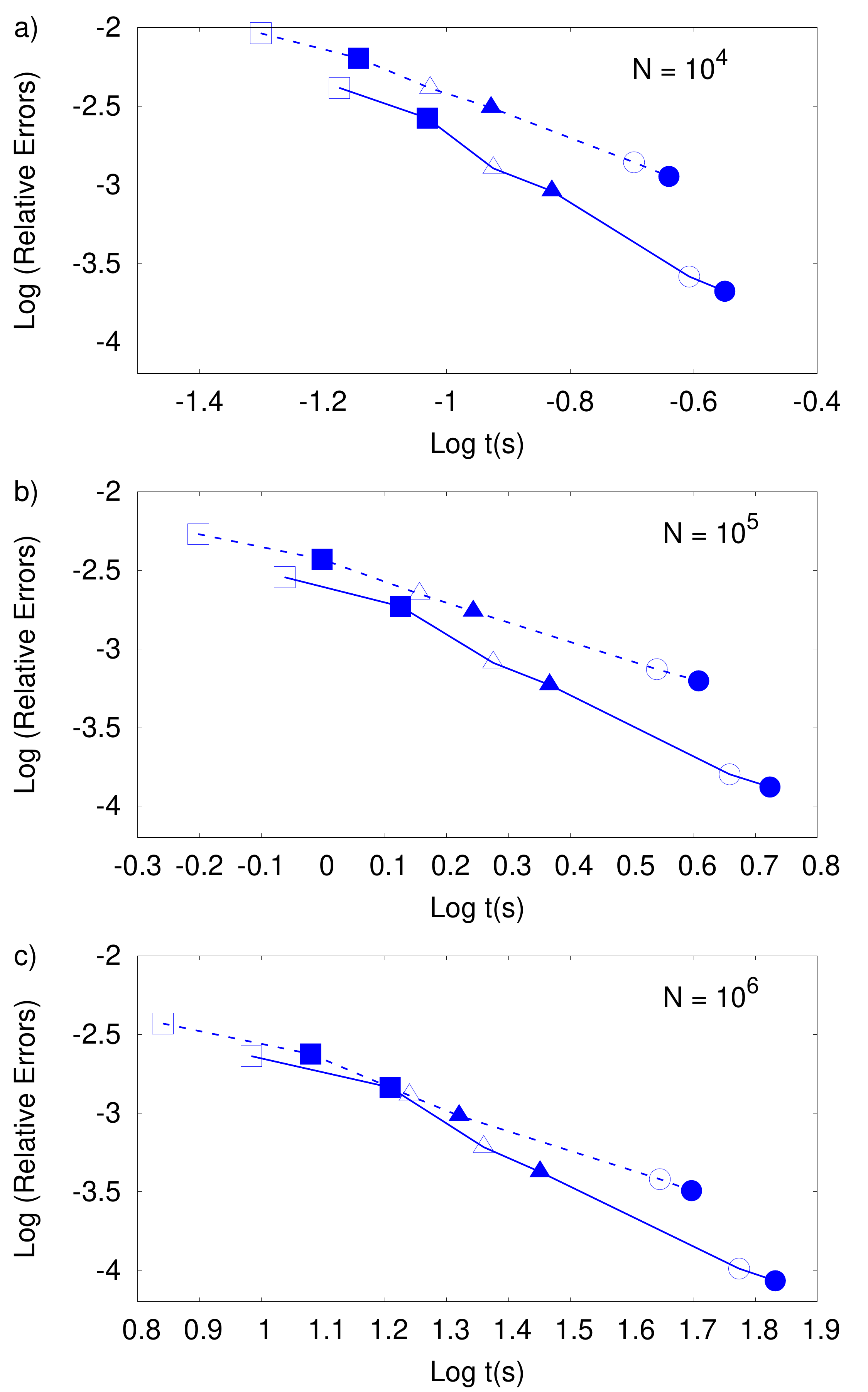} } 
\caption{Tree code relative errors $Err(a_k)$ (averaged over all the three Cartesian coordinates) for the gravitational field computation, in function of the CPU-time.
The data are illustrated for different particle numbers ($N=10^4$, $N=10^5$, $N=10^6$) respectively in panel a, b and c.
In each panel, the full line connects the points related to a computation of the gravitational field made with the quadrupole approximation, while the dashed line refers to computations made by using just the monopole term.
The shape of the `void' markers discriminates different choices of $\theta$ with the ``standard'' criterion (Eq. \eqref{eq_sph_2}): $\theta=0.8$ (void squares), $\theta=0.6$ (void triangles), $\theta=0.4$ (void circles).
Results for different opening angles with the ``opening law'' (Eq. \eqref{eq_sph_23}) criterion are also marked: $\theta=0.8$ (solid squares), $\theta=0.6$ (solid triangles), $\theta=0.4$ (solid circles).
}
\label{fig3_4B} 
\end{figure}

Smaller angles should provide higher precision at the cost of a longer computational time.
On the other hand, using larger angles we have less direct point-to-point interactions and we gain in efficiency, but we expect a lower accuracy.
We evaluated the accuracy of our tree code by measuring, according to the prescription suggested by \cite{hernquist87}, a ``mean relative error'' in computing the accelerations, with different conditions of particle number N and opening angle $\theta$.
The prescription consists into a comparison of the 3 components of the acceleration vector as computed by means of the tree scheme, $a_k^{TREE},\quad k=1,2,3$, with the ``exact'' value $a_k^{NBODY}$ computed by direct summation.
A  mean error $\langle\delta a_{k}\rangle = \frac{1}{N} \sum\limits_i \left( a_{k}^{(i) TREE} - a_{k}^{(i) NBODY} \right)$ is computed by averaging over all the N particles.
Then, the relative error is computed as follows:
\begin{equation}\label{eq_treecode_error}
 Err(a_k) = \dfrac{ \sum\limits_{i=1}^N  |~ {a_{k}^{(i) TREE} - a_{k}^{(i) NBODY}  - \langle{\delta a_k}\rangle~ |}}{ \sum\limits_i |a_{k}^{(i) NBODY}|}  
\end{equation} 
Figure \ref{fig3_4B} shows these relative errors, obtained with \GaSPH, in function of the CPU-time.
The figure does not illustrate the relative errors for each single component but it limits to show the mean values, computed by the simple average $ \frac{1}{3} \sum \limits_{k=1}^3  Err(a_k)$.
Results for several setup configurations are illustrated.
The figure shows the results in three different panels, according to the value of $N$ (respectively, $N=10^4$, $N=10^5$, $N=10^6$).
For each value of $N$, we used different combinations of parameter $\theta$ (0.4, 0.6, 0.8) with different opening criteria (\eqref{eq_sph_2} or \eqref{eq_sph_23}) and different multipole approximations (only monopole term or inclusion, also, of the quadrupole term). 
As data show, the approximation of the field with the quadrupole moment represents always an optimal choice in terms of performance since, at the same error, it requests a smaller amount of CPU time compared to the monopole approximation.
On the other hand, the choice of the new opening criterion gives a smaller improvement of the error with respect to the benefits obtained by switching from monopole to quadrupole term.

For lower particle numbers, a better computational performance without loss of accuracy is obtainable by the inclusion of quadrupole approximation and the (more expensive) opening criterion given \eqref{eq_sph_23}, together with a suitable change of the $theta$ angle.
Let's focus, for example, on the simulation setups characterized by $\theta=0.6$ with whatever opening criterion, with monopole approximation, with $N=10^4$ or $N=10^5$. 
The change $\theta=0.6 \rightarrow \theta=0.8$, together with the use of the criterion \eqref{eq_sph_23}, represents a good choice providing more performant simulations without degrading the precision of the algorithm.
We obtain the same advantage if we want to pass, similarly, from $\theta=0.4$ to $\theta=0.6$.
On the other hand, for $N=10^6$ (Fig. \ref{fig3_4B} - panel c) the results related to the approximation with monopole and with quadrupole have smaller differences, compared to the other cases with different $N$.
Hence, the choice of a larger opening angle (passing from $\theta=0.6$ to $\theta=0.8$ or passing from $\theta=0.4$ to $\theta=0.6$) together with the use of the new opening criterion, can give better performance despite the accuracy gets slightly worse.

In any case, if we want to preserve the high efficiency of the tree-code by keeping the CPU-time to scale as $N \log N$, an angle $\theta \geq 0.3$ must be chosen \citep{hernquist87}.
The choice of $\theta = 0.6$ in quadrupole approximation or the choice of $\theta = 0.4$ in monopole approximation, together with the criterion \eqref{eq_sph_23} represents a satisfying option, since it provides relative errors of the order of, at most, $10^{-3}$.

We now add $N_{ob}=20$ stars to the SPH distribution.
As explained in the previous section, a 14-th order explicit method is applied to calculate the evolution of such objects and, giving their low number in comparison to N, a scarce extra CPU-time is expected.
Table \ref{table:1} reports the percentage or workload related to the main relevant subroutines for the new gas+stars system: tree-building + particle sorting routine, density computation routine, acceleration routine, and the star evolution routine.
In addition, we report the rest of time needed for the basic operations (such as $\vec{v}$, $u$ and $\vec{r}$ updating, energies computation, time-steps computation).
Different work balances are shown for several values of N.
The percentage of workload related to the tree building routine is stable to the order of $6\%$ at different N.
On the contrary, the work needed by the density routine becomes less and less relevant as $N$ increases, while the gravity+SPH computation acquire more and more essential.
The computational effort to treat the evolution of stars, together with their interaction with the gas, is due both to the pure N-body RK coupling, expected to scale as $N_{ob}^2$, and the time for coupling each star with each SPH particle, expected to be linear in $N$.
Nevertheless, the table \ref{table:1} shows that 20 stars give a paltry contribution to the total CPU-time. 
For the specific purposes of our current work, the number of stars used is less or equal than 2 and thus their contribution on the code effort is far less than the $2\% \div 3 \%$.

\begin{table}
\centering                                      
\begin{tabular}{c c c c c c }          
\hline\hline                        
 
  \\
\      & Tree & neigh. & GRAV +& Stars \& & other \\ 
\      & build. & search &  HYDRO  & gas & oper. \\
$N$    &  &   & acc.  & inter. &   \\ \\
\hline                                   

\\$ ~~~~~ 10^4 $        & 8.3 & 19.7 & 66.9 & 2.5 & 2.5  \\  
\\$ 2 \times 10^4 $      & 6.0 & 18.2 & 71.1 & 2.4 & 2.3  \\ 
\\$ 5 \times 10^4 $      & 6.3 & 16.9 & 71.9 & 2.2 & 2.8  \\  
\\$ ~~~~~ 10^5 $        & 6.0 & 16.7 & 72.7 & 2.0 & 2.5  \\ 
\\$ 2 \times 10^5 $      & 6.0 & 15.7 & 74.2 & 2.2 & 2.0  \\ 
\\$ 5 \times 10^5 $      & 5.9 & 15.0 & 75.2 & 1.9 & 2.0  \\ 
\\$ ~~~~~ 10^6 $        & 6.0 & 14.5 & 75.9 & 2.0 & 1.7  \\ 
\\$ 2 \times 10^6 $      & 6.2 & 14.3 & 75.8 & 1.8 & 1.9  \\ 
\\$ 5 \times 10^6 $      & 6.3 & 13.7 & 76.2 & 1.7 & 2.1  \\ 
\\
\hline\hline                                             
\ \\
\end{tabular}
\caption{Work profiling (in percentage respect to the total) of \GaSPH, tested on a Plummer gas distribution with the addition of 20 stars and different numbers of SPH ($1st$ column).
The opening angle is $\theta = 0.6$,and the choice of  $N' = 60 $ neighbours particles was done. }              
\label{table:1}      

\end{table}

\section{Protoplanetary disks}
 
\subsection{Protoplanetary disks around one star}  

\subsubsection{Disk model}

Here we illustrate the general setup we use to model a protoplanetary disk in equilibrium around a  star of mass $M_s = 1$ M$_\sun$.
According to the classical \textit{Flared-Disk} model \citep[see for example][]{garcia11,armitage11}, we let the disk revolve around the central object with a rough Keplerian frequency $\Omega_k \approx \sqrt{  M_s G / R^3}$ (given $R$ the cylindrical cohordinate $R=\sqrt{x^2+y^2}$ in the reference frame centered in the central object).
The disk evolution is essentially driven by secular viscous dissipation.
According to the well-known $\alpha$-disk model \citep{shakura&sunyaev73}, the internal disk turbulence is schematized by means of a pseudo viscosity of the following form:
\begin{equation}\label{eq_alphadisk}
   \nu = \alpha_{SS}~c_s H .   
\end{equation}
Such a kinematic viscosity perturbs the fluid equations by leading to a net transport of matter inward and an outward flux of angular momentum.
$\alpha_{SS}$ represents a characteristic efficiency coefficient for the momentum transport, while $H =  c_s / \Omega_k$ represents a characteristic vertical pressure disk scale height.
The viscous evolution is usually much slower than the dynamical evolution ( the characteristic secular time-scale is $\propto r^2 / \nu$, typically 2 or 3 orders of magnitude larger than $\Omega_k^{-1}$).
Such modelization of the turbulence is basically dimensional and is made by mainly taking into account dynamical turbulence processes.
Thus, $\alpha_{SS}$ ranges over a wide range of variability (typically $10^{-4}$ and $10^{-2}$). 
When the disk self-gravity is stronger enough, another important effect arises, due to the gravitational perturbations.
Several works \citep[see for example][]{mayer&al02,boss98,boss03} gave a numerical estimation of the gravitational timescales in a protoplanetary disk, being of the same order of its dynamical time.
They have shown that, under certain conditions, matter can undergo instabilities and eventually condense forming clumps in $10^3 \div 10^4  ~\textup{yr}$, potentially destinated to give rise to gaseous planets.
It can been shown that disk keep their equilibrium state against collapse according to the Toomre's criterion: 
\begin{equation}\label{eq_toomre_factor}
  Q = \dfrac {c_s \Omega_e }{\pi G \Sigma} > 1.5    
\end{equation}
where $\Omega_e$ represents the epicyclic frequency, approximatively equivalent to $\Omega_k$ for Keplerian disks \citep[see ][ for a detailed study] {binney&tremaine87, toomre64}.
The Toomre's factor is a general coefficient which quantifies the predominance of the gravitational processes over the typical thermal and dynamical actions.

We let initially revolve our disk with an azimuthal velocity $v_k \equiv v_{\phi}(R) = \sqrt{\frac{G (M_s+M(R))}{R}}$ which depends both on the mass of the central star $M_s$ and on the internal mass of the disk itself $M(R)= \int \limits_0^R \Sigma(R) 2\pi R dR$.
The cumulative mass $M(R)$ can be neglected only in case of low disk masses $M_D << M_s$.
The shape of the disk along the direction perpendicular to the revolving midplane depends on the vertical pressure scale height H, such that pressure and density scale with a gaussian profile $\exp(- z^2 / 2H^2)$.
Here a local vertically isothermal approximation is used, assuming that any radiative input energy from the star is efficiently dissipated away: the cooling times are far shorter than the dynamical time-scales.
The disk is thus vertically isothermal and the temperature depends only on the radial distance from the central star. 

We set the disk thermal profile according to the well-known flared disk model, for which the ratio $H/R$ increases with $R$ (see \citet {garcia11}, chap.~2, and \citet{dullemond&al07} for a full clarification).
The disk temperature thus follows the profile:
\begin{equation}\label{eq_fldisk_01}
 T = T_0 \left( \dfrac{R}{R_0} \right) ^{-q} ,
\end{equation}
which is commonly used by setting $q=1/2$, while $R_0$ represents a scale length.
We use a slightly different slope $q = 3/7$, adopted by \citet{d_alessio&al99} by making the assumption that the thermal processes in the inner layers of the disk don't affect its dynamical stability.

Due to the use of the above temperature profile (independent of both $t$ and $z$), the gas pressure follows a barotropic equation of state $P = c_s^2 ~\rho$. 
This choice represents a rough approximation of the cooling processes, and allows us to model self-gravitating disks in equilibrium only for cases in which $Q>2$, excluding the models of disks in a state of marginal stability ($Q \approx 1 $).
In a realistic model of disk without the isothermal approximation, when the Toomre parameter approaches the unity, the loss of thermal energy due to radiative cooling processes leads to a matter aggregation which in turns causes shock waves that heat up again the gas.
If the disk is capable to retain a sufficient amount of the extra thermal energy generated, the collapse gives rise only to some spiral instabilities which don't grow up exponentially.
The collapse process is thus arrested and the disk reaches a meta-stable state in which every time that a gravitational instability occurs, it is further dissipated by the heat back production.
For a good treatment, see for example \cite{kratter&lodato16}.
On the contrary, the isothermal equation adopted by our model forces the system to cool down at an infinitely high efficient rate, expelling outwards all the extra thermal energy generated by the compression of matter.
Thus, in regions where $1 \leq Q < 2$, the density increases without any production of heat opposing the collapse process.
Our model of a disk in equilibrium is thus limited to masses $M_D$ for which the self-gravity guarantees the condition $Q \geq2$.

We use $\bar{\mu} = 2.33$ as mean molecular weight for the gas.
It represents a parameter commonly adopted to model protoplanetary disks \citep[see for example][]{kratter&al08,liu&al17}, since it is an average mean weight for a gas composed by \textup{H$_2$} and \textup{He}, based on the observed cosmic abundance of the elements.

The effects of the Shakura-Sunyaev viscosity associated to Keplerian disks can be emulated by means of the SPH artificial viscosity.
\cite{meglicki&al93} found that the SPH viscosity coefficient  $\alpha$ provides a viscous acceleration with an effective kinematic viscosity containing a similar form with shear component plus a bulk viscosity.
Provided a cubic spline function is used for the kernel function, they have shown that $\nu$ assumes the following form:
\begin{equation}\label{eq_fldisk_02}  
\nu \propto \alpha ~ c_s ~ h.
\end{equation}

In several works (as in \cite{artymowicz&lubow94}, or \cite{nelson&al98}) the viscosity term appearing in the expression \eqref{eq_3_20_tesi} is used under peculiar conditions: it acts no more only for approaching particles, but also for points which move out (with $\vec{r}_{ij} \cdot \vec{v}_{ij} > 0$).
It tourns out that $\nu = 0.1 ~ \alpha ~ c_s ~ h$ \citep[for a comprehensive explanation, see ][]{meru&bate12}.
We thus have the following law which connects the Shakura-Sunyaev viscosity coefficient to the $\alpha$ parameter used in SPH:
\begin{equation}\label{eq_fldisk_03} 
\alpha_{SS} = \dfrac{1}{10} \alpha  \dfrac{h}{H}.
\end{equation}

Such modification on the SPH formalism provides a more realistic prediction of the effect given by a kinematic viscosity since it acts both under compression and under gas expansion.
Such a prescription is reliable as far as we do not deal with strong velocity gradients, i.e. shock waves due to strong compressions.
In that case, the classical \cite{morris&monaghan97} amplification law (equation \eqref{eq_3_23_tesi} in this paper) would generate high dissipative forces even in expansion regions.

Protoplanetary disks are usually modeled as quiet systems and are not expected to undergo such huge compressions to let strong shock waves arise.

Using the expression \eqref{eq_3_20_tesi} for the viscosity and, consequently,  activating dissipation only for  particles approaching each other, the law \eqref{eq_fldisk_03} can be modified and improved by considering, also, the effects of the $\beta$ coefficient on the kinematic viscosity \citep{meru&bate12,picogna&marzari13}.
It follows that, for Keplerian disks:
\begin{equation}\label{eq_fldisk_03_b} 
\alpha_{SS} = \dfrac{31}{525} \alpha \dfrac{h}{H} + \dfrac{9}{70 \pi} \beta \dfrac{h^2}{H^2}
\end{equation}

The relations \eqref{eq_fldisk_03} and \eqref{eq_fldisk_03_b} formally don't contain any effect of the Balsara switch to compensate the false sharing attenuation (equation \ref{eq_3_21_tesi}).
We use, as done in \cite{picogna&marzari13}, the disk artificial viscosity term \eqref{eq_3_20_tesi}, and multiply the factor $\mu_{ij}$ by the term of \cite{balsara95} $\bar{f}$.

\subsubsection{Viscous Disk evolution}\label{sec_disk_lyndenbell}
We have conducted a test about the response of our model with respect to long time scale dissipative processes characteristic of viscous turbulent disks.
We started from the well-known disk model due to \cite{lynden_Bell&pringle74} (see also \cite{pringle81,hartmann&al98}).
It consists in a thin (H/R <<1) non self gravitating disk, subjected to a power law dissipative turbulent viscosity.
The surface density evolution is described by the following equation (see \cite{pringle81,hartmann&al98}):
\begin{equation}\label{eq_fldisk_01_c}
\dfrac{\partial \Sigma(R,t)}{\partial t} = \dfrac{3}{R}~ \dfrac{\partial}{\partial R}  ~ { \left[ ~ R^{1/2} \dfrac{\partial}{\partial R} \left( R^{1/2}~ \nu ~\Sigma(R,t) \right) ~ \right]}
\end{equation}
 If the disk is perturbed by a radial power-scaling kinematic viscosity $\nu \propto R^{~\phi}$, it has been shown that the differential equation admits the following similarity solution (see \cite{lynden_Bell&pringle74} and \cite{hartmann&al98}):
\begin{equation}\label{eq_fldisk_01_d}
 \Sigma(R,t) = \Sigma_0 \left(\dfrac{R}{R_1}\right)^{-\phi} \left( \dfrac{t}{\tau_\nu} + 1 \right)^{-\gamma} \exp\left[-\left( \dfrac{t}{\tau_\nu} + 1 \right)^{-1}\left(\dfrac{R}{R_1}\right)^{2-\phi} \right],
\end{equation}
where $\gamma = \frac{5/2-\phi}{2-\phi}$. 
$R_1$ is a characteristic radial scale as the one containing about the 68\% of the total disk mass, while $\Sigma_0$ is a normalization scale density.
In Eq. \eqref{eq_fldisk_01_d} $\tau_\nu = \frac{R_1^2}{3~(2-\phi)^2~\nu_1 }$ represents the characteristic viscous time scale of the disk, and it is proportional to the inverse of the viscosity evaluated in correspondence to the scale radius ($\nu_1  =\nu(R_1)$).
 
We sampled a disk, made of $20,000$ particles, which followed the thermal law of Eq. \eqref{eq_fldisk_01} described in the previous section, where we assumed $R_0= 10$ AU for scaling and $T_0 = 25 ~\textup{K}$.
The SPH particle distribution was done by sampling the initial (t=0) radial density profile as from eq. \eqref{eq_fldisk_01_d}:
\begin{equation}\label{eq_fldisk_01_e}
 \Sigma(R,0) = \Sigma_0 \left(\dfrac{R}{R_1}\right)^{-\phi} \exp\left[-\left(\dfrac{R}{R_1}\right)^{2-\phi} \right],
\end{equation}
where $R_1$ was set here to 50 AU.
We used a constant value for the viscosity coefficient, $\alpha_{SS}=10^{-2}$.
Given the proportionality of the kinematic viscosity to the sound speed and the vertical scale height (Eq.  \eqref{eq_alphadisk}), and assuming the exponent $q=3/7$ in the power law \eqref{eq_fldisk_01}, it turns out that $\nu$ follows a radial power law with a positive exponent $\phi = 15/14 \approx 1.07 $.
In order to reproduce the effects of such a radial viscosity law, we impose a specific prescription for the artificial viscosity in the code.
Taking into account the law \eqref{eq_fldisk_03_b}, considering that we use $\beta = 2\alpha$, we obtain the following expression for the coefficient $\alpha$:
\begin{equation}\label{eq_alpha_SPH}
    \alpha = \alpha_{SS} ~ \dfrac{H}{h} ~\dfrac{1}{31/525 + \left(9/35\pi\right)\left(h/H\right)}
\end{equation}
with constant $\alpha_{SS} = 10^{-2}$.
To apply such form, we insert the expression above in the artificial viscosity term \eqref{eq_3_20_tesi}, which is the one used in all our disk simulations, without considering the \cite{morris&monaghan97} variation law.

The (infinite) disk has been truncated at $R_{out} = 8 R_1 = 400 AU$.
The distribution has been also truncated at an inner cut-off $R_{in} = R_1/5 = 10 AU$.
The gravitational softening radius of the central star, together with its sink radius, were set equal to $R_{in}$.
Actually, the inner border condition we set is that the gas particles crossing the $R_{in}$ radius are absorbed by the star and consequently excluded from time integration.
The mass of the sank SPH particles was considered to grow the mass of the central star.
 
The ratio $R_{out} / R_{in} \approx 40$ can be considered to be large enough to match as accurately as possible the infinite extension of the analytical density profile, reducing the external radial boundary discontinuity.
\cite{hartmann&al98} indeed point out that the inward flux of matter (which constitutes the most important process in guiding the evolution of an $\alpha$ disk) depends considerably on the outwards angular momentum trasportation and, thus, on the disk expansion through the external shells.

The boundary conditions at $R_{in} > 0$ may affect the disk evolution along the whole spatial extension.
As pointed out in the work of \cite{lynden_Bell&pringle74}, and remarked by \cite{hartmann&al98}, formally a viscous disk may extend to $R \rightarrow 0 $ but there exists a critical radius, of the same order of the radius of the star, in which both the torque and, consequenly, the viscosity $\nu$ go to zero.
Actually, computational efficiency purposes don't allow to model a disk by introducing a very small cutoff radius.
Anyway, we imposed a zero viscosity $\nu$ by varying the coefficient $\alpha$ within $R=3R_{in}$ down to zero, assumed at $R=R_{in}$.

The parameter $\alpha$ in the equation \eqref{eq_alpha_SPH} varies with time, since the disk evolution processes lead to a time variation of the ratio between $h$ and the height scale $H$.
Initially, for $R = R_1/2$ the disk has an average $h/H \approx 2$, while such ratio reaches a minimum value of about $1.5$, in correspondence of $R=2R_1$, providing respectively a value of $\alpha \approx 0.02$ and $\alpha \approx 0.04$.
As the gas is captured by the central star, we expect the density to decrease and the $h$-to-$H$ ratio to increase.
As we will show further, the surface density will vary quickly during an initial phase, and will evolve at a relatively lower rate during later times, consequently letting $h/H$ to follow the same cadence.
At later stages (after about 1.6 Myr), the disk will be indeed integrated with smaller values of $\alpha$, approximately ranging from $0.01$ to $0.03$.

Our disk is virtually non-selfgravitating, since $Q>>2$ over all its surface (the minimum value is about 20), despite we formally take into account the disk mass in setting the azimuthal velocity $v_{\phi}(R)$.
For $R=R_1$, it has a vertical aspect ratio $H/R \approx 0.05$.
With such a setup, according to the analytical model, the disk should have a viscous evolution time scale $\tau_\nu \approx 840,000$ yr.
 
\begin{figure}   
\resizebox{\hsize}{!}{ \includegraphics{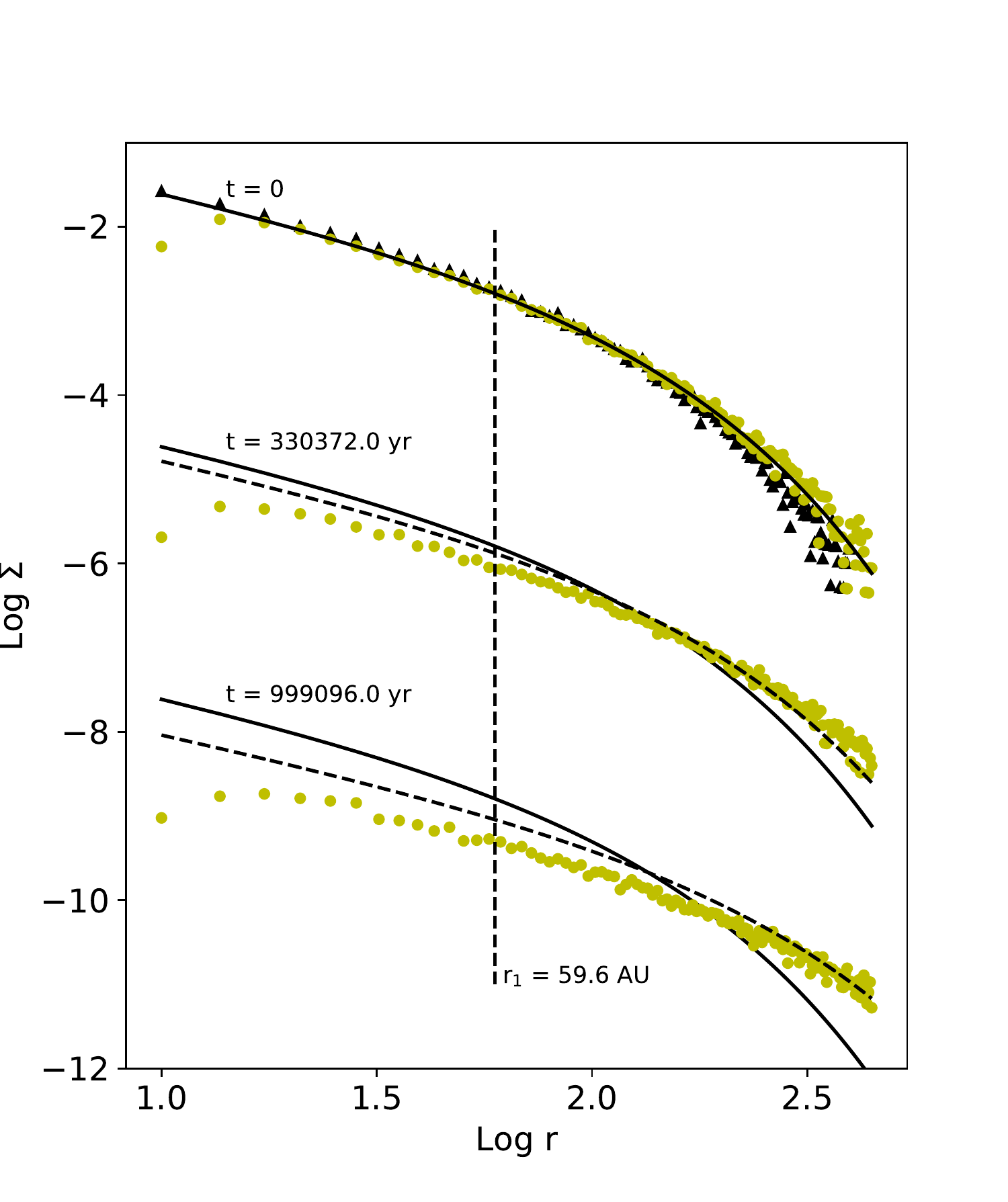} }
\caption{ 
Surface density profiles $\Sigma(R)$ of the disk at different times (t=0 corresponds to the disk state after it has been relaxed up to a time of 50,000 yr).
The dots represent the numerical results, while the lines refer to the analytical model (Eq. \eqref{eq_fldisk_01_d}).
The actual initial disk setting (black triangles) is reported for the sake of comparison with the relaxed state.
The abscissa is in AU while density is expressed in $10^{-2}$ M$_{\sun}$ AU$^{-2}$.
For the purpose of a clear visualization of the plots, in this graph the evolved density profiles at t>0 ($t=330,000$ yr and $t=1,000,000$ yr) are shifted down by $-3$ and $-6$ in logarithm, respectively.
The fitting curve of the density profile at t=0 is also plotted (full lines).
}\label{fig_sigmaconfrontobis} %
\end{figure}
\begin{figure}   %
\resizebox{\hsize}{!}{ \includegraphics{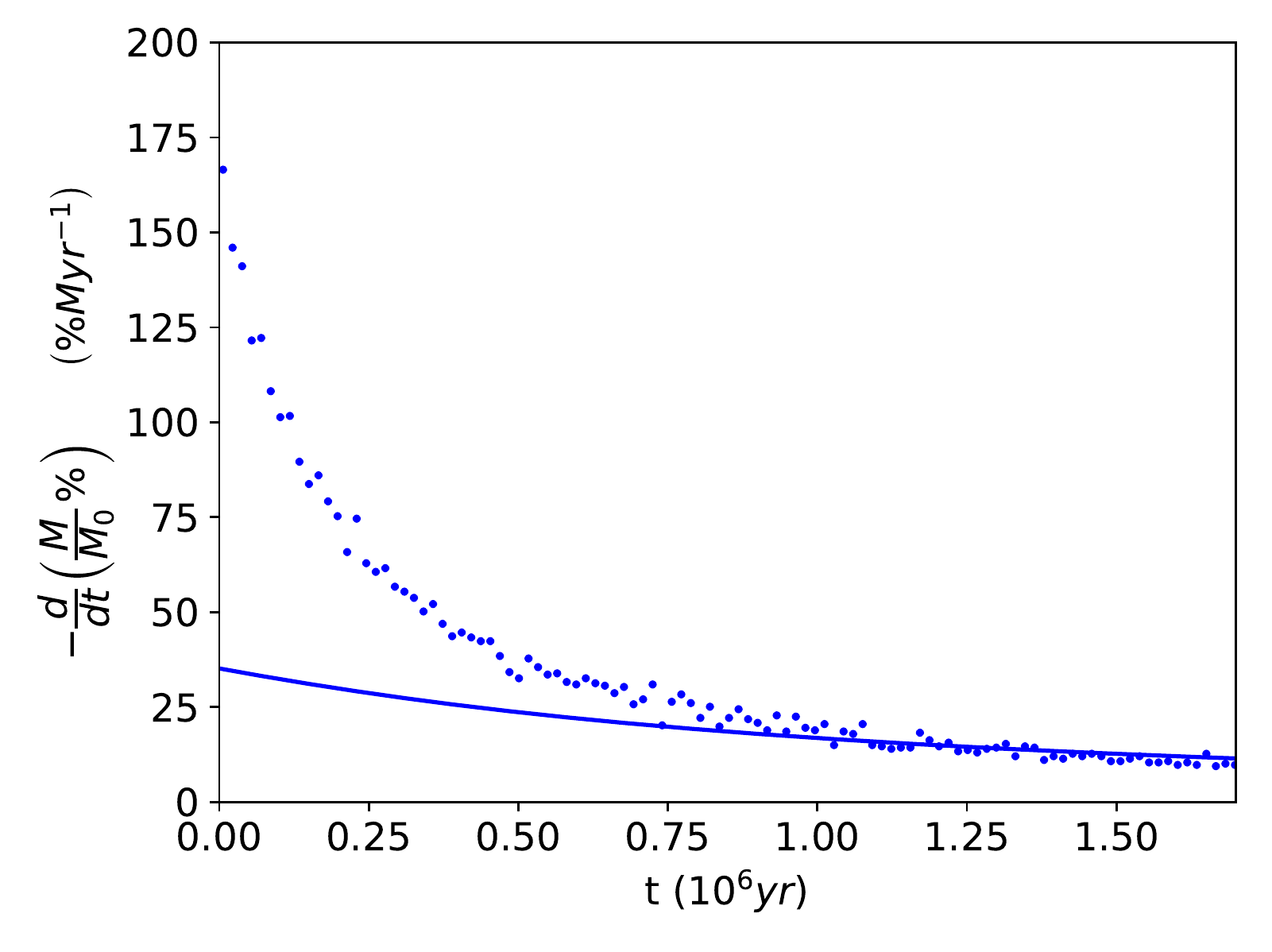} }
\caption{The relative (percentage) mass loss rate of the disk $ -\dfrac{d}{dt} \left(\dfrac{M(t)}{M_0}\right)$, expressed in units of Myr$^{-1}$ and time is in Myr.
The results of our numerical simulation (dots) are compared with the theoretical behaviour (full line).}
\label{fig_massa_disco_bis}
\end{figure}

Due to the inner cutoff, during the earlier phases of integration the disk experiences a fast relaxation in which the internal density discontinuity is smoothed out and it fades out near the inner border.
We have thus allowed the system to relax for a time of about $50,000$ yr (about 900 keplerian orbits at $R=R_1$), far shorter than $\tau_\nu$ and sufficient to obtain a steady state.
From this time, the only expected changes in the disk density profile are the ones due to the secular viscous evolution.
Thus, after 50,000 yr, the disk quickly arranges in a configuration slightly different from the initial one, characterized by a density distribution which follows the density law \eqref{eq_fldisk_01_e} but now with a larger radius $R_1 \approx$ 60 AU. 
With this new radius, the disk has the evolution time-scale $\tau_{\nu} \approx 10^6$ yr.
The properties of self-similarity owned by the equation \eqref{eq_fldisk_01_d} indeed guarantee that the surface density profile maintains the same analytical form for every time.
Thus, at every-time the surface density of a disk can be considered an initial solution of a new disk, described by the equation \eqref{eq_fldisk_01_e}, with a different parameter $R_1$ and thus a different viscosity timescale $\tau_{\nu}$.
The initial profile and the density profile after 50,000 yr (which conventionally we set as the instant t=0) are shown in figure \ref{fig_massa_disco_bis} on the top.
The initial t=0 state fits (full line) with the disk profile \eqref{eq_fldisk_01_e} with $R_1 \approx$ 60 AU.
We considered the disk evolution starting from this configuration and plotted the density profile for various times ($t= \tau_{\nu}/3\simeq 330,000$ yr and $t \simeq 1,000,000$ yr), making also a comparison with the analytical predictions obtained by the \eqref{eq_fldisk_01_d} (dashed lines).
We found a non-match between results and model, since the numerical disk appears to evolve faster than the one predicted by the analytical theory.
For $t\simeq 330,000$ yr the discrepancy between the analytical density was about 45\% higher than the numerical result at $R=R_1$, while for  $t\simeq 1,000,000$ yr the discrepancy increased up to the 95\%.
\begin{figure}   %
\resizebox{\hsize}{!}{ \includegraphics{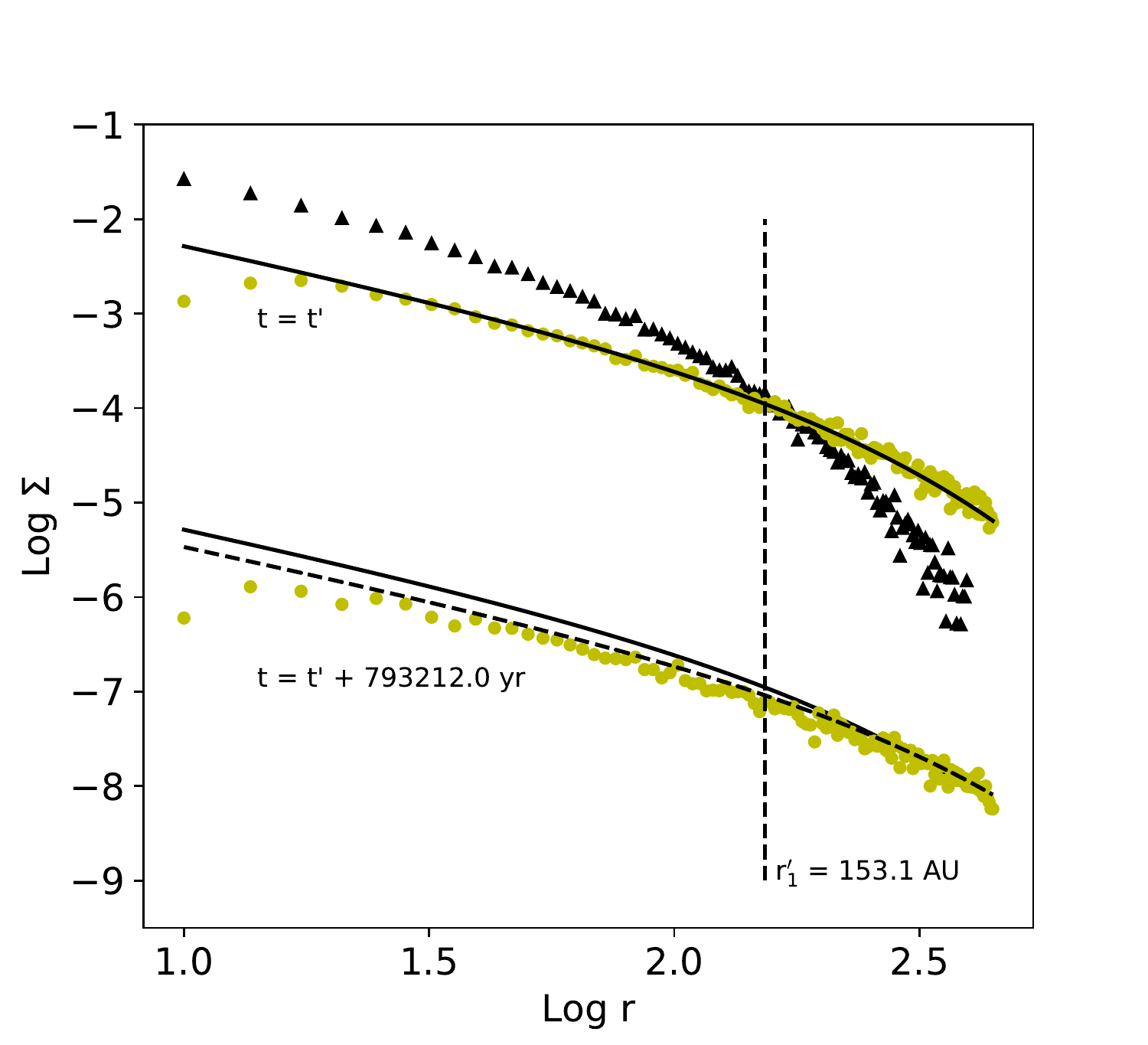} }
\caption{ 
Logarithm of the numerical disk density radial profile at $t'=850,000$ yr (dots).
The density profile at the beginning of the simulation is also plotted (black triangles).
The density at t=t' matches the self-similar solution (Eq. \eqref{eq_fldisk_01_e}) with a parameter $R_1' \approx 153$ AU and a viscosity time-scale $\tau_{\nu}' = 2.4 \times 10^6$ yr.
The evolution of such disk after a time $\tau_{\nu}'/3$ is also plotted (bottom, artificially shifted by $-3$) together with the analytical prediction (dashed line).
For a clearer comparison, the fitted density profile at t=t' is plotted twice (both in the top and in the bottom curves) in full line.
}\label{fig_sigmaconfrontobisbis}
\end{figure}
The disk thus looses mass more quickly than it would be expected from the analytical theory, and thus its density decreases too rapidly.
This excessive loss can be ascribed to the inner border which is absent in the analytical theory, where the disk matter flows onto the star in a single point.
To check how wrongly the SPH disk depleted its mass, we considered the radial-cumulative disk mass, predicted at a given time by the analytical theory, which is strictly connected to the viscosity time-scale:
\begin{equation}\label{eq_fldisk_01_f}
 M(\leq R,t) =  M_D(0) \left( \dfrac{t}{\tau_\nu} + 1\right)^{\frac{-1}{2(2-\phi)}}\left[1 - \exp \left( \left(\dfrac{R}{R_1}\right)^{2-\phi}\left( \dfrac{t}{\tau_\nu} + 1\right)^{-1}\right)\right] 
\end{equation}
given $M_D(0)$ the starting disk mass at $t=0$.
In Fig. \ref{fig_massa_disco_bis} we illustrate the fractional disk mass loss rate $ -\dfrac{d}{dt} \left(\dfrac{M(t)}{M(0)}\right)$ in function of time.
The result obtained with our model is compared with the theoretical rate, where the mass in function of time was considered as the integral of the analytical surface density \eqref{eq_fldisk_01_d}, from $R=10$ AU and $R=\infty$, i.e. $M(t)=\int\limits_{10~AU}^\infty{\Sigma(R,t)~ 2\pi R dR} = M_D(t) - M(\leq 10AU)$, being $M_D(t)$ the disk total mass at a given time.
The SPH-disk mass loss rate tends to reach the analytical curve only for $t \rightarrow \tau_\nu$), while for earlier stages we have a huge mass loss rate which decreases as time approaches $\tau_{\nu}$.

To make a more substantial comparison we moved out from the early stages and focus to the later phases at $t'=850,000$, where the mass loss rate illustrated in Fig. \ref{fig_massa_disco_bis} approached the analytic value within 10\%.
For that time, we considered our model density distribution as the starting state of a new disk and studied its following evolution.
We summarized such evolution in the Fig. \ref{fig_sigmaconfrontobisbis}.
As the figure shows indeed, at $t'$ the disk density surface corresponds to a Lynden-Bell solution of the same form of the starting one but with a different characteristic radius $R_1' \approx 153~AU \simeq 2.5 R_1$.
This radius corresponds to a viscous time-scale $\tau_{\nu}' \approx 2.4 \times 10^6$ yr.
In the same figure we show the evolution of the new surface density after a time of $\tau_{_\nu}' / 3 \approx 790,000 yr$, together with the theoretical expected value (in dashed line).
This has been done by analogy with the figure \ref{fig_sigmaconfrontobis}, where the result at $t = \tau_{\nu} / 3 \approx 330,000$ yr is shown (plotted in the middle).
Comparing the two curves, we note that during the later stages the density of the SPH disk shows a less deviation from the analytical prediction (within an error of 20\% at $R=R_1'$), although the discrepancies are not negligible.
The discrepancies turn out to be smaller also in internal regions at $R=20 ~AU << R_1'$, close to the inner border.
To relate the discrepancy between numerical density and its analytical prediction, we made a further verification by adopting a smaller inner border radius, i.e. reducing the star sink  from 10 AU to 5 AU, and we indeed observed a substantial reduction of the inner mass loss rate of the disk in the initial phases of the integration.
We have also noted that the gap of the surface density within the star sink radius is reduced in amplitude, when a smaller inner boundary is chosen.

\subsection{Self-gravitating disk in a binary system.}\label{sec_disk_in_binary_system}  
We tested our code on a more complex dynamical system by treating the evolution of a self-gravitating disk interacting with a binary star.
The system is characterized by a circumprimary disk around a 1 M$_\odot $ star, truncated by the gravitational field of  a 0.4 M${_\odot}$ external companion star.
This topic has been treated in a relevant work by \cite{marzari&al09}, who integrated the time evolution of such configuration and studied the effects of the stars on the orbital disk parameters: eccentricity and periastron argument.
The authors used the well-known eulerian code \fargo, implemented with a full scheme for the self-gravity (see \cite{masset00,baruteau&masset08}), performing a 2D simulation. 
We investigated with \GaSPH, with a 3D model, the evolution of such system, for the particular case of a binary with eccentricity 0.4. 

\subsubsection{System Setup.}\label{sec_disk_in_binary_system_setup}  
The binary system has an orbit characterized by an eccentricity $e_b=0.4$ and a semimajor axis of 30 AU.
As it was performed in the paper of \cite{marzari&al09}, since we are focused on the gravitational effects of the two stars on the disk, we keep their orbit fixed during the integration, i.e. the dynamics of the binary star is not affected by the gas feedback.
The disk initial configuration adopted in the original model is characterized by a radial surface density $\Sigma \propto R^{-1/2} $ extended from 0.5 AU to 11 AU from the central primary star. 
Farther than 11 AU, the density quickly fades out; the total disk mass is M$_D$ = 0.04 M$_\odot$.  
Rather than using a typical flared-shaped disk, a flat disk is adopted, by setting a linear vertical scale height H = 0.05 R.\\
The choice adopted by  \cite{marzari&al09}, concerning both the disk shape and the viscosity law, leads to a schematization slightly different compared to the model we used in section \ref{sec_disk_lyndenbell}, for which we needed to set carefully the parameters inside our SPH 3D code.
In fact, $\alpha_{SS}$ turns out not to be constant, but in their investigation the authors use rather a constant kinematic viscosity $\nu$.
Furthermore, the constant value of the aspect ratio $H/R$ leads the speed of sound to scale as $c_s \propto R^{-1/2}$.
As a matter of fact, given $\nu = cost.$ and given the alpha-disk law \eqref{eq_alphadisk},  we have that $\alpha_
{SS} \propto H^{-1} c_s^{-1} = H^{-2}\Omega^{-1} \propto R^{-1/2}$.
The coefficient $\alpha_{SS}$  was indeed set by the above authors by calibrating it to correspond to the value $\alpha_{SS} = 2.5 \times 10^{-3}$ in the central regions about $5$ AU within the disk.
Thus, we can deduce it as follows:
\begin{equation}\label{eq_alphass_marzari}
\alpha_{SS} = 2.5 \times 10^{-3} \left(\frac{R}{R_{ref}}\right) ^ {-1/2},
\end{equation}with $R_{ref}=5AU$.
We applied the artificial viscosity term by setting the SPH $\alpha$ parameter according to the same expression \eqref{eq_alpha_SPH} used in the previous section, but now using the non-constant coefficient $\alpha_{SS}$ with the radial profile \eqref{eq_alphass_marzari} illustrated above.

The disk is coplanar with the star orbit, and we built it by confining a set of SPH particles between R = 0.5 AU and R = 11 AU.
We integrated the system for about 3000 yr. 
All the gas particles flown across the inner border were excluded from the integration.
Three runs for the same model have been made with different particles number, $N=20,000$, $N=50,000$, and $N=100,000$.

As in the case of protoplanetary disk around one star discussed in the previous section, the ratio $h/H$, and consequently $\alpha$, are not constant in time.
For $N=20,000$, after an initial quick relaxation phase, the disk acquires a steady state configuration where $h/H$ has a rather slow evolution along the timescale of a binary rotation period.
After about $500$ yr, the disk acquires an average ratio $h/H \approx 6.5$ ($\alpha \approx 0.001$) in the inner regions ($R=1 AU$), which reaches a minimum of about $1$  ($\alpha \approx 0.011$) in the middle regions ($R=5.5. AU$).
Similarly, the disk with $N=50,000$ has an average ratio $h/H\approx 3.7$ in the inner regions, with a minimum of $h/H \approx 0.7$, corresponding respectively to $\alpha \approx 0.004$ and $\alpha \approx 0.02$.
The disk with $N=100,000$ has $\alpha \approx 0.008$ in the internal regions and a maximum $\alpha \approx 0.03$ in the intermediate radial regions.
Whatever the resolution, after an integration time of 3000 yr, the $h$-to-$H$ ratio slightly changes in the inner regions while its minimum substantially increases, giving rise to a maximum value of $\alpha$ of about $0.008$, $0.015$ and $0.02$, respectively for the disk with the lower, the middle, and the higher particle numbers.
 
\subsubsection{Disk deformation and gravitational feedback to the stars.}  
The gravitational field of the stars affects the disk configuration by altering its average orbital parameters such as the eccentricity and the periastron argument.
In order to describe the disk evolution, we calculate the mean eccentricity and the mean periastron argument by averaging over the disk surface, with the same prescription adopted by \cite{pierens&nelson07}:
\begin{equation}\label{eq_binary02}  
\begin{aligned}
e_{\textup{disk}}  = \dfrac{1}{M_{\textup{D}}} \sum\limits_{\textup{i}}  e_{\textup{i}}(R,\phi)~ m_{\textup{i}} ~~~~~,~~~~~ R_A \leq R \leq R_B    \\
\omega_{\textup{disk}} = \dfrac{1}{M_{\textup{D}}} \sum\limits_{\textup{i}}  \omega_{\textup{i}}(R,\phi)~ m_{\textup{i}} ~~~~~,~~~~~ R_A \leq R \leq R_B   
\end{aligned}
\end{equation}
 Where $M_{\textup{D}}$ is the disk mass included within $R_{\textup{A}}$ and $R_{\textup{B}}$ (the latter being a quantity of the order of the effective radius $R_{\textup{D}}$).
 The disk radius is defined by the following expression:
 \begin{equation}\label{eq_diskradius} 
R_{\textup{D}} ~ \equiv ~ <R_{\textup{D}}>_{\textup{L}} ~ \propto ~ \left( \dfrac{L}{M_{\textup{D}}} \right)^2
\end{equation}
and is computed as the radial distance containing the total angular momentum L of the disk, with $M_D$ its total mass.
$R_D$ turns out not to be so different from the half-mass radius.

The local orbital parameters are evaluated by using the eccentricity vector:
\begin{equation}\label{eq_eccentricity}
\vec{e} = \dfrac{\vec{r} ~ x ~ \vec{l} } {G M_{\textup{c}} } - \vec{\hat{r}}
\end{equation}
where $\vec{l} = \vec{r}~x~\vec{v}$ represents the angular momentum per unit mass, while $M_c$ is the mass of the central star. 
The vector $\vec{e}$ characterizes the orbit described by the position $\vec{r}$ of a point about the center of mass of the binary, assuming that it corresponds to an elliptical trajectory.
The absolute value $e = |\vec{e}|$ corresponds to the orbital eccentricity.
Moreover, $\vec{e}$ is always parallel to the semi-major axis, thus, its normalized components $e_{\textup{x}} / e = \cos(\omega)$ and $e_{\textup{y}} / e = \sin(\omega )$ provide the local periastron argument $\omega$ and, finally, the local semi-major axis.
In the \eqref{eq_binary02}, only the particles with orbits tied to the central primary star are considered part of the disk and thus are included in the summation.
\begin{figure}   %
\resizebox{\hsize}{!}{ \includegraphics{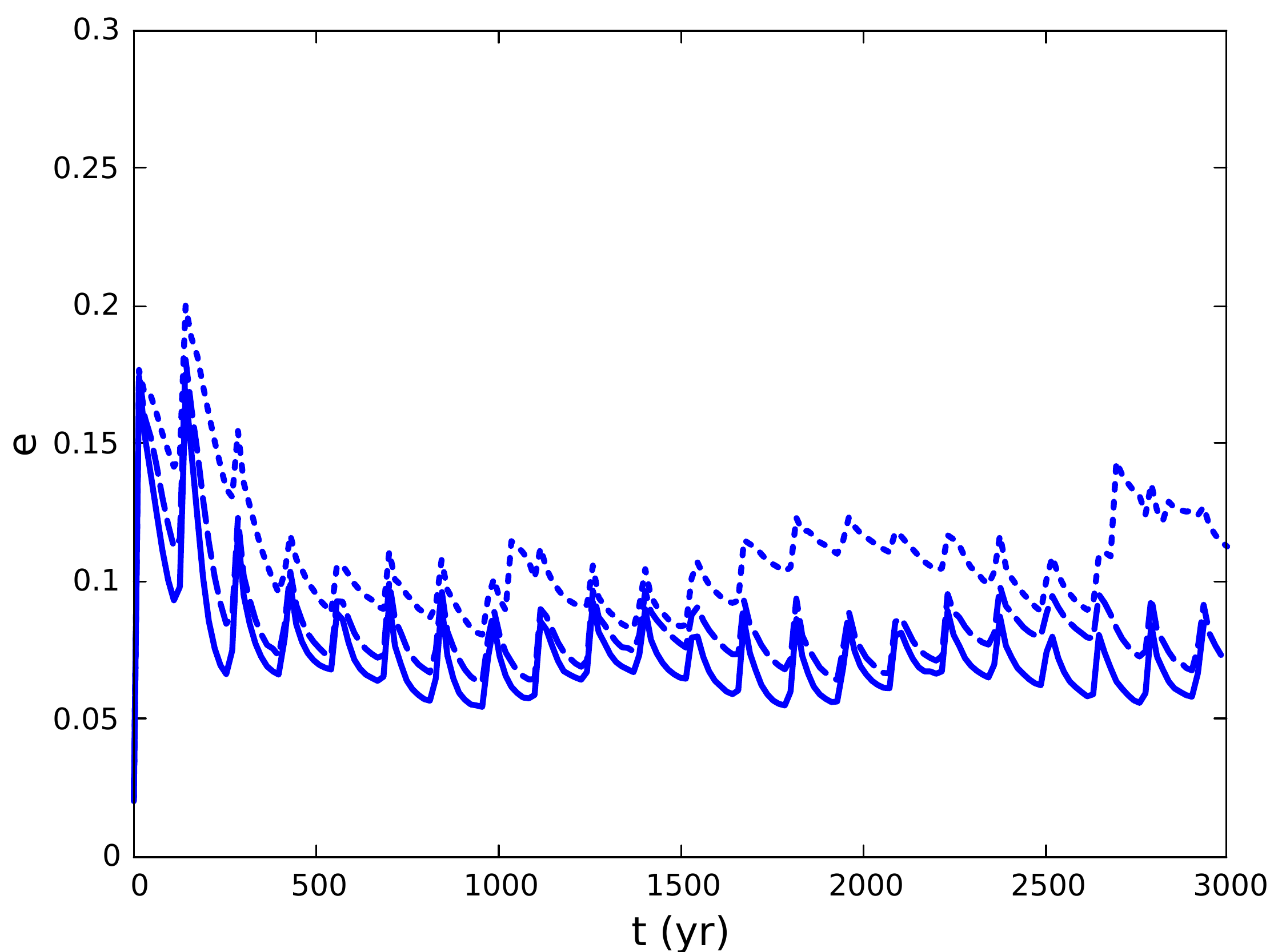} }
\caption{
Disk eccentricity, $e_{disk}$, evolution under the perturbation of a binary system of $e= 0.4$.
 Values referring to different simulations are plotted:$N=20,000$ (dotted line); $N=50,000$ (dashed line); $N=100,000$ (full line).
}\label{fig_binaria_e}
\end{figure}

\begin{figure}   %
\resizebox{\hsize}{!}{ \includegraphics{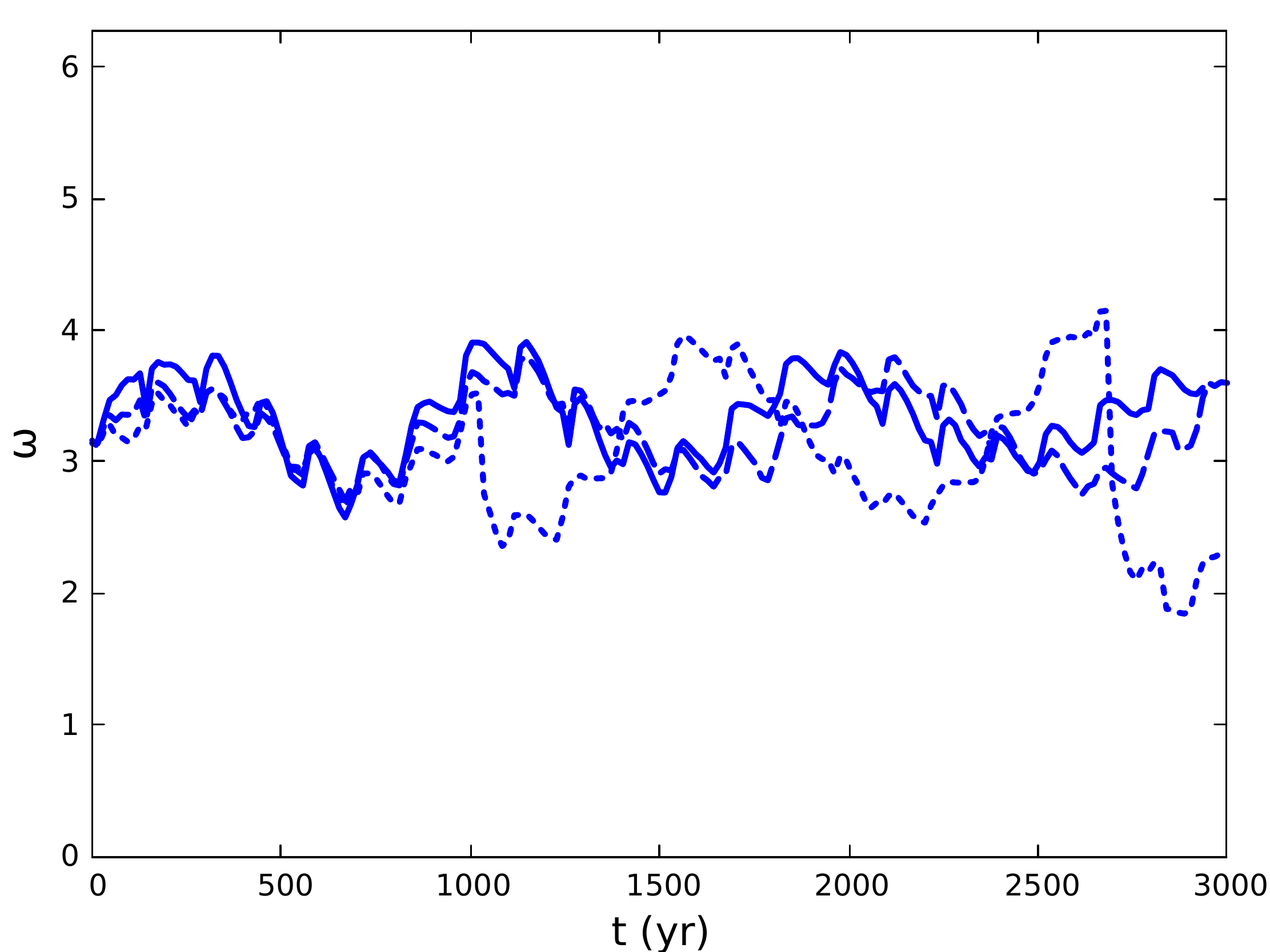} }
\caption{
Time evolution of periastron argument, $\omega_{disk}$, in an eccentric ($e=0.4$) binary system.
Like in Fig. \ref{fig_binaria_e}, results for different simulations are illustrated: $N=20,000$ (dotted line); $N=50,000$ (dashed line); $N=100,000$ (full line).
}  
\label{fig_binaria_omega}
\end{figure}

The choice of a 3D model introduces some new degrees of freedom with respect to a 2D scheme since the disk self-gravity and the stars gravity act even along the vertical direction.
In particular, the angular momentum, whose flux across the disk plays a crucial role on the disk evolution itself, spreads out also in the vertical direction.

Figures \ref{fig_binaria_e} and \ref{fig_binaria_omega} show, respectively, the evolution of the disk eccentricity and of the angle $\omega_{disk}$.
Initially, after a few orbital periods, the secondary star truncates the disk and a chaotic phase arises during which the mean eccentricity increases abruptly.
In a second phase, eccentricity stabilizes around an average value similar to that obtained by \cite{marzari&al09}, which is $e_{disk}\approx 0.075$.
Moreover, commonly with the other investigation, figure \ref{fig_binaria_e} shows that $e_{disk}$ makes some little oscillations modulated with the binary period ($P_{bin}\approx 134$ yr), and ascribable to the strong variation of the gravitational field of the companion star at periastron. 
 
Similarly, Fig. \ref{fig_binaria_omega} shows the mean inclination of the disk semimajor axis oscillations around the initial value (which conventionally was taken as $ \pi$.
Even for the disk periastron argument, a convergence can be observed near the higher resoluted simulations.

\section{Summary and conclusions} 
{ 
The primary intent of this paper is the presentation, testing and preliminary application of our new SPH code, GaSPH, which is thought as a multi-purpose code, applicable to a variety of astrophysical, multi phase, self-gravitating environments.} 

Let us briefly summarize the main points:
  
\begin{itemize}
 
\item we presented and discussed in some details the characteristics of our code, which, at the moment, does not deal with treatment of radiative transfer, but takes in proper account the internal gas gravity and the gas-star mutual gravity;
\item the code fully overcomes the classic tests in both slowly varying situations, assessing its stability, and violent cases, well reproducing the Sedov-Taylor blast wave;
\item the code shows good numerical performance (speed), stability and quality, as we discussed in Sect. \ref{subsec_3_2}
\item the capability of the code in treating the evolution of a protoplanetary disks both interacting with a single star and a binary star has been tested.
 
\end{itemize}
{
In a near future, we aim to a much better resolution, achievable with an MPI parallel version of our code, that would allow us to study disks on the smaller, planetary, scale. 
Further scientific applications of our code will deal with the evolution of protoplanetary disks in a star cluster environment, in order to study the star-to-disk feedback.
}
 
\begin{acknowledgements}
 
We express our gratitude to Fabrizio Capaccioni of INAF-IAPS-Istituto di Astrofisica e Planetologia Spaziali (Rome, Italy), for the precious support and useful discussions along the preparation of this work. 
We want, also, to thank an anonymous referee for his suggestions and comments which helped a lot in the improvement of the paper.

\end{acknowledgements} 
 
\bibliographystyle{aa} 
\bibliography{bibliografia,bibl_da_libri,bibl_secondaria,bibliografia2} 

\begin{thebibliography}{66}
\expandafter\ifx\csname natexlab\endcsname\relax\def\natexlab#1{#1}\fi

\bibitem[{{Aarseth}(1999)}]{aarseth99}
{Aarseth}, S.~J. 1999, \pasp, 111, 1333

\bibitem[{Allen \& Tildesley(1989)}]{allen&tildesley89}
Allen, M.~P. \& Tildesley, D.~J. 1989, Computer Simulation of Liquids (New
  York, NY, USA: Clarendon Press)

\bibitem[{{Andersen}(1983)}]{andersen83}
{Andersen}, H.~C. 1983, Journal of Computational Physics, 52, 24

\bibitem[{Armitage(2011)}]{armitage11}
Armitage, P.~J. 2011, Annual Review of Astronomy and Astrophysics, 49, 195

\bibitem[{{Artymowicz} \& {Lubow}(1994)}]{artymowicz&lubow94}
{Artymowicz}, P. \& {Lubow}, S.~H. 1994, \apj, 421, 651

\bibitem[{{Attwood} {et~al.}(2007){Attwood}, {Goodwin}, \&
  {Whitworth}}]{attwood&al07}
{Attwood}, R.~E., {Goodwin}, S.~P., \& {Whitworth}, A.~P. 2007, \aap, 464, 447

\bibitem[{{Balsara}(1995)}]{balsara95}
{Balsara}, D.~S. 1995, Journal of Computational Physics, 121, 357

\bibitem[{{Barnes} \& {Hut}(1986)}]{barnes&hut86}
{Barnes}, J. \& {Hut}, P. 1986, \nat, 324, 446

\bibitem[{Barnes(1986)}]{barnes86}
Barnes, J.~E. 1986, The Use of Supercomputers in Stellar Dynamics: Proceedings
  of a Workshop Held at the Institute for Advanced Study Princeton, USA, June
  2--4, 1986 (Berlin, Heidelberg: Springer Berlin Heidelberg), 175--180

\bibitem[{{Baruteau} \& {Masset}(2008)}]{baruteau&masset08}
{Baruteau}, C. \& {Masset}, F. 2008, \apj, 678, 483

\bibitem[{{Binney} \& {Tremaine}(1987)}]{binney&tremaine87}
{Binney}, J. \& {Tremaine}, S. 1987, Galactic dynamics (Princeton, NJ:
  Princeton University Press)

\bibitem[{{Boss}(1998)}]{boss98}
{Boss}, A.~P. 1998, \apj, 503, 923

\bibitem[{{Boss}(2003)}]{boss03}
{Boss}, A.~P. 2003, \apj, 599, 577

\bibitem[{{Capuzzo-Dolcetta} \& {Miocchi}(1998)}]{capuzzo&miocchi98}
{Capuzzo-Dolcetta}, R. \& {Miocchi}, P. 1998, Journal of Computational Physics,
  143, 29

\bibitem[{Capuzzo-Dolcetta {et~al.}(2013)Capuzzo-Dolcetta, Spera, \&
  Punzo}]{capuzzo&spera&punzo13}
Capuzzo-Dolcetta, R., Spera, M., \& Punzo, D. 2013, Journal of Computational
  Physics, 236, 580

\bibitem[{{Chandrasekhar}(1958)}]{chandrasekhar58}
{Chandrasekhar}, S. 1958, An introduction to the study of Stellar Structure.
  (New York, US: Dover Publications, Inc.)

\bibitem[{{Chapman} {et~al.}(2007){Chapman}, {Jost}, \& {Van Der
  Pas}}]{chapman&al07}
{Chapman}, B., {Jost}, G., \& {Van Der Pas}, R. 2007, {Using OpenMP –
  Portable Shared Memory Parallel Programming} (One Rogers Street, Cambridge,
  MA 02142-1209, USA: The MIT Press)

\bibitem[{{Ciardi} {et~al.}(2018){Ciardi}, {Crossfield}, {Feinstein},
  {Schlieder}, {Petigura}, {David}, {Bristow}, {Patel}, {Arnold}, {Benneke},
  {Christiansen}, {Dressing}, {Fulton}, {Howard}, {Isaacson}, {Sinukoff}, \&
  {Thackeray}}]{ciardi&al18}
{Ciardi}, D.~R., {Crossfield}, I. J.~M., {Feinstein}, A.~D., {et~al.} 2018,
  \aj, 155, 10

\bibitem[{{D'Alessio} {et~al.}(1999){D'Alessio}, {Calvet}, {Hartmann},
  {Lizano}, \& {Cant{\'o}}}]{d_alessio&al99}
{D'Alessio}, P., {Calvet}, N., {Hartmann}, L., {Lizano}, S., \& {Cant{\'o}}, J.
  1999, \apj, 527, 893

\bibitem[{{Dullemond} {et~al.}(2007){Dullemond}, {Hollenbach}, {Kamp}, \&
  {D'Alessio}}]{dullemond&al07}
{Dullemond}, C.~P., {Hollenbach}, D., {Kamp}, I., \& {D'Alessio}, P. 2007,
  Protostars and Planets V, 555

\bibitem[{Feagin(2012)}]{feagin12}
Feagin, T. 2012, Neural, Parallel and Scientific Computations, 20, 437

\bibitem[{{Garcia}(2011)}]{garcia11}
{Garcia}, P.~J.~V. 2011, {Physical Processes in Circumstellar Disks around
  Young Stars} (The University of Chicago Press)

\bibitem[{{Gingold} \& {Monaghan}(1977)}]{gingold&monaghan77}
{Gingold}, R.~A. \& {Monaghan}, J.~J. 1977, \mnras, 181, 375

\bibitem[{{Handy}(1998)}]{handy98}
{Handy}, J. 1998, {The Cache Memory book} (125, Sixth Avenue, San Diego, CA,
  92101-4311, USA: Academic Press, inc.)

\bibitem[{{Hartmann} {et~al.}(1998){Hartmann}, {Calvet}, {Gullbring}, \&
  {D'Alessio}}]{hartmann&al98}
{Hartmann}, L., {Calvet}, N., {Gullbring}, E., \& {D'Alessio}, P. 1998, \apj,
  495, 385

\bibitem[{{Hernquist}(1987)}]{hernquist87}
{Hernquist}, L. 1987, \apjs, 64, 715

\bibitem[{{Hernquist} \& {Katz}(1989)}]{hernquist&katz89}
{Hernquist}, L. \& {Katz}, N. 1989, \apjs, 70, 419

\bibitem[{{Hockney} \& {Eastwood}(1981)}]{hockney&eastwood81}
{Hockney}, R.~W. \& {Eastwood}, J.~W. 1981, Computer Simulation Using Particles
  (New York: McGraw-Hill)

\bibitem[{{Hockney} \& {Eastwood}(1988)}]{hockney&eastwood88}
{Hockney}, R.~W. \& {Eastwood}, J.~W. 1988, {Computer simulation using
  particles} (Bristol: Hilger)

\bibitem[{Hosono {et~al.}(2016)Hosono, Saitoh, \& Makino}]{hosono&al16}
Hosono, N., Saitoh, T.~R., \& Makino, J. 2016, The Astrophysical Journal
  Supplement Series, 224, 32

\bibitem[{{Hubber} {et~al.}(2013){Hubber}, {Allison}, {Smith}, \&
  {Goodwin}}]{hubber&al13}
{Hubber}, D.~A., {Allison}, R.~J., {Smith}, R., \& {Goodwin}, S.~P. 2013,
  \mnras, 430, 1599

\bibitem[{{Hut} {et~al.}(1995){Hut}, {Makino}, \& {McMillan}}]{hut&al95}
{Hut}, P., {Makino}, J., \& {McMillan}, S. 1995, \apjl, 443, L93

\bibitem[{Kratter \& Lodato(2016)}]{kratter&lodato16}
Kratter, K. \& Lodato, G. 2016, Annual Review of Astronomy and Astrophysics,
  54, 271

\bibitem[{Kratter {et~al.}(2008)Kratter, Matzner, \& Krumholz}]{kratter&al08}
Kratter, K.~M., Matzner, C.~D., \& Krumholz, M.~R. 2008, The Astrophysical
  Journal, 681, 375

\bibitem[{{Liu} {et~al.}(2017){Liu}, {Yao}, \& {Ding}}]{liu&al17}
{Liu}, C.-J., {Yao}, Z., \& {Ding}, W.-B. 2017, Research in Astronomy and
  Astrophysics, 17, 078

\bibitem[{{Lucy}(1977)}]{lucy77}
{Lucy}, L.~B. 1977, \aj, 82, 1013

\bibitem[{{Lynden-Bell} \& {Pringle}(1974)}]{lynden_Bell&pringle74}
{Lynden-Bell}, D. \& {Pringle}, J.~E. 1974, \mnras, 168, 603

\bibitem[{{Marzari} {et~al.}(2009){Marzari}, {Scholl}, {Thébault}, \&
  {Baruteau}}]{marzari&al09}
{Marzari}, F., {Scholl}, H., {Thébault}, P., \& {Baruteau}, C. 2009, \aap,
  508, 1493

\bibitem[{{Masset}(2000)}]{masset00}
{Masset}, F. 2000, Astronomy and Astrophysics Supplement Series, 141, 165

\bibitem[{{Mayer} {et~al.}(2002){Mayer}, {Quinn}, {Wadsley}, \&
  {Stadel}}]{mayer&al02}
{Mayer}, L., {Quinn}, T., {Wadsley}, J., \& {Stadel}, J. 2002, Science, 298,
  1756

\bibitem[{{Meglicki} {et~al.}(1993){Meglicki}, {Wickramasinghe}, \&
  {Bicknell}}]{meglicki&al93}
{Meglicki}, Z., {Wickramasinghe}, D., \& {Bicknell}, G.~V. 1993, \mnras, 264,
  691

\bibitem[{{Meru} \& {Bate}(2012)}]{meru&bate12}
{Meru}, F. \& {Bate}, M.~R. 2012, \mnras, 427, 2022

\bibitem[{{Miocchi} \& {Capuzzo-Dolcetta}(2002)}]{miocchi&capuzzo02}
{Miocchi}, P. \& {Capuzzo-Dolcetta}, R. 2002, \aap, 382, 758

\bibitem[{{Monaghan}(1988)}]{monaghan88}
{Monaghan}, J.~J. 1988, Computer Physics Communications, 48, 89

\bibitem[{{Monaghan}(1989)}]{monaghan89}
{Monaghan}, J.~J. 1989, Journal of Computational Physics, 82, 1

\bibitem[{{Monaghan}(1992)}]{monaghan92}
{Monaghan}, J.~J. 1992, \araa, 30, 543

\bibitem[{{Monaghan}(2005)}]{monaghan05}
{Monaghan}, J.~J. 2005, Reports on Progress in Physics, 68, 1703

\bibitem[{{Monaghan} \& {Lattanzio}(1985)}]{monaghan&lattanzio85}
{Monaghan}, J.~J. \& {Lattanzio}, J.~C. 1985, \aap, 149, 135

\bibitem[{{Morris} \& {Monaghan}(1997)}]{morris&monaghan97}
{Morris}, J.~P. \& {Monaghan}, J.~J. 1997, Journal of Computational Physics,
  136, 41

\bibitem[{{Nelson} {et~al.}(1998){Nelson}, {Benz}, {Adams}, \&
  {Arnett}}]{nelson&al98}
{Nelson}, A.~F., {Benz}, W., {Adams}, F.~C., \& {Arnett}, D. 1998, \apj, 502,
  342

\bibitem[{{Picogna, G.} \& {Marzari, F.}(2013)}]{picogna&marzari13}
{Picogna, G.} \& {Marzari, F.} 2013, A\&A, 556, A148

\bibitem[{{Pierens} \& {Nelson}(2007)}]{pierens&nelson07}
{Pierens}, A. \& {Nelson}, R.~P. 2007, \aap, 472, 993

\bibitem[{{Plummer}(1911)}]{plummer11}
{Plummer}, H.~C. 1911, \mnras, 71, 460

\bibitem[{{Price} \& {Monaghan}(2007)}]{price&monaghan07}
{Price}, D.~J. \& {Monaghan}, J.~J. 2007, \mnras, 374, 1347

\bibitem[{{Pringle}(1981)}]{pringle81}
{Pringle}, J.~E. 1981, \araa, 19, 137

\bibitem[{{Quinn} {et~al.}(1997){Quinn}, {Katz}, {Stadel}, \&
  {Lake}}]{quinn&al97}
{Quinn}, T., {Katz}, N., {Stadel}, J., \& {Lake}, G. 1997, ArXiv Astrophysics
  e-prints [\eprint{astro-ph/9710043}]

\bibitem[{{Rosswog} {et~al.}(2000){Rosswog}, {Davies}, {Thielemann}, \&
  {Piran}}]{rosswog&al00}
{Rosswog}, S., {Davies}, M.~B., {Thielemann}, F.-K., \& {Piran}, T. 2000, \aap,
  360, 171

\bibitem[{{Rosswog} \& {Price}(2007)}]{rosswog&price07}
{Rosswog}, S. \& {Price}, D. 2007, \mnras, 379, 915

\bibitem[{Saitoh \& Makino(2009)}]{saitoh&makino09}
Saitoh, T.~R. \& Makino, J. 2009, The Astrophysical Journal Letters, 697, L99

\bibitem[{{Sedov}(1959)}]{sedov59}
{Sedov}, L. 1959, {Similarity and Dimensional Methods in Mechanics} (CRC Press,
  Inc.)

\bibitem[{{Shakura} \& {Sunyaev}(1973)}]{shakura&sunyaev73}
{Shakura}, N.~I. \& {Sunyaev}, R.~A. 1973, \aap, 24, 337

\bibitem[{{Springel}(2005)}]{springel05}
{Springel}, V. 2005, \mnras, 364, 1105

\bibitem[{{Springel} \& {Hernquist}(2002)}]{springel&hernquist02}
{Springel}, V. \& {Hernquist}, L. 2002, \mnras, 333, 649

\bibitem[{Stamatellos {et~al.}(2011)Stamatellos, Maury, Whitworth, \&
  André}]{stamatellos11}
Stamatellos, D., Maury, A., Whitworth, A., \& André, P. 2011, Monthly Notices
  of the Royal Astronomical Society, 413, 1787

\bibitem[{{Tasker} {et~al.}(2008){Tasker}, {Brunino}, {Mitchell}, {Michielsen},
  {Hopton}, {Pearce}, {Bryan}, \& {Theuns}}]{tasker&al08}
{Tasker}, E.~J., {Brunino}, R., {Mitchell}, N.~L., {et~al.} 2008, \mnras, 390,
  1267

\bibitem[{{Toomre}(1964)}]{toomre64}
{Toomre}, A. 1964, \apj, 139, 1217

\end{thebibliography}

\appendix
\section{Numerical method details} \label{sec_appendiceA}
\subsection{Velocity-Verlet method}\label{appendiceA1}
Each time iteration, $\vec{a}$ and $\dot{u}$ are evaluated twice, in correspondence to the current step $n$ and the next $n+1$. 
Starting from a generic n-th time iteration, we use  $\vec{a}^{[n]}$ and $\dot{u}^{[n]}$ to predict the velocity and the energy:

\begin{equation}\label{eq_appendice01}
\begin{cases}
\vec{v^*}^{[n+1]} = \vec{v}^{[n]} + \vec{a}^{[n]}  \Delta t  \\
  u ^{*[n+1]} = u^{[n]} + \dot{u}^{[n]}  \Delta t  
\end{cases}
\end{equation}
 
 \ \\ while the position is directly updated to the next step $n+1$, without any prediction: 
\begin{equation}\label{eq_sph_22_A} 
 \vec{r}^{[n+1]} = \vec{r}^{[n]}  +   \vec{v}^{[n]} \Delta t +   \dfrac{1}{2} \vec{a}^{[n]} \Delta t^2  
\end{equation}

\ \\ With such new quantities, a new calculation is performed for $\vec{a}$ and $\dot{u}$;  we thus have :
\begin{equation}\label{eq_appendice01_B}
\begin{cases} 
\vec{a^*}^{[n+1]} = \vec{a} \left( \vec{r}^{[n+1]},\vec{v^*}^{[n+1]},u^{*[n+1]} \right) \\
\dot{u^*}^{[n+1]} = \dot{u} \left( \vec{r}^{[n+1]},\vec{v^*}^{[n+1]},u^{*[n+1]} \right) 
\end{cases}
\end{equation}

\ \\ which we can use to correct velocity and energy :
\begin{equation} \label{eq_sph_22_B} 
\begin{cases}
\vec{v}^{[n+1]} = \vec{v}^{[n]}  + \left( \vec{a}^{[n]} + \vec{a}^{*[n+1]} \right)  \dfrac{\Delta t}{2} \\
u^{[n+1]} = u^{[n]} + \left( \dot{u}^{[n]} + \dot{u}^{*[n+1]} \right) \dfrac{\Delta t}{2}   
\end{cases}
\end{equation}
 
It can be straightforwardly shown that, when acceleration does depend only on the positions, i.e. in a Newtonian problem without the involvement of the hydrodynamics, the Verlet method described above is equivalent to a standard 2nd order \textit{Kick-Drift-Kick} (KDK) Leap-Frog method.
As a matter of fact, if the acceleration does not depend on the velocity field nor on the internal energy, the quantity $\vec{a^*}^{[n+1]}$  corresponds to the actual acceleration $\vec{a}^{[n+1]}$ related to the next step.
The numerical method can thus be rewritten in the following way:
\begin{equation}\label{eq_verlet_leapfrog}
\begin{cases}
\vec{r}^{[n+1]} = \vec{r}^{[n]}  +  \left( \vec{v}^{[n]} +   \dfrac{1}{2} \vec{a}^{[n]} \Delta t \right) \Delta t = \vec{r}^{[n]} + \vec{v}^{[n+ 1/2]} \Delta t   \\
\vec{v}^{[n+1]} = \vec{v}^{[n]} + \dfrac{1}{2} \vec{a}^{[n]} \Delta t +  \dfrac{1}{2} \vec{a}^{[n]} \Delta t = \vec{v}^{[n+1/2]} +  \vec{a}^{[n]} \dfrac{\Delta t}{2}
\end{cases}
\end{equation}
which represents indeed the standard expression of a KDK Leap-Frog integrator that requires just one force calculation per time-step \citep[see][for leap-frog methods and further improvements]{hockney&eastwood88,hut&al95,quinn&al97}.
\begin{figure}   
\resizebox{\hsize}{!}{ \includegraphics{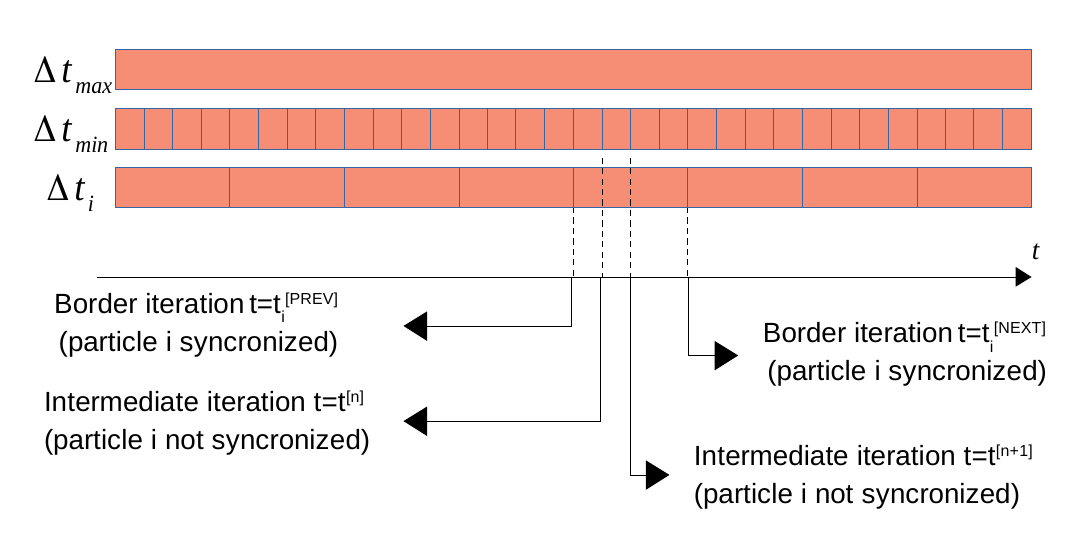} }
\caption{
Sketch of the hierarchical time-step subdivision.
A generic individual time-step, $\Delta t_i$ is a power OF $2$ multiple of the minimum time-step $\Delta t_{min}$, and a power of $2$ submultiple of the maximum ONE, $\Delta t_{max}$.
The integration is performed by means of elementary iterations from $t^{[n]}$ to $t^{[n+1]} = t^{[n]} + \Delta t_{min}$, while the updating of the particle is performed from $t^{[PREV]}$ to $t^{[NEXT]} = t^{[PREV]} + \Delta t_i$. 
The hydrodynamics and force routines are activated only for particles syncronized at $t^{[n]}=t^{[PREV]}$ or at $t^{[n+1]}=t^{[NEXT]}$.
} \label{fig_diagramA}
\end{figure}
\begin{figure}   
\resizebox{\hsize}{!}{ \includegraphics{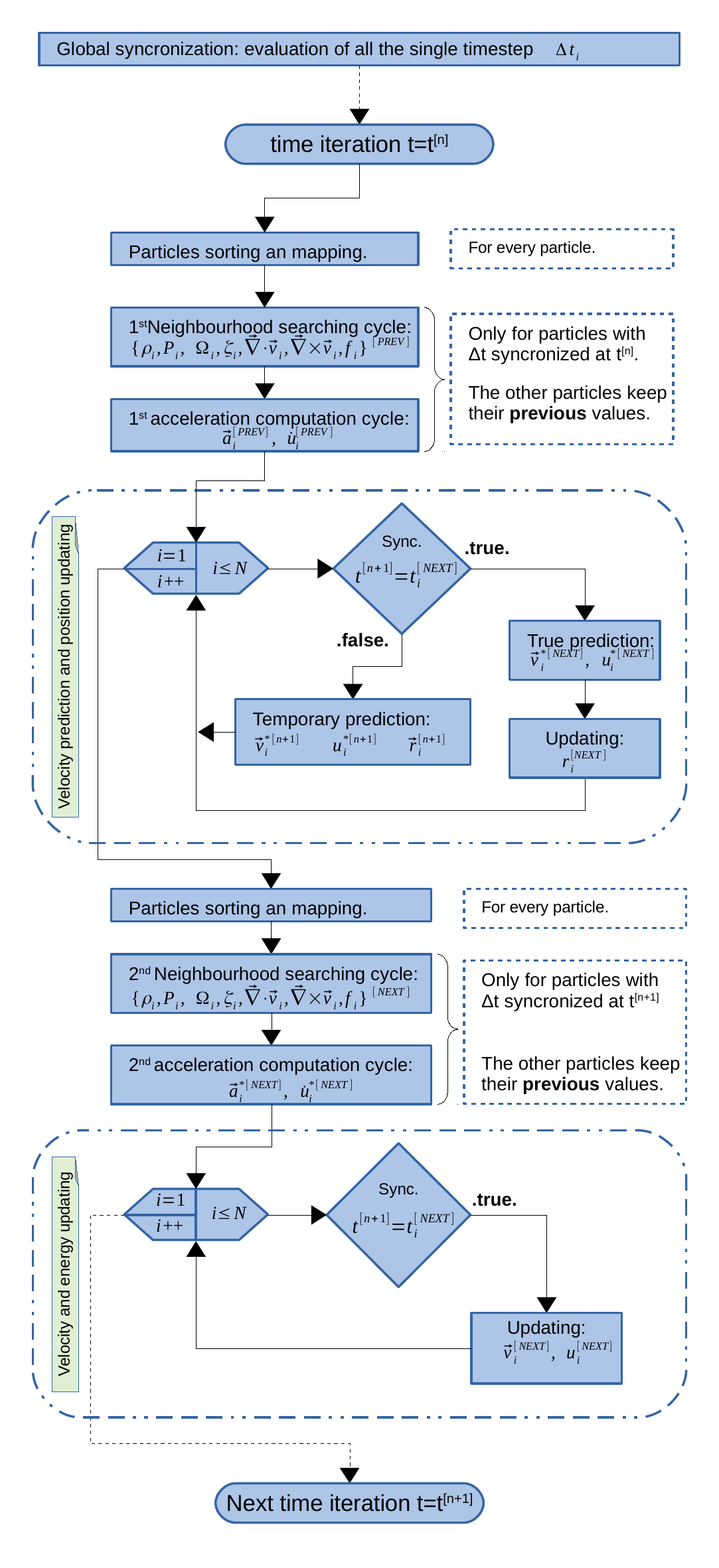} } 
\caption{
Flow chart showing the main scheme of a single time iteration.
} \label{fig_diagramB}
\end{figure}

Each particles has its own individual time-steps, sorted by the code as sub-multiples of the maximum time-step $\Delta t_{max}$.
For a generic particle of index i, the time-step is calculated according to the criteria expressed by the \eqref{eq_sph_22_A1} and the \eqref{eq_sph_22_A2} and approximated to the nearest value $\Delta t_i = 2^{-P} \cdot \Delta t_{max}$, where P is a positive integer number. 
Like the simple scheme in figure \ref{fig_diagramA} shows, a single time iteration between two consecutive steps n and n+1 is performed between the time $t^{[n]}$ and $t^{[n+1]} = t^{[n]} + \Delta t_{min}$, while the update of a generic particle of index i is performed between two characteristic times: $t_i^{[PREV]}$ and $t_i^{NEXT} = t_i^{[PREV]} + \Delta t_i$. 
The routines dedicated to the neighbour searching, to the hydrodynamic forces and to the gas self-gravity, will be activated for the particle i only in case its time-step is synchronized, i.e. if the conditions $t^{[n]} = t_i^{[PREV]}$ or $t^{[n+1]} = t_i^{[NEXT]}$ are satisfied.
In the figure \ref{fig_diagramB}, a scheme of a single time iteration is illustrated.
Firstly, the code sorts and maps all the particles by building the tree and calculating the quadrupole momentum and all the other key quantities related to the cubes.
Such an operation is thus independent on the particles time-step.
Then, the code runs the main cycles of neighbour searching and acceleration computation.
For each \textit{i-th} particle, it verifies whether the particle is syncronized at $t=t^{[n]}$, i.e. whether the condition $ t^{[n]} = t_i^{[PREV]}$ is satisfied.
In that case, the algorithm computes the density $\rho_i^{[PREV]}$, the pressure $P_i^{[PREV]}$, the velocity gradient ${\vec \nabla \cdot \vec v}_i^{[PREV]}$ and the velocity rotor ${\vec \nabla \times \vec v}_i^{[PREV]}$, the switching coefficient \eqref{eq_3_21_tesi}, the quantities $\omega_i^{[PREV]}, \zeta_i^{[PREV]}$, the hydro-gravitational acceleration $\vec a_i^{[PREV]}$ and the time variation of internal energy $\dot{u}_i^{[PREV]}$.
Such quantities are suitably stored in memory, to be used in further phases of the integration and in further iterations, too.
For the remaining non-synchronized particles, the algorithm indeed uses the quantities calculated in previous stages.

If some stars are included in the simulation, $a_i^{[PREV]}$ is incremented by a contribution due to the star-gas interaction, for every \textit{i-th} particle.

After that, we make the time updating of $\vec v$, $u$ and $\vec r$.
For particles synchronized at $t=t^{[n+1]}$, so that $t_i^{[NEXT]}=t^{[n+1]}$, the velocity and the energy are updated with the same predictor scheme as expressed by the equations \eqref{eq_appendice01}, thus, we have :

\begin{equation}\label{eq_appendice02}
\begin{cases}
\vec{v_i^*}^{[n+1]} = \vec{v_i^*}^{[NEXT]} = \vec{v_i}^{[PREV]} + \vec{a_i}^{[PREV]}  \Delta t_i  \\
   u_i^{*[n+1]} = u_i^{*[NEXT]} = u_i^{[PREV]} + \dot{u_i}^{[PREV]}  \Delta t_i  
\end{cases}
\end{equation}
The position is updated as well, in the same manner indicated by equation \eqref{eq_sph_22_A}:
\begin{equation}\label{eq_appendice03} 
 \vec{r_i}^{[n+1]} = \vec{r_i}^{[NEXT]} = \vec{r_i}^{[PREV]}  +   \vec{v_i}^{[PREV]} \Delta t_i +   \dfrac{1}{2} \vec{a_i}^{[PREV]} \Delta t_i^2  
\end{equation}

\ \\  

On the other hand, for non-synchronized points, we don't make any update but just estimate the quantities $\vec{v^*}^{[n+1]}$,  ${u^*}^{[n+1]}$ and $\vec{r}^{[n+1]}$ by means of the following predictor scheme:

\begin{equation}\label{eq_appendice05}
\begin{cases}
\vec{v_i^*}^{[n+1]} = \vec{v_i}^{[PREV]} + \vec{a_i}^{[PREDICT]} \delta t_i  \\
u_i^{*[n+1]} = u_i^{[PREV]} + \dot{u_i}^{[PREV]}  \delta t_i  \\
\vec{r_i}^{[n+1]} = \vec{r_i}^{[PREV]} + \vec{v_i}^{[PREDICT]} \delta t_i   + \dfrac{1}{2}\vec{a_i}^{[PREDICT]} \delta t_i^2, 
\end{cases}
\end{equation}
containing the following quantities: 
\begin{equation*}
\begin{aligned}
\delta t_i =  t^{[n+1]} - t_i^{[PREV]} \\
\vec a_i^{[PREDICT]} = \frac{1}{2}\left(\vec{a_i}^{[PREV]} + \vec{a_i}^{[n]}\right) \\
\vec v_i^{[PREDICT]} = \frac{1}{2}\left(\vec{v_i}^{[PREV]} + \vec{v_i}^{*[n+1]}\right)
\end{aligned}
\end{equation*}

$\vec{a_i}^{[n]}$ is exactly equal to the old value $\vec a_i^{[PREV]}$ in simulations with pure gas, since the algorithm does not compute the acceleration at the current iteration.
While if one or more stars interact with the gas, the two accelerations are different since they contain different contributions due to the gas-star interaction, which is calculated every step independently on the particle synchronization.

Basing on the new velocities, energies and space coordinates, a second phase of recalculation of the hydrodynamical variables and the accelerations occurs, of course, limiting to the synchrhonized points at $t=t^{[n+1]}$.
Thus, each synchronized particle \textit{i} owns the following updated hydrodynamical quantities $\rho_i^{[NEXT]}$, $P_i^{[NEXT]}$, ${\vec \nabla \cdot \vec v}_i^{[NEXT]}$, ${\vec \nabla \times \vec v}_i^{[NEXT]}$, $f_i^{[NEXT]}$, $\omega_i^{[NEXT]}, \zeta_i^{[NEXT]}$.
The accelerations and energy can be thus computed in the same manner as shown by the \eqref{eq_appendice01_B}:
\begin{equation}\label{eq_appendice06}
\begin{cases} 
\vec{a_i^*}^{[n+1]} = \vec{a_i}^{*[NEXT]} = \vec a \left( \vec{r_i}^{[NEXT]},\vec{v_i^*}^{[NEXT]},u_i^{*[NEXT]} \right) \\
\dot{u^*}^{[n+1]} = \dot{u^*}^{[NEXT]} = \dot{u}  \left( \vec{r_i}^{[NEXT]},\vec{v_i^*}^{[NEXT]},u_i^{*[NEXT]} \right) 
\end{cases}
\end{equation}
Then, by applying the same formalism as expressed by equation \eqref{eq_sph_22_B}, energy and velocities can be finally updated:
\begin{equation} \label{eq_appendice07} 
\begin{cases}
\vec{v_i}^{[n+1]} = \vec{v_i}^{[NEXT]} = \vec{v}^{[PREV]}  + \left( \vec{a_i}^{[PREV]} + \vec{a_i}^{*[NEXT]} \right)  \dfrac{\Delta t_i}{2} \\
u_i^{[n+1]} = u_i^{[NEXT]} = u_i^{[PREV]} + \left( \dot{u_i}^{[PREV]} + \dot{u_i}^{*[NEXT]} \right) \dfrac{\Delta t_i}{2}   
\end{cases}
\end{equation}

\subsection{14th order Runge-Kutta method}\label{appendiceA2}
For a generic set of $N_{ob}$ objects we want to integrate the following differential equations associated with a generic object $i$ :
\begin{equation}\label{eq_differenziali_moto}
\dfrac{d \vec{r_i}}{dt} = \vec{v}_i  ~~~~~~~~~~;~~~~~~~~~~   \dfrac{d \vec{v_i}}{dt} = \vec{f}_i  
\end{equation}

 \  \\
The method begins with a first estimation of the explicit derivatives at the iteration $n$:
\begin{equation}\label{eq_rungekutta14_A}
\begin{cases}
  \vec{Kv}_{1}^{(i)} = \vec{f} ( \vec{r}_1^{[n]},\vec{r}_2^{[n]}, ... , \vec{r}_i^{[n]}, ... , \vec{r}_N^{[n]}  )    \\
  \vec{Kr}_{1}^{(i)} = \vec{v}_i^{[n]}  
\end{cases}
\end{equation}

\ \\ which can be used to estimate the further quantities corresponding to a second sub-step $n+c_2$:
\begin{equation}\label{eq_rungekutta14_B}
\begin{aligned}
\vec{Kv}_{2}^{(i)} =  \vec{f} (  \vec{r}_1^{[n+c2]}, \vec{r}_2^{[n+c2]}, ... , \vec{r}_i^{[n+c2]}, ... , \vec{r}_N^{[n+c2]} )    \\
\vec{Kr}_{2}^{(i)}  = \vec{v}_i^{[n]} + a_{21} \vec{Kv}_{1}^{(i)}  \Delta t  ,
\end{aligned}
\end{equation}
where $ \vec{r}_i^{[n+c2]} = \vec{r}_i^{[n]}  + a_{21} \vec{Kr}_{1}^{(i)} \Delta t ~~ $ represents the i-th particle position updated to an intermediate time $t + c_2 \Delta t$. 
At the same manner, further consecutive estimations of $\beta$-th terms can be performed:  

\begin{equation}\label{eq_sph_22_C} 
\begin{aligned}
 \vec{Kv}_{\beta}^{(i)} = \vec{f} (  \vec{r}_1^{[n+c\beta]}, \vec{r}_2^{[n+c\beta]}, ... , \vec{r}_i^{[n+c\beta]}, ... , \vec{r}_N^{[n+c\beta]} )    \\
  \vec{Kr}_{\beta}^{(i)} = \vec{r}_i^{[n]} + \sum\limits_{\gamma=1}^{\beta-1} a_{\beta\gamma}  \vec{Kv}_{\gamma}^{(i)}~\Delta t  
\end{aligned}
\end{equation}
with $\vec{r}_i^{[n+c\beta]}  =  \vec{r}_i^{[n]}  + \sum\limits _{\gamma=1}^{\beta -1} a_{\beta\gamma} \vec{Kr}_{\gamma }^{(i)} \Delta t ~~$ the vector position of particle i at a generic intermediate time $t + c_{\beta} \Delta t$. 
Since $\gamma < \beta$, every quantity depends explicitly on previous estimations. 
The coefficients $a_{\beta\gamma} $ are the elements of a 35x34 matrix, while $b_{\beta}$ and $c_{\beta}$ represent two arrays of 35 elements.
The matrix $a$ requires the following restriction:
\begin{equation}\label{eq_rungekutta_coefficients} 
\sum\limits_{\gamma=1}^{34} a_{\beta\gamma}  = c_{\beta}
\end{equation}

\ \\
Since we are dealing with a full explicit method, $a$ is triangular, and $a_{\beta\gamma} = 0$, for $\gamma > \beta$. 
In total, each star will have 35 velocity RK coefficients  $\vec{Kr}_{\beta}$  and 35 acceleration RK coefficients $\vec{Kv}_{\beta} $, the resultant velocity and position at the next time-step will be given by:

\begin{equation}\label{eq_sph_22_D}  
\begin{aligned}
\vec{v}_i^{[n+1]} = \vec{v}_i^{[n]} + \sum\limits_{\beta=1}^{35} b_{\beta} \vec{Kv}_{\beta}^{(i)} \Delta t   \\
\vec{r}_i^{[n+1]} = \vec{r}_i^{[n]} +  \sum\limits_{\beta=1}^{35} b_{\beta} \vec{Kr}_{\beta} ^{(i)} \Delta t
\end{aligned}
\end{equation}
 \ \\
To integrate the time evolution of a system composed of both gas and stars, we couple the Verlet and the RK integration methods as follows.
At the starting iteration $n$, the gas particles feel the gravity field from the stars, the SPH mutual interactions and eventually its self-gravity.
Then, $\vec{v^*}^{[n+1]}$ and $u^{*[n+1]}$ are predicted and their positions are thus updated according to the \eqref{eq_sph_22_A}, 
At the same time, stars positions and velocities are first updated with the Runge-Kutta method, then, the explicit force contributions due to the SPH particles $\vec{a}_{part}^{[n]}$ are added to the \eqref{eq_sph_22_D}.
Stars and SPH particles are coupled by direct point-to-point interaction, without any approximation for the gravitational field. 
Finally, we correct the gas positions and velocity according to the \eqref{eq_sph_22_B}, by recalculating the accelerations and the energy rates at the new stage $n+1$. 

\section{Technical features}\label{sec_appendiceB}
\subsection{Neighbour searching and acceleration updating.}\label{sec_appendiceB1}

Before being employed to compute gravitational interactions, the tree-grid is also used to support the operations related to the nearest neighbour searching.
To calculate the density and the smoothing length, each $i-th$ particle starts with a first guess value of $h_i$ and  makes a tree walk by adopting an opening criterion slightly different from the \eqref{eq_sph_2} and \eqref{eq_sph_23}.
Given a generic particle of index i, in order to find the other points enclosed in its SPH kernel support, we check, for each cube, the overlapping with a sphere of radius $2h_i$ centered onto the point $r_i = (x_i, y_i, z_i)$.
A box is opened only if the following three conditions are valid:
\begin{equation}\label{eq_sph_24}
\begin{aligned}
 |x_i - x_{\textup{A}}| ~ < ~ 2 h_i + \dfrac{D_{\textup{L}}}{2} \\
 |y_i - y_{\textup{A}}| ~ < ~ 2 h_i + \dfrac{D_{\textup{L}}}{2} \\
 |z_i - z_{\textup{A}}| ~ < ~ 2 h_i + \dfrac{D_{\textup{L}}}{2} 
\end{aligned}
\end{equation}
where $(x_{\textup{A}}, y_{\textup{A}}, z_{\textup{A}})$ are the coordinates of the cube geometrical center and $D_{\textup{L}}$ is its side length.
The tree walking keeps on by opening the further cubes according to the last rule, until the single particles are reached and the neighbourhood is thus determined.
At the end of the walk, a temporary value of density $\rho_i$ is calculated.
If the number of neighbours encountered differs from the one expected (with the difference exceeding the tolerance number), $h_i$ is updated according to the formula \eqref{eq_sph_2_03}.
Then, a new tree walk is performed, and a new value for the density is computed.
The tree walk and the $h_i$, $\rho_i$ updating are executed cyclically until the neighbours number converges to the desired value.
During the tree walk, the various quantities  $P_i$, $\Omega_i$, $\zeta_i$, $f$, $\vec{\nabla} \cdot \vec{v_i}$ and $\vec{\nabla} \times \vec{v_i}$ are computed at the same time together with the density.

After such a preliminary phase, the acceleration $\vec{a}$ and the thermal energy rate $\dot{u}$ can be computed by performing, for each particle, a further tree walk is performed.
During the walk, the hydrodynamics contribution to the acceleration and the gravity field are computed at the same time.
For each box, before applying the criterion \eqref{eq_sph_2} (or the \ref{eq_sph_23}) the code primarily checks the condition \eqref{eq_sph_24} to discriminate cubes that could contain some neighbour particles for the SPH interpolation.
If it is satisfied, then the box is opened, otherwise the particles inside will not contribute at all to hydrodynamics.
After such primary check, the algorithm makes a secondary control by applying the opening criterion \eqref{eq_sph_2} (or the \ref{eq_sph_23}) to decide if the multipole approximation can be applied to the Newtonian field.
As a result, practically, a generic $i-th$ particle will interact with its local SPH neighbourhood following the equations \eqref{eq_sph_18a} and \eqref{eq_sph_19a}, experiencing the Newtonian force through a direct particle-to-particle interaction.
The remaining particles, lying outside the SPH domain, will contribute to the accelerations with or without multipole approximation, according to the basic criteria \eqref{eq_sph_2} or \eqref{eq_sph_23}.

During the computation of acceleration terms, $\vec{a}_i$, the neighbour domain searching process may suffer of some technical issues.
If we give a look at the interpolation of the density (Eq. \ref{eq_hrho}) together with the interpolations \eqref{eq_omega} \eqref{eq_zeta}, \eqref{eq_3_21_tesi} and \eqref{eq_3_22_tesi} we find them rather straightforward to be performed since they require a sum over a domain only dependent on the local $h_i$.
On the other hand, the equation \eqref{eq_sph_18a} contains sums extended over a more complex region, since the $W$ and $g_{soft}$ functions depend not only on the local smoothing length, but also on the $h_j$, which characterizes the domain extension of a surrounding particle $j$.
The condition \eqref{eq_sph_24} guarantees that we can find all the nearest $j-th$ particles with a smoothing length $h_j \leq h_i$.
Nevertheless, there exist also particles having $h_j > h_i$ with $2h_j$ smaller than the mutual distance $r_{ij}$.
They give a non-null contribution to the $i-th$ acceleration, but they are excluded from the neighbour search and the code doesn't take them into account, leading to an incorrect implementation of the SPH equations.
It is rather easy to fix such problem when the algorithm works with a uniform time-step.
In fact, \eqref{eq_sph_18a} contains couples of terms symmetric with respect to the swap of indexes $i-j$; moreover, each couple has one term dependent only on $h_i$, while the other depends only on $h_j$.
So, when particle $i$ walks along the tree and finds particle $j$ within the interpolation domain $2h_i$, the code may add to $a_i$ the quantity $-\frac{1}{2} g_{soft}(r,h_i)  {\vec{r}}/{r} - (\chi_{Ai}+\chi_{Bi}+\chi_{Ci}) \vec{\nabla_i} W(h_i)$, and may add to $a_j$ the same quantity with the opposite sign.
Here we have used the following expressions, for a generic particle of index $i$:
\begin{equation}\label{eq_chi_factors}
\begin{aligned}
\chi_{Ai} = m_i \frac{G}{2}  \frac{\zeta_i}{\Omega_i} \\
\chi_{Bi} = m_i \frac{P_i}{ \rho_i^2 \Omega_i}  \\
\chi_{Ci} = \frac{1}{2} m_i \Pi_{ij}
\end{aligned} 
\end{equation}

When, in its turn, it is the particle $j$ that makes the tree walk, the code can find the particle $i$ within the domain $2h_j$.
In that case, the quantity $-\frac{1}{2} g_{soft}(r,h_j)  {\vec{r}}/{r} - (\chi_{Aj}+\chi_{Bj}+\chi_{Cj}) \vec{\nabla_j} W(h_j)$ is added to $a_j$ and, the same quantity with opposite sign is added to $a_i$.

Such a technique cannot be applied in cases in which an individual time-step is assigned to each particle, since during the force computation some particles are inactive and the symmetry of the equation \eqref{eq_sph_18a} is thus broken. 
Supposing the particle $i$ active and the particle $j$ being inactive and lying outside the radius $2h_i$, to take into account the contributions of $j$ the algorithm adds to $a_i$ the full set of terms in equation \eqref{eq_sph_18a}, and increases the neighbour searching radius.
So, only during the second-step tree walk, the code applies the criterion \eqref{eq_sph_24} with a slight modification of the radius by using a value larger than $2h_i$, estimated as follows.
The algorithm calculates a local interpolation of the gradient of the softening length, estimated as:
\begin{equation}\label{eq_grad_h} 
\vec{\nabla} h_i = \dfrac{\partial h_i}{\partial \rho_i} \vec{\nabla}  \rho_i = - \dfrac{1}{3} \dfrac{h_i}{\rho_i} \sum\limits_j m_j \vec{\nabla} W\left( r_{ij},h_i\right)
\end{equation}
which is computed in the same loop as that used for the density. 
Then, in place of $h_i$, the code uses the quantity $h_i \cdot \max \left( \delta_h h_i^{[MAX]}/h_i ~,~ 1 + |\vec{\nabla}h_i|\right)$, with $\delta_h$ a suitable constant that we set to $1.3$.
The quantity $h_i^{[MAX]}$ represents an estimation of the maximum local value of the softening length.
It tries to probe all the neighbours $j$, including the points outside the radius $2h_i$ that can make an SPH interaction with $i$, i.e such that $h_j>h_i \vee 2h_j>r_{ij}$.
Hence, in an earlier time-step, when the particle $j$ is active, its length $h_j$ is checked if it is greater or not than the local maximum length around $i$.
Such a scheme increases the CPU-time  by increasing the direct point-to-point interactions.
Nevertheless, it involves just local interactions, and as N increases the CPU efforts become less and less relevant if compared with the non-local gravity computations performed with the tree scheme.
Such a method  represents empirically a correct technique to find the neighbours of a point in case of individual time-steps, despite it cannot be mathematically proven  that it is able to find the 100\% of the effective $j$ particles in all the feasible density configurations.
As a matter of fact, we made a series of tests, even employing extreme density contrasts like in the case of a Sedov-Taylor blast wave profile, and found out that the 100\% of the effective neighbour particles were found whenever $\delta_h$ was above 1.2.
\ \\

\subsection{Tree-code memory optimization}\label{par_2_3_1}
The development of faster and faster RAM memory architectures during the last years looks promising. 
The CPU clock speed is not anymore the only one parameter which affects substantially the performance of a program. 
Indeed, in order to write efficient algorithms, one has both to minimize the amount of CPU operations, and to suitably store the data in memory to be red as fast as possible. 
Moreover, modern architectures support a huge amount of \textit{Cache Memory}, whose order of magnitude ranges from 1 MB to $10^2$ MB.
The Cache represents a refined and fast-readable level of memory close to the CPU \citep[for an exhaustive essay on cache memory architectures, it can be read the book written by ][]{handy98}.
Each time a system needs to manipulate some data, a little chunk of memory, in which the relative variables are contained, is gathered from the RAM and copied onto the cache, from which the processor can operate very quickly. 
For such reasons, the efficiency of an algorithm is strictly connect to its ability to make several consecutive operations using variables stored very close in memory.
In this way, the data are loaded once onto the cache and the CPU can make directly the computations by minimizing the memory traffic with the RAM.
For such purposes, it is straightforward to write efficient tree codes and, more specifically for our purposes, SPH algorithms, provided that the informations related to both the particles and the cubes are suitably ordered inside the RAM.
\cite{barnes86}, remarked that the particles should be ordered in memory according to their Cartesian positions coordinates.
The sorting criterion should accurately follow the same arrangement in which the boxes are mapped inside the RAM. 
Following such prescription, the closer the particles, the closer the areas of memory in which they are stored, and the closer the informations of the related cubes.

\subsection{OMP parallelization}
By exploiting the \omp ~libraries designed for Fortran-90, the code can even run with shared memory multi-core CPUs \citep[see][for a modern treatment of the \ompsemplice~paradigm]{chapman&al07}.
We implemented a parallelization both for the density evaluation routine and for the acceleration field routine, so that different threads perform calculations on different particles.
Given a generic point $i$, it may happen that the algorithm needs to update the quantities $\vec{a}_i$ and  $\dot{u}_i$ by summing two or more different contributions at the same time.
The so-called \textit{Data-Race} problem arises: two or more threads get access and update the same memory location at the same time, leading errors in storing the correct values. 
To overcome such issues, each thread is provided with a private array to store partial values of acceleration and internal energy rate.
After the tree descent, such temporary arrays are summed. 
  
\ \\  

\end{document}